\documentclass[twocolumn]{iopart}

 \usepackage{graphicx} 
\usepackage[ansinew,latin1,utf8]{inputenc}
 \usepackage{enumerate}
\usepackage{amssymb,amsfonts} 
  \usepackage{xcolor} 

\usepackage{mathrsfs}
\usepackage[square,sort&compress,comma,numbers]{natbib}
\usepackage{hyperref}
\hypersetup{colorlinks=true, citecolor=blue, urlcolor=blue, linkcolor=blue}
\usepackage[english]{babel}
\usepackage[utf8]{inputenc}
\usepackage{graphicx}
\usepackage[caption=false]{subfig}
\usepackage{amsfonts}
\usepackage{amssymb}
\usepackage{doi}

\newcommand{\1}{\begin{equation}}
\newcommand{\2}{\end{equation}}
\newcommand{\ea}{\begin{eqnarray}} 
\newcommand{\ee}{\end{eqnarray}}
\newcommand{\4}[2]{{\frac{#1}{#2}}}

\newcommand{\Sum}[2]{{\sum\limits_{#1}^{#2}}}

\newcommand{\erw}[1]{\left\langle\, #1\,\right\rangle} 

\definecolor{dgreen}{rgb}{0,0.7,0}
\newcommand{\vs}{{\vskip 0.8cm \noindent}}
\newcommand{\vsb}{{\vskip 0.3cm \noindent}}
\renewcommand{\contentsname}{Inhaltsverzeichnis}

\begin{document}

\topical[Interactions in Active Colloids]{
\\Interactions in Active Colloids}

\author{B. Liebchen, A. K. Mukhopadhyay}
\address{Institute of Condensed Matter Physics, Technische Universit{\"a}t Darmstadt, 64289 Darmstadt, Germany}
\ead{benno.liebchen@pkm.tu-darmstadt.de}
\vspace{10pt}

\begin{abstract}
	The past two decades have seen a remarkable progress in the development of 
	synthetic colloidal agents which are 
	capable of creating directed motion in an unbiased environment at the microscale. 
	These self-propelling particles are often praised 
	for their enormous potential to self-organize into dynamic nonequilibrium structures 
	such as living clusters, synchronized super-rotor structures or self-propelling molecules featuring a complexity which is rarely found outside of the living world. 
	However, the 
	precise mechanisms underlying the 
	formation and dynamics 
	of many of these structures are still barely understood, which 
	is likely to hinge on the gaps in our understanding of how active colloids interact. 
	In particular, besides showing comparatively short-ranged interactions which are well known from passive colloids (Van der Waals, electrostatic etc.), active colloids show novel hydrodynamic interactions as well as phoretic and substrate-mediated ``osmotic'' cross-interactions which hinge on the action of the
	phoretic field gradients which are induced by the colloids on other colloids in the system. 
	The present article discusses 
	the complexity and the intriguing properties of these interactions
	which in general are long-ranged, non-instantaneous, non-pairwise and non-reciprocal
	and which may serve as key ingredients for the design of future nonequilibrium colloidal materials. 
	Besides providing a brief overview on the state of the art of our understanding of these interactions a key aim of this review is to 
	emphasize open key questions and corresponding open challenges. 
\end{abstract}

\tableofcontents

\section{Introduction: Cross-interactions in autophoretic Janus colloids}Colloidal suspensions play an important role in our everyday life, where they occur e.g. as milk, paint, blood or corn starch as well as in various high-tech industrial applications.
The macroscopic properties of these suspensions are largely determined by their intrinsic structure and dynamics which in turn largely hinge on the cross-interactions of the individual ``colloidal particles'' within a suspension. Accordingly, these interactions have been explored in detail for many decades, which has led to a substantial knowledge filling specialized books 
\cite{verwey1948theory,derjaguin1987surface,langbein1974theory,mahanty1976dispersion,napper1983polymeric,farinato1999colloid,parsegian2005van,butt2010surface,lekkerkerker2011depletion,israelachvili2015intermolecular} and significant parts of many general textbooks on colloids
\cite{russel1991colloidal,evans1999colloidal,hunter2001foundations,morrison2002colloidal,lyklema2005fundamentals,de2005food,everett2007basic,berg2010introduction}. %
\vsb 
More recently, active colloidal suspensions \cite{paxton2004catalytic,marchetti2013hydrodynamics,bechinger2016active} have been developed which comprise micron sized particles (commonly also referred to as ``colloids'') which are capable of self-propulsion. That is, besides moving stochastically (Brownian motion), these particles also move ballistically on a characteristic persistence time $\tau_p=1/D_r$ where $D_r$ is the rotational diffusion coefficient and $t_p$ is on the order of seconds in many experiments. 
(e.g. temperature, chemical concentration or electric potential) across their own surface, e.g. by catalyzing a certain chemical reaction on part of their surface,  
to which they respond by diffusio-, thermo- or electrophoresis or a combination thereof. 
That is, in the simplest case the gradients in the phoretic field create forces on the solvent within a thin interfacial layer of a colloid only, leading to a 
(diffusio-, thermo- or electro-osmotic) solvent flow across the colloid's boundary layer. This flow in turn induces a 
net motion of the colloid in the direction opposite to the surface-averaged solvent flow, such that the overall momentum of the colloid and the solvent is conserved. 
\vsb 
\begin{figure}
	\begin{center}
		\includegraphics[width=1.0\textwidth]{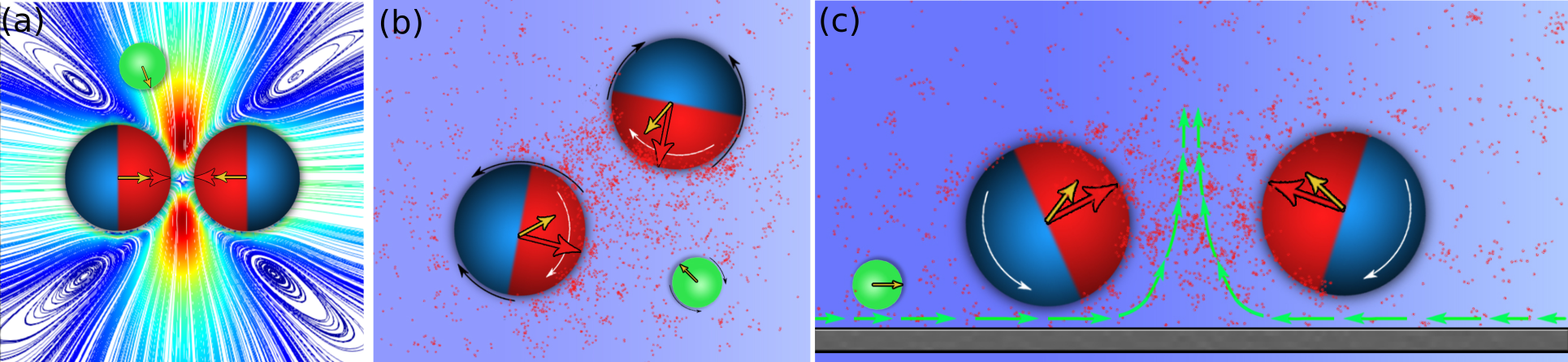}
		\caption{Schematic representations of hydrodynamic (a), phoretic (b), osmotic (c)
			cross-interactions among Janus colloids and tracers.
			The Janus colloids induce gradients in a phoretic field e.g. by
			catalyzing the production of solutes (small red particles) on their
			catalytic hemisphere (red) but not at their inactive hemisphere (blue).
			The gradient as induced by each Janus colloid induces a solvent flow
			along the colloids' own surfaces leading to self-propulsion (red arrows)
			and to a perturbation of the solvent flow field which in turn advects
			other Janus colloids and tracers (green) in the system leading to
			hydrodynamic cross-interactions (a).
			The same gradients which lead to self-propulsion decay slowly in space			and act induce a solvent flow in the interfacial layer
			of other colloids and of external walls.
			The former induces a phoretic motion of other colloids, leading to
			phoretic-cross interactions (b), whereas the latter flow advects other			particles parallely to the substrate leading to osmotic
			cross-interactions (c).
			These interactions feature a number of remarkable properties and can in
			particular be non-reciprocal (Janus colloids phoretically attract
			tracers but not vice versa), non-instantaneous (the phoretic field does
			not instantaneously reach its steady state) and non-pairwise (colloids			displace the phoretic field due to other ones).}
		\label{all_int}
	\end{center}
\end{figure}
As a remarkable byproduct of their self-propulsion mechanism, 
autophoretic colloidal microswimmers show spectacular interactions which crucially differ from those occurring in passive colloids. 
While (force free) passive colloids typically interact over distances of up to a few nanometers, active colloids show much longer ranged interactions which may be truly long-ranged or may be screened at scales far beyond the Debye length. These interactions 
feature a number of remarkable properties: they are in general non-instantaneous, non-pairwise, non-reciprocal and non-isotropic and can even change from attractive to repulsive as the interparticle distance or the relative orientation of the involved colloids changes.
While some of these features are of course impossible for fundamental interactions, such as e.g. for those acting between the individual solvent molecules and the atoms which make up the colloids, they can be observed in simulations and experiments at the level of the colloids show various effective interactions which are all mediated by the solvent and are influenced by confinement. These interactions represent the focus of this review. 
Before discussing the present understanding of these interactions in detail in the following sections, let us briefly describe 
their origin and introduce some terminology. 
\begin{enumerate}[i)]
	\item \underline{Hydrodynamic interactions:} 
	\\These interactions are caused by the action of the solvent flow due an active particle on other particles in the system. In microswimmers the flow field perturbations can either occur through particle-shape deformations as relevant for many biological microswimmers or by inducing a gradient in a phoretic field (often temperature, chemical concentration or electric potential) causing a flow within the particles \emph{own} interfacial layer as relevant for autophoretic active colloids.  
	This flow decays slowly in space and advects other colloids in the system, leading to hydrodynamic cross-interactions 
	among force-free microswimmers, which fundamentally differ from the 
	hydrodynamic interactions in (colloidal) particles experiencing a net force (see Fig.~\ref{all_int}(a)). In far-field, the strength of hydrodynamic interactions in microswimmers typically scales as $1/r^2$ with the interparticle distance $r$, or sometimes as $1/r^3$ in far field, rather than as $1/r$ as e.g. for sedimenting and other externally forced colloids. 	In addition, microswimmers can also experience hydrodynamic self-interactions which occur when the flow field due to an individual microswimmer is influenced by interfaces or walls and acts back on that microswimmer itself.
	Note that for swimmers comprising several body-parts hydrodynamic interactions of course occur also between these parts and are sometimes also referred to as ``self-interactions''. 
	\item \underline{Phoretic interactions:}
	\\Autophoretic self-propulsion hinges on the action of a gradient in a phoretic field on the solvent in the vicinity of the particle which induces this gradient (see Fig.~\ref{all_int}(b)).  
	Phoretic (cross-)interactions are based on the action of gradients in the same phoretic field, which decays slowly in space (ideally as $1/r$ with the distance $r$ from the colloid), on the fluid in the interfacial layer of other particles in the system. 
	This leads to a phoretic motion of a test colloid towards or away from a second one, the strength of which typically scales as $1/r^2$ with the interparticle distance $r$ in far-field. 
	Similarly as for hydrodynamic interactions, the presence of walls, interfaces or external potentials deforming (or displacing) the phoretic field due to an active colloid leads 
	to additional phoretic self-interactions (and modifies phoretic cross-interactions). 
	\\Effectively, phoretic (cross-)interactions resemble chemical-interactions (based on chemotaxis or quorum sensing)
	in microorganisms \cite{eisenbach2007chemotaxis} which self-produce attractants (or repellents) to which other cells respond, as seen e.g. in Dictyostelium cells which are attracted by self-produced cAMP \cite{gerisch1982chemotaxis,van1982signal,devreotes1988chemotaxis,nichols2015chemotaxis,king2009chemotaxis} and in E.coli bacteria
	\cite{mesibov1972chemotaxis,tso1974negative,berg2008coli}
	which respond to excreted aspartate leading to pattern formation \cite{budrene1991complex,budrene1995dynamics} and autoinducer II signals
	\cite{laganenka2016chemotaxis,laganenka2018autoinducer}) leading to a collapse similar as in Dictyostelium. 
	This analogy between phoretic interactions and chemical interactions \cite{liebchen2018synthetic} -  or the the way insects and microorganisms communicate - 
	has recently been praised as a unique advantage of chemically powered active colloids
	\cite{wang2020practical}.
	\item \underline{Osmotic interactions:} 
	\\In the presence of external walls or other confinement, the gradients of the phoretic field due to an (active) particle also create forces in the interfacial layers of the walls leading to an osmotic flow along these walls. These flows can advect other particles in the system leading to wall-induced cross-interactions 
	to which we will refer to as \emph{``osmotic cross-interactions''} (see Fig.~\ref{all_int}(c)).\footnote{We use this term to distinguish these interactions from phoretic and hydrodynamic interactions. So far, different parts of the literature have interpreted these osmotic cross-interactions as 
		phoretic interactions (because they are cross-interactions caused by the phoretic field due to a phoretically active (Janus) particle) or as hydrodynamic interactions, because they ultimately cause a flow which advects other particles (as for hydrodynamic interactions).}
	Thus, phoretic cross-interactions are based on the creation of osmotic flows along the inner surfaces of the system (surfaces of other colloids), whereas osmotic cross-interactions are based on the creation of osmotic flows along the outer walls of the system (boundaries). 
	Besides osmotic cross-interactions, there are also osmotic self-interactions because the osmotic flows along external walls which are caused by the phoretic field gradients due to a certain Janus colloid act back onto that colloid, 
	leading in particular to a reorientation of the colloid. 
	Osmotic interactions can be tuned through the properties of the substrate e.g. by the choice of the substrate material or by functionalizing it e.g. with polymer brushes. 
\end{enumerate}

\section{Phoretic and osmotic interactions\label{phoros}}
\subsection{Chemotaxis and phoretic interactions} 
Synthetic active colloids can not only self-propel like insects and bacteria
but they also mimic 
how they communicate. 
Just like their biological counterparts, they show chemical (or more generally phoretic) interactions which lead to a rich panorama of (dynamical) structures 
including clusters, swarms, waves and even superstructures resembling aspects of the complexity and order which we encounter throughout the biological world. 
Interestingly, this analogy between biological and synthetic microswimmers shows up also formally, albeit the underlying ``sensory'' mechanisms of course fundamentally differ from each other. 
To stress the analogy between chemical communication in microorganisms and phoretic interactions in synthetic active colloids, we begin with an effective minimal model
serving as a generic starting
point to discuss interactions in 
biological active colloids
like bacteria \cite{poon2013clarkia} and synthetic ones.
\vsb
\underline{Minimal model for chemically interacting particles:}
\\Let us consider an ensemble of $N$ overdamped Brownian point particles representing microorganisms or colloids responding to 
the gradient of a concentration field $c(\vec r,t)$ either 
by chemotaxis (microorganisms) \cite{eisenbach2007chemotaxis,tindall2008overview,liebchen2020modeling} or by (diffusio)phoresis (synthetic colloids) \cite{anderson1989colloid}. The 
position of the $i$-th particle evolves in time as
\1
\dot {\vec r}_i(t) = \beta \nabla c(\vec r_i,t) + \sqrt{2D}\vec \eta_i(t) \label{langevin}
\2
where $D$ is the diffusion coefficient representing the effect of Brownian motion and where a positive sensitivity coefficient $\beta>0$ represents a tendency of the particles to move up the local gradient of the concentration field and for $\beta<0$ they tend to move down the gradient.
\\If the particles self-produce the concentration field (e.g. by secreting chemoattractant or -repellent molecules, autoinducers or by catalyzing a chemical reaction on their surface) with a rate $k_0$ we obtain a two-way coupling between the particle dynamics and the dynamics of the concentration field, which evolve according to the 
diffusion equation for the chemical/phoretic field:  
\1\dot c(\vec r,t)=D_c \nabla^2 c(\vec r,t) + k_0 \Sum{i=1}{N}\delta(\vec r - \vec r_i(t)) -k_d c(\vec r,t) \label{chemdiff}\2
Here, $D_c$ is the diffusion coefficient of the relevant field and $k_d$ is the degradation rate of the field which may be zero, which may apply e.g. to self-thermophoretic microswimmers where $c$ is the temperature field, or may take a finite value, e.g. if the environment is enzymatically active or if the relevant chemicals which emerge in surface-catalyzed chemical reactions 
participate in association-dissociation or other bulk reactions.
Similar models for 
the collective dynamics of active particles (or ``active walkers'') based on Langevin-equations have been formulated 
e.g. in refs. \cite{schweitzer1994clustering,ben1997snowflake,romanczuk2008beyond} (see also the review \cite{romanczuk2012active}). Ref. \cite{meyer2014active} in particular discusses the case of a non-constant sensitivity $\beta(c)$ leading to a range of patterns. 
Earlier models based on random walk models include refs. \cite{patlak1953random,  alt1980biased,stevens1997aggregation,othmer2000diffusion,othmer2002diffusion}; 
see also the review \cite{horstmann20031970} for further references.


\vsb\underline{Autochemotaxis:}
\\For $N=1$, Eqs. (\ref{langevin},\ref{chemdiff}) describe an auto-chemotactic agent interacting with its self-produced (or self-consumed) chemical field. While the concentration field due to a stationary agent at position $\vec r_0$ is isotropic and reads (if assuming that the production is instantaneously switched on at time zero) \cite{liebchen2020modeling,Sengupta_PRE_Chemotactic_Predatorprey_2011}:
\1
c(\vec{r}, t)=\int_{0}^{t} \mathrm{~d} t^{\prime} \frac{k_0}{\left(4 \pi D_{c}\left|t-t^{\prime}\right|\right)^{\frac{d}{2}}} \exp \left(-\frac{\left(\vec{r}-\vec{r}_0\right)^{2}}{4 D_{c}\left|t-t^{\prime}\right|}-k_d\left|t-t^{\prime}\right|\right), \label{tdfield}
\2
a motile agent leaves a ``trail'' in the concentration field behind and creates a nonuniform field distribution which biases its own motion. This self-interaction is called autochemotaxis and is based on memory effects due to a non-instantaneous relaxation of the concentration field to its steady state for a given particle position. 
For $\beta>0$ this can lead to self-trapping (or self-localization) which was originally thought to be permanent (for $k_d=0$) in 1D and 2D \cite{tsori2004self} 
and has later been shown to occur only transiently (both for $k_d=0$ and $k_d>0$) such that the long-time dynamics is diffusive, with the transient trapping being longer in low dimensions \cite{grima2005strong,sengupta2009dynamics}.
A similar model for an active particle in 2D which self-propels at a constant speed (following $\dot {\vec r}_i(t)=
v_0 \vec{p}_i(t) + \sqrt{2D}\vec \eta_i(t)$ where $\hat p_i(t)=(\cos\theta_i(t),\sin\theta_i(t))$ and $\dot \theta_i=\beta \hat p_i(t) \times \nabla c(\vec r_i) $) and aligns its swimming direction with the local gradient (topotaxis \cite{eisenbach2007chemotaxis} with respect to the self-produced field) also leads to diffusive behavior in the long-time limit, even for very strong chemotactic coupling \cite{taktikos2011modeling}. 
\vsb\underline{Cross-interactions and the $1/r^2$ law:\label{sec1r2law}}
\\Two particles interact chemically even if memory effects and hence autochemotaxis is unimportant. 
This is because each of the two particles experiences a gradient due to the contribution of the other particle to the concentration field.
In 3D the steady state solution of equation (\ref{chemdiff}) for $k_d=0$ and constant $\vec r_i(t)=\vec r_i$
reads (see e.g. \cite{liebchen2020modeling} for variants of this solution and considerations in other dimensions):
\1 c(\vec r) = \4{k_0}{4\pi D_c}\Sum{i=1}{N} \4{1}{|\vec r-\vec r_i|} \2
Using (\ref{langevin}) and averaging over noise quantifies the average motion of a particle with 
position vector $\vec r_i$ due to $N-1$ other particles in the system read
\1
\dot{\vec r}_i = \beta \nabla c(\vec r_i) =  \4{\beta k_0}{4\pi D_c} \Sum{j=1,j\neq i}{N}\4{(\vec r_j-\vec r_i)}{|\vec r_j-\vec r_i|^3} \label{1r2law}
\2
Thus, for $k_d=0$ the relative velocity of one particle due to a second one decays as 
$1/r^2$ where $r$ is the interparticle distance. We call this the $1/r^2$ law and discuss below to which extend it applies to Janus colloids (section \ref{r2janus}).
Clearly, in the considered special case chemical/phoretic interactions are equivalent  to Coulomb/gravitational interactions. We can define 
an effective potential 
$ U_{\textrm{eff}}(\vec r) = \4{-\beta k_0 \gamma}{4\pi D_c}\Sum{i=1}{N}\4{1}{|\vec r - \vec r_i|}$ where $\gamma$ is the Stokes drag coefficient, leading to the 
Langevin-equations 
\1 \dot {\vec r}_i = - \4{1}{\gamma}\nabla_{\vec r_i} U_{\textrm{eff}} + \sqrt{2D}\vec \eta_i\2 
where $i=1,2..N$. Here we have eliminated the concentration field, resulting in a particle-only model. 
\\Note that unlike for Janus colloids and synthetic droplet swimmers where $\beta$ is typically constant, for microorganisms $\beta$ is often a $c$-dependent function \cite{murray2001mathematical} which in some cases leads to a different scaling of the interactions. 
The logarithmic sensing law $\beta=\beta_0/c$ with some constant $\beta_0$ \cite{keller1971traveling,murray2001mathematical} for example is 
relevant for E.coli \cite{kalinin2009logarithmic} and leads to a decay of the interactions with $1/r$. The popular
receptor law $\beta\propto 1/(\kappa+c)^2$ where $\kappa$ is a constant \cite{lapidus1976model}, in turn, 
is relevant e.g. for some (other) bacteria \cite{dahlquist1972quantitative,murray2001mathematical} and for neuron chemotaxis \cite{segev2000generic}. Here the $1/r^2$ law applies in far field, whereas for close interparticle distances the attraction converges to a constant. See 
\cite{ford1991analysis} for an overview on sensitivity functions and their relations to experimental data \cite{ford1991measurement1,ford1991measurement2}. 

\vsb 
\underline{Effective screening\label{effsc}:}
\\When $k_d > 0$ in Eq. (\ref{chemdiff}) representing e.g. a degradation in microorganisms, or an effective ``decay" of the chemical species which are relevant for phoretic interactions in autophoretic Janus colloids, due to bulk reactions (including association-dissociation reactions), the steady state solution of 
Eq.~(\ref{chemdiff}) reads \cite{liebchen2020modeling}: 
\1 
c(\vec r) = \4{k_0 {\rm e}^{-\kappa |\vec r|}}{4\pi D_c |\vec r|} \label{csol}
\2
Here $1/\kappa=1/\sqrt{\4{k_d}{D_c}}$ is an effective screening length which defines the range of the phoretic interactions in the presence of effective decay-processes of the relevant phoretic field. 
The average phoretic drift of a particle with position $\vec r_i$ due to the field due to other particles in the system then 
reads 
\1 \dot {\vec r}_i = \beta \nabla c(\vec r_i) = \4{\beta k_0}{4\pi D_c} \nabla_{\vec r_i}\4{{\rm e}^{-\kappa |\vec r_j-\vec r_i|}}{|\vec r_j-\vec r_i|} \label{yuk}\2
representing a Yukawa (or screened Coulomb) pair interaction.
Eq. (\ref{yuk}) fits well to experimental data for the time-evolution of the ensemble-averaged relative distance between active Janus-colloids and passive tracers and suggests that the screening length
might be on the order of a few particle diameters in Janus-colloids \cite{liebchen2019interactions,hauke2020clustering}. 
However, in the future a more detailed modeling of the effect of bulk reactions within a multi-species framework would be desirable to understand effective screening in more detail (see open challenges at the end of the article). 
\\A similar 
effective screening can also be caused by the ions which are created in 
association-dissociation reactions 
such as $\mathrm{H}_{2} \mathrm{O}_{2} \rightleftharpoons \mathrm{H}^{+}+\mathrm{HO}_{2}$
which are supported by polar solvents like water. Since the 
relevant ion concentrations in such reactions are determined by the system specific dynamic equilibrium condition such ions are expected to be present 
even if the surface reactions involve only neutral molecules. 
Besides allowing for an electrophoretic contribution to self-propulsion \cite{brown2014ionic,ebbens2014electrokinetic}, the presence of the ions  can result in a kind of exponential `reactive-screening' \cite{brown2017ionic,banigan2016self}, limiting the range of phoretic interactions.
\vsb
\underline{Properties of chemical interactions beyond the $1/r^2$-law:}
\\Despite the discussed analogy between chemical/phoretic interactions and Coulomb/gravitational/Yukawa interactions it is important to keep in mind that the former ones are effective interactions, which allows for a range of unusual properties: 
\begin{enumerate}[i)]
	\item Non-instantaneous: The concentration field evolves in time such that chemical/phoretic interactions are in general not instantaneous. For a single point particle this leads to autochemotaxis and for $N>1$ it additionally leads to non-instantaneous cross-interactions.
	\item Non-pairwise: Due to the linearity of Eq. (\ref{chemdiff}) the chemical interactions in a system of $N$ point particles are generally pairwise. However, this changes if the particles have a finite size and 
	displace the concentration field from their interior. Then, 
	the presence of a third particle displaces the concentration field which governs the pair-interaction of two particles making the chemical interactions non-pairwise.
	Thus, for particles with a finite size the $1/r^2$-law applies only in far-field. 
	\item Non-reciprocal: Unlike the fundamental Coulomb/gravitational interactions which are of course reciprocal 
	and obey Newton's action reaction principle, the coefficients controlling the contribution of a particle to the chemical/phoretic field and its response to this field are independent (i.e. there is nothing analogous to the equality of heavy and inertial mass for chemical/phoretic interactions). 
	Thus, for nonidentical particles (or for non-symmetric configurations of anisotropic particles) 
	chemical/phoretic interactions are typically nonreciprocal 
	violating the action-reaction principle (and hence momentum conservation) at the level of the particles (colloids, microorganisms). Since, the underlying microscopic interactions are of course reciprocal and any 
	net momentum which is generated at the level of the particles (colloids, microorganisms) is balanced by the surrounding medium leading e.g. to a 	counter propagating solvent flow. 
\end{enumerate}
\vsb 
\vs
\underline{The Keller-Segel continuum model:}
\\A popular generic model to study the collective behavior of chemically communicating particles is the Keller-Segel model
which has been derived from many different starting points both phenomenologically and microscopically. 
In its original form it has been derived phenomenologically by Keller and Segel (1970) \cite{keller1970initiation,keller1971model} to describe the collective behavior of 
Dictyostelium \cite{keller1970initiation} (1970) 
and of E. coli \cite{adler1966chemotaxis,adler1966effect} in particular also regarding the formation of chemotactic 
bands \cite{keller1971traveling} seen in experiments \cite{adler1966chemotaxis}.
Microscopic derivations have been performed already by 
Patlak (1953) \cite{patlak1953random}.
More recent popular works discussing microscopic approaches based on random walk models (or jump processes) to
taxis and chemotactic aggregation and microscopic derivations of the 
Keller-Segel model include 
\cite{alt1980biased,stevens1997aggregation,othmer2000diffusion,othmer2002diffusion}; see the review \cite{horstmann20031970} for more further references. 
Recently it has also been derived for quorum sensing run-and-tumble particles \cite{rein2016collective}. 
\\The Keller-Segel model can be viewed as a 
mean-field continuum model of Eqs. (\ref{langevin}), (\ref{chemdiff}) as we will discuss in the following. Using 
Dean's approach \cite{dean1996langevin} and following ref. \cite{chavanis2010stochastic} one readily obtains the  ``stochastic Keller-Segel model'' \cite{chavanis2010stochastic}
\1
\begin{array}{l}
	\frac{\partial \rho}{\partial t}=-\beta \nabla \cdot\left(\rho \nabla c\right)+D \nabla^2 \rho +  \nabla \cdot\left(\sqrt{2 D \rho} \vec{R}\right) \\
	\frac{\partial c}{\partial t}=D_{c} \nabla^2 c-k_d c+k_0 \rho \label{stochKS}
\end{array}
\2
where $\rho=\Sum{i=1}{N}\delta(\vec r - \vec r_i(t))$ and where $\vec R(\vec r,t)$ represents a Gaussian white noise field with zero mean and unit variance. Notice that the $N$-particle density is subject to multiplicative noise which can play an important role e.g. when calculating correlation functions. 
One approach to do this has been developed in  
ref. \cite{newman2004many} which also starts with Eqs.~(\ref{langevin},\ref{chemdiff}) and develops a many-body theory and a 
diagrammatic framework to account for statistical correlations. Notably, unlike for many 
applications of diagrammatic methods in 
condensed matter physics and quantum field theory, chemical interactions lead to memory effects, because the concentration field (\ref{chemdiff}) which mediates the interactions between the particles evolves in time. 
\\The standard, or minimal Keller-Segel model can be readily obtained by ignoring the multiplicative noise term in Eqs. (\ref{stochKS})
and interpreting $\rho(\vec r,t)$ as a coarse grained (mean) density field. Now  
choosing length and time units as $x_u=\sqrt{D_c/k_d}$ and $t_u=1/k_d$ and defining $u:=k_0 \rho/k_d$ (for constant 
$\beta,D,D_c$) we directly obtain the following continuum model, which is called the 
(minimal) Keller-Segel model \cite{tindall2008overview}: 
\ea
\dot u &=& - \chi \nabla \cdot (u \nabla c) + \mathcal{D} \nabla^2 u \\\dot c &=& \nabla^2 c + u - c \label{KSm}
\ee
Here we have introduced the dimensionless numbers 
$\chi:=\beta/D_c$ and $\mathcal{D}=D/D_c$. 
This is an advection-reaction-diffusion system comprising an 
advection-diffusion equation for the (average) density which is coupled to  
a reaction-diffusion equation describing the dynamics of the chemical/phoretic field. 
\\Specifically, if the chemical dynamics adiabatically follows that of the agents, i.e. for $\dot c \approx 0$, and in the absence of degradation ($k_d=0$), Eq. (\ref{KSm}) reduces to the Smoluchowski-Poisson system of overdamped self-gravitating Brownian particles in mean field \cite{chavanis2010stochastic,chavanis2007kinetic,lushnikov2010critical}.
\\One obvious solution of Eqs. (\ref{KSm}) is $(u,c)=(u_0,u_0)$ representing the disordered uniform state. When $\chi$ is sufficiently large compared to $\mathcal{D}$ this uniform (disordered) state can become unstable such that small initial fluctuations can amplify as we will discuss in the next section.

\begin{figure}
	\begin{center}
		\includegraphics[width=0.50\textwidth]{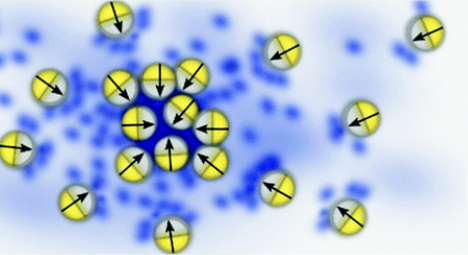}
		\caption{Schematic illustration of the aggregation of chemoattractive Janus colloids following the Keller-Segel instability. Adapted with permission from ref. \cite{liebchen2016pattern}}
		\label{liebchen2016}
	\end{center}
\end{figure}

\begin{figure}
	\begin{center}
		\includegraphics[width=1.0\textwidth]{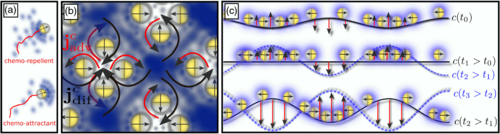}
		\caption{(a) Schematic representation of colloidal Janus particles (spheres) half-coated with a catalytic material (yellow) that produces a chemical species (blue); the colloids self-propel by autophoresis in the resulting gradient. 
		(b) Schematic of the anisotropy-induced Janus instability leading to clusters of chemorepulsive Janus colloids with a self-limiting size. (c) Schematic of the delay-induced instability, which can lead to traveling wave patterns in ensembles of chemorepulsive Janus colloids. Figure reproduced with permission 
		from \cite{liebchen2015clustering}.}
		\label{liebchen2015}
	\end{center}
\end{figure}

\vsb 
\underline{Aggregation (Chemotactic collapse):}
\\To understand structure formation in the Keller-Segel model, one can perform 
a linear stability analysis of the uniform state $(u,c)=(u_0,u_0)$ by making the Ansatz $u(t)=u_0+\delta u(t)$, $c(t)=c_0+\delta c(t)$, linearizing (\ref{KSm}) in $\delta u,\delta c$ and solving the resulting equations in Fourier space. This reveals a long-wavelength linear 
instability, i.e. an exponential growth of $\delta u, \delta c$ at early times and a band of sufficiently long wavelength if  
$u_0 \chi/\mathcal{D}>1$ (or $\beta k_0 \rho_0/(D k_d)>1$). This instability is called the Keller-Segel instability and leads to aggregation of the chemotactic agents (see Fig.~\ref{liebchen2016} for a schematic illustration of the Keller-Segel instability for chemoattractive Janus colloids). (Note that the diffusion coefficient of the chemical does not occur in the instability criterion.)
The instability is 
caused by a positive feedback loop: 
a small fluctuation-induced initial accumulation of particles 
leads to an enhanced local production of the concentration field and hence of the steepness of the gradient, which in turn leads to an enhanced chemoattraction of other particles. The latter further enhances the local particle density, further enhancing the 
steepness of the gradient and so on. Following the calculated instability criterion, this feedback loop is effective if if the production of the attractant, the cell density and the sensitivity are large enough and the diffusivity of the attractant as well as the decay rate are sufficiently small. 
\\The dynamics of the Keller-Segel model at later times (after fluctuations have grown beyond the linear regime), can be explored e.g. based on 
numerical simulations starting with the uniform solution and additional small fluctuations. The simulations reveal the emergence of clusters at short time, which grow (or coarsen) as time evolves (Ostwald ripening) leading to a 
``chemotactic collapse'' for $t\to \infty$ \cite{herrero1996chemotactic}.This aggregation closely resembles the behavior of Dictyostelium discoideum cells which start producing cAMP signaling molecules when starving \cite{gerisch1982chemotaxis,levine1998dynamics} as well as the aggregation of E.coli bacteria interacting via self-produced autoinducers \cite{laganenka2016chemotaxis}
as well as of chemically interacting leukocytes at infectious tissue parts (in vivo) modeled in \cite{lauffenburger1979effects} or by bacteria in a densiometric assay \cite{dahlquist1972quantitative,nossal1973analysis}. 
The same aggregation mechanism is also relevant for the commonly observed aggregation of autophoretic Janus colloids at low density \cite{theurkauff2012dynamic,palacci2013living}; however, the clusters which occur in these systems often do not coarsen beyond a certain size but dynamically break up, which is probably due to the more complicated interactions in these systems \cite{liebchen2019interactions}, which we will discuss in more detail below.
\vsb 
\underline{Generalizations of the Keller-Segel model:}
\\The Keller-Segel model has been generalized in many ways. We refer the reader to the reviews \cite{horstmann20031970,tindall2008overview,hillen2009user,painter2019mathematical,arumugam2021keller} for an overview. The very recent review 
\cite{arumugam2021keller} in particular contains a long list of the existing variants of the Keller-Segel model and corresponding references.
\\To describe chemical interactions in Janus particles with a catalytic and an inert hemisphere, the  
Keller-Segel model has been phenomenologically generalized in ref. 
 \cite{liebchen2015clustering}. Unlike the minimal Keller-Segel model this variant accounts for the asymmetry of the particles which allows them to align their 
self-propulsion direction with the gradient of the phoretic field and leads to 
an asymmetric production of the latter. 
This model predicts a phase diagram well beyond that of the minimal Keller-Segel model: besides the chemotactic collapse
which occurs for sufficiently strong chemoattraction, the model predicts two new phases occurring when the chemical interactions are repulsive. The first one is caused by the anisotropic production of the phoretic field and has therefore been called the Janus instability. It 
leads to particle clusters which do not coarsen but have a self-limiting size. Schematically, one can imagine clusters of Janus colloids creating a shell of chemicals surrounding them. These shells keep the chemorepulsive colloids in the cluster together and other Janus colloids away from the cluster (see Fig.~\ref{liebchen2015}). 
The 
second instability is caused by memory effects and has therefore been called the delay-induced instability. It 
leads (in combination with the anisotropic production) to traveling waves, traveling lattices or continuously evolving amorphous patterns (see Fig.~\ref{liebchen2015}). 

Another very recent generalization of the Keller-Segel model which accounts 
for effects of the particle orientation (``polarity``) on the response of chemotactic agents (cells) to the collectively produced concentration field has been introduced in 
ref. \cite{mahdisoltani2021nonequilibrium}. This work has developed a many-body theory  
leading to a new nonlinear coupling term in the stochastic Keller-Segel model (\ref{stochKS}) (in the instantaneous limit $\dot c\to 0$).
A dynamical renormalization group approach predicts critical exponents revealing superdiffusive density fluctuations and non-Poissonian number fluctuations. 
\\Another recent generalization of the Keller-Segel model has been provided in ref.~\cite{gelimson2015collective} 
which has studied the combined role of a logistic growth of the cell density and non-number conserving fluctuations through 
a dynamical renormalization group analysis yielding critical exponents suggesting the presence of long-range correlations and anomalous diffusion.

\subsection{Far-field interactions in Janus colloids\label{r2janus}}
We have seen above that Eqs. (\ref{langevin},\ref{chemdiff}) lead to long-range $1/r^2$ Coulomb/gravitational interactions between chemically interacting point particles 
for $k_d=0$ and to Screened Coulomb/Yukawa interactions for $k_d\neq 0$. 
These interactions provide a useful starting point to describe the far-field interactions in a broad range of systems, ranging from 
chemically interacting 
microorganisms with a suitable sensitivity function and ``all reacting and dissolving particles" \cite{moerman2017solute} such as various synthetic droplet swimmers to 
autophoretic Janus colloids. 
In this section, we focus on the latter and discuss in more detail when this simple form of the phoretic interactions applies, quantify their strength and discuss additional alignment effects occurring for Janus colloids with a nonuniform surface mobility. 
\vsb
\underline{When does the $1/r^2$-law (not) apply to Janus colloids?}
\\The simple $1/r^2$ law can be expected to hold true only 
\begin{enumerate}
 \item for large interparticle distances $r\gg R$ (far-field), where $R$ is the particle radius. This allows describing the particles as point sources and ignoring the impact of their non-uniform surface mobility for the response to the concentration field.
\item when phoretic interactions can be described by one effective phoretic field. This condition is met by self-thermophoretic colloidal microswimmers \cite{jiang2010active,kroy2016hot} where the temperature field is the relevant phoretic field. For diffusiophoretic active colloids at least one fuel and one product species is relevant \cite{saha2014clusters} and the usage of one effective phoretic field to 
describe their combined effect on the interactions of the colloids is in general an approximation. 
Finally, for many microswimmers (ionic-)diffusiophoresis, thermophoresis and electrophoresis can act together so that the usage of one effective phoretic field is clearly an approximation. 
For photo-catalytic microswimmers the relevant species are often not even known \cite{wang2020practical}. 
\item when effective screening e.g. due to bulk reactions (including association-dissociation reactions) is unimportant. If the net effect of such reactions can be modeled by a decay (or evaporation) of the effective phoretic field ($k_d\neq 0$ in Eq. \ref{chemdiff}), their impact is simple and the colloids are expected to show 
effective Yukawa (Screened-Coulomb) interactions rather than $1/r^2$-interactions (as discussed in section \ref{effsc}). 
\item If the particles move sufficiently slowly. 
For a single point source moving with a constant velocity $\vec r_1(t)={\vec v} t$, the solution of Eq. (\ref{chemdiff})
reads instead \cite{liebchen2020modeling}
\1
c(\vec{r})=\frac{\lambda_{e}}{4 \pi D_{c} r} \exp \left(-\bar{\kappa} r-\frac{\vec{v} \cdot \vec{r}}{2 D_{c}}\right)
\label{movingsource}
\2
where $\tilde{\kappa}=\sqrt{\frac{k_d}{D_{c}}+\frac{\vec{v}^{2}}{4 D_{c}^{2}}}$.
Thus, even for $k_d=0$ for a moving source the $1/r^2$ law approximately applies only at distances $r \ll D_c/v$. At larger distances the interactions are effective screened for most configurations. 
For particles which do not move with a constant speed but are sufficiently slow, one can approximately replace 
$\vec v$ by their instantaneous speed $\dot {\vec r}(t)$ in Eq. (\ref{movingsource}).
\item when 
memory-effects can be neglected i.e. when the chemical concentration field reaches its steady state configuration before the colloids move much.
This is often justified by arguing that the relevant ``chemical" species is much smaller than the colloids and therefore features a much larger diffusion coefficient making it relax quasi-instantaneously to its steady state, but the validity of this argument is not generally obvious even in far-field,
because the sensitivity coefficient of the active colloids is proportional to the chemical diffusion equation \cite{liebchen2017phoretic}.  
\item when the impact of the advection due to the self-produced flow field of the Janus colloids on the phoretic field is unimportant. 
This will be the case if the P\'eclet number of the chemical is small ${{Pe}}_c=\4{u R}{D_c} \ll 1$, which measures the relative importance of advection and diffusion of phoretic field and where $u$ is the characteristic velocity of the flow field. 
\item If the colloids are near a substrate or interface, the phoretic fields are deformed (or displaced) 
\cite{uspal2015self}. This effect is expected to merely change the coefficient of the $1/r^2$-law not its form. However, the action of a phoretic field gradient due to a Janus colloid also induces osmotic flows along the substrate which not only advect other particles in the system, leading to additional osmotic cross-interactions; see section \ref{exps} \cite{liebchen2018unraveling,thutupalli2018flow},
but which also support solute advection and can therefore have a direct impact on phoretic cross-interactions. 
\end{enumerate}
While this list seems to be rather restrictive, the $1/r^2$-law fits relatively well to 
experimental measurements of the relative velocity of remote Janus colloids (or Janus colloids and passive particles) at large interparticle distances as we will discuss in section \ref{exps}. 
\vsb
\underline{Sensitivity coefficient -- How strong are phoretic interactions?\label{sensitivity}}
\\For phoretically moving (Janus) colloids, the sensitivity coefficient 
$\beta$ is determined by the 
surface mobility of the (two hemispheres of) the colloid and is typically independent of the strength of the phoretic field (unlike for most microorganisms). 
Let us first focus on the important case of colloids where the 
relevant phoretic fields induce forces on the fluid only within a boundary layer which is thin compared to the particle size, which leads to a slip-layer outside of which the fluid is essentially force free. 
For such colloids the translational 
velocity reads $\vec v=-\langle \vec v_s(\vec r_s)\rangle $
\cite{anderson1989colloid} where $\vec v_s(\vec r_s)$ is the slip velocity at a point $\vec r_s$ on the surface (or of the slip layer). If the slip-velocity is caused by a single phoretic field $c$, it reads $\vec v_s(\vec r_s)=\mu(\vec r_s)\xi \nabla_{\parallel} c(\vec r_s)$ where $\mu(\vec r_s)$ is the phoretic surface mobility at a point $\vec r_s$ on the surface of the colloids, $\nabla_\parallel c$ is the projection of the gradient of $c$ onto the tangential plane of the colloid
and $\xi$ accounts for the deformation of the external phoretic field due to the presence of the Janus particle. 
For diffusiophoretic colloids we have 
$\xi=3/2$ and for thermophoretic ones $\xi=3\kappa_s/(2\kappa_s+\kappa_p$ with $\kappa_{s,p}$ being the heat conductivity of the solvent and the particle respectively \cite{bickel2014polarization}. 
Specifically for spherical Janus colloids with phoretic mobilities $\mu_1,\mu_2$ on their two hemispheres in an external phoretic field which does not change much on the scale of the colloid ($|\nabla c| \ll c/R$), 
it can be shown that $\beta = \xi \4{\mu_1+\mu_2}{3}$ \cite{bickel2013flow,bickel2014polarization}. 
The values of the surface mobilities $\mu_1,\mu_2$ depend on the relevant phoretic mechanism. Although there is a significant body of theory to link them with the surface properties of the colloids and the properties of the solvent 
\cite{anderson1989colloid}, unfortunately, their values are (completely) unknown in many experiments with Janus colloids. 
Fortunately however, for active Janus colloids the self-propulsion velocity depends on the same coefficients. 
For half-coated self-diffusiophoretic Janus colloids with a constant local production rate of 
$\sigma=k_0/(2\pi R^2)$, it reads \cite{golestanian2007designing}
\1
\vec{v}_{0}=-\frac{k_{0}\left(\mu_{\mathrm{1}}+\mu_{\mathrm{2}}\right)}{16 \pi \mathrm{R}^{2} \mathrm{D}_{\mathrm{c}}} \hat {e} \label{selfpr}
\2
While this result can not be used to predict the value of the self-propulsion velocity of Janus colloids if their surface mobilities are unknown, it is of great use to express $\mu_1+\mu_2$ through other parameters and to 
quantify the strength of phoretic far-field interactions. 
Using this result one readily obtains the phoretic velocity of a 
half-coated Janus-colloid due to the presence of $N$ other remote Janus colloids of the same type as \cite{liebchen2019interactions}
\1
\dot {\vec r}_i = \vec{v}(\vec r_i)=\frac{\pm 4 v_{0} R^{2}}{3} \nabla_{\vec r_i} \Sum{j=1,j\neq i}{N} \frac{1}{|\vec r_j-\vec r_i|} \label{farf}
\2
where $\pm = - \textrm{sgn}(\mu_1+\mu_2)$. This equation expresses the strength of phoretic far-field interactions among Janus colloids through the self-propulsion velocity and the particle radius, which are known in basically all experiments with active Janus colloids. The result suggests, for example, that at an interparticle distance of say $4R$ two Janus would phoretically move with a relative velocity of $v_0/6$ towards or away from each other (if the far-field expression approximately works at such a distance). More generally, one can use (\ref{farf}) to define an effective potential $U=\pm\epsilon \4{R}{r}$ where $\epsilon=\4{4}{3}\gamma v_0 R$, yielding $\4{\epsilon}{k_B T}=\4{4}{3}\4{v_0 R}{D}=Pe$ where the P\'eclet number is defined as $Pe=\4{v_0}{D_R R}$ with $D_R$ being the rotational diffusion coefficient of the colloid which is related to the translational diffusion coefficient $D=k_B T/\gamma$ via the relation $D_R=\4{3D}{4R^2}$. Since $v_0/D_r$ is the run-length of the colloids, the P\'eclet number $Pe$ can be interpreted as the number of particle radii over which a colloid moves on average before its orientation is randomized due to rotational diffusion. This shows that the strength of the phoretic cross-interaction is controlled by the same parameter which controls the self-propulsion, reflecting the intrinsic connection between these two \cite{liebchen2017phoretic}. For colloids with a radius of $R\sim 1\mu m$, a self-propulsion speed of $v_0\sim 5\mu m/s$ and a rotational diffusion coefficient of $D_R\sim 0.2/s$, the coefficient of the phoretic far-field interactions has a strength of $\epsilon\sim 25 k_B T$ which in reality might of course be weakened in some experiments by effective screening.
In addition, while these considerations provide us with a concrete idea on how strong phoretic (far-field) interactions can be, in experiments with microswimmers involving several chemical species (which are sometimes even unknown e.g. for photocatalytic swimmers \cite{wang2020practical}) phoretic fields and substrates they may of course take a significantly more complicated form. This applies in particular to active colloids not featuring a thin-boundary layer outside of which the fluid is essentially force free 
and hence the concept of a slip-velocity is not well justified \cite{samin2015self,gomez2017tuning,eloul2020reactive}. 
\\A generic approach to determine the phoretic mobilities which applies in principle even to active colloids not featuring a thin boundary layer has been recently developed in refs. 
\cite{burelbach2018unified,burelbach2019determining}. To achieve this \cite{burelbach2018unified}
has developed a description of phoretic motion within the framework of non-equilibrium thermodynamics based on 
Onsager's reciprocal relations. In ref. \cite{burelbach2019determining} this approach has been used to determine 
the electrophoretic and thermophoretic mobility of weakly charged colloids in aqueous electrolyte solution for arbitrary boundary conditions and boundary layers which may by thick. 
(See also refs. \cite{piazza2004thermal,dhont2008single} for previous approaches beyond the boundary-layer approximation for the special case of stick boundary-conditions.)
\vsb
Let us finally note that expressions for 
the sensitivity or mobility function/coefficient are also available for various microorganisms (section \ref{sec1r2law})
and for synthetic droplets at interfaces \cite{moerman2017solute}.
\vsb
%
\underline{Alignment:}
\\Following the heterogeneity of their surface, 
Janus colloids in general also align (or anti-align) with 
the local gradient of the 
phoretic field induced by other particles in the system (or with an imposed gradient). 
Physically, this is because the phoretic field gradient causes an osmotic flow 
at the fluid-solid interface at the surface of a Janus colloid the strength of which depends on the surface material and properties. 
Hence, for colloids with a thin boundary layer, we have two different slip-velocities at the 
two hemispheres of a Janus colloid, i.e. a net rotational flow. 
This leads to a counter-rotation of the colloid 
(angular momentum conservation) until the symmetry axis of the Janus colloid is aligned with the 
external gradient and the osmotic flow pattern across its surface is symmetric. The angular velocity (alignment rate) reads 
\1
\vec \Omega = \4{3}{2R}\erw{\vec v_s(\vec r_s) \times \hat e}
\2
where $\erw{.}$ represents the average over the surface of the colloid, $R$ is the radius of the colloid and $\hat e$ points along the symmetry axis of the colloid. 
In the simplest case of a spherical Janus colloid at a position $\vec r$ in a slowly varying field $c(\vec r)$
we obtain \cite{bickel2014polarization}: 
\1 \vec \Omega = - \hat e \times \frac{3 \xi\left(\mu_2-\mu_1\right)}{8 R} \nabla c(\vec {r})\2
where $\hat e$ points along the symmetry axis of the colloid, from the hemisphere with surface mobility $\mu_1$ to the one with surface mobility $\mu_2$ and the factor $\xi$ accounts for the deformation of the field due to the presence of the Janus particle \cite{bickel2014polarization}.
Also here, one can use the expression for the self-propulsion velocity 
(\ref{selfpr}) to simplify the coefficients, which shows that the angular velocity of a half-coated Janus colloid at position $\vec{r}_i$ due to the presence of other $N-1$ other 
Janus colloids at positions $\{\vec{r}_j\}$ scales as
\1 |\vec \Omega_i| \sim \left|\mu_r R v_0 \hat e \times \left(\Sum{j=1,j\neq i}{N}\nabla_{{\vec r}_i}\4{1}{|\vec r_j-\vec r_i|}\right) \right|\2
where $|\mu_r|=|(\mu_2-\mu_1)/(\mu_2 + \mu_1)|$ \cite{liebchen2019interactions}. Thus the turning rate of a Janus colloid in the field produced by other Janus colloids 
is proportional to its self-propulsion velocity and maximally strong if the Janus colloid is oriented perpendicular to the local gradient of the field 
and vanishes if its surface mobility is uniform. 
\vsb
Particle based models studying the impact of such alignment effects on the collective behavior of Janus colloids have been considered e.g. in references 
\cite{taktikos2012collective,saha2014clusters,pohl2015self,liebchen2017phoretic}.
\\Generalizations of phoretic interactions beyond leading order in far-field have been discussed very recently e.g. in refs. \cite{saha2019pairing,Varma_PRF_Modeling_Chemohydrodynamic_2019,nasouri2020exact} mostly based on the method of reflections. These works have considered also hydrodynamic interactions. 
In particular, 
for two active colloids beyond the far-field limit, it was shown that although self-propulsion is the dominant term governing the translational velocities of the particles, their angular velocities originate from the interactions alone \cite{saha2019pairing}. This can lead to a separation of timescales between the dynamics of the radial position and the orientation of the polar axis of the particles which can induce either bound or scattering states. Additionally, orbiting states have been predicted for situations where 
this separation of timescales does not occur such that the linear and angular degrees of freedom are significantly coupled to each other \cite{saha2019pairing}.
\vsb
\noindent\underline{Osmotic interactions:}
\\Most existing experiments with active Janus colloids have been performed near a substrates or in between two parallel plates. As discussed in the 
introduction, besides leading to 
self-propulsion and phoretic interactions in such setups the phoretic field due to a Janus colloid 
also induces an osmotic flow along the substrate which may point towards (or away) from the catalytic cap, but has to come off the substrate near the colloids because of the incompressibility of the flow. 
This flow (i) acts back on the Janus colloid itself yielding osmotic self-interactions and (ii) it also advects other (active and passive) colloids in the system, leading to interface-mediated cross-interactions (``osmotic (cross-)interactions").
\\Osmotic self-interactions play a key role for colloidal surfers \cite{palacci2013living}
and have been explored e.g. in refs. \cite{uspal2016guiding,popescu2017chemically} where it has also been discussed that the particles can be guided by chemically patterning the relevant walls. In refs. \cite{heidari2020self,heidari2021}
it has been shown that the measured self-propulsion velocity of self-thermophoretic active colloids strongly depends on the surface functionalization, which has been attributed to the self-induced thermo-osmotic flows which reorient the colloids in a way which reduces their speed component parallel to the wall. 
\\Osmotic cross-interactions are likely to play a role in many experiments with active colloids. Their net effect is similar as for phoretic interactions. They are (almost) isotropic at large interparticle distances and they can be 
reciprocal (e.g. for two Janus colloids in a ``symmetric'' configuration) or non-reciprocal as e.g. for the interactions of a Janus-colloid and a tracer and they are even expected to show the same scaling in most cases. The latter is because the osmotic solvent speed along the substrate is expected to scale as \1 v_s \sim |\nabla_\parallel c| \2 in most cases \cite{anderson1989colloid} where $\nabla_\parallel c$ is now the gradient of the phoretic field along the substrate, 
suggesting together with Eq. (\ref{1r2law}) and Eq. (\ref{yuk}) that the osmotic cross-interactions scale as 
$1/r^2$ if effective screening effects of the relevant field are absent and Yukawa-like if the relevant phoretic field ``decays'' e.g. due to bulk reactions. Their strength differs in general however from that of phoretic interactions and is controlled by the surface properties of the substrate (and of other walls or interfaces if present) rather than by the surface properties of the colloids. 
To discriminate between phoretic and osmotic interactions in a given experiment one could therefore 
changing or functionalizing the substrate, which is expected to have an impact only on the latter type of interactions. Alternatively, one could also explore the impact of an additional top-substrate at various distances, which is expected to have a strong effect on osmotic cross-interactions because it hinders the (incompressible) solvent to escape into the third direction near the colloids.

\subsection{Measurements\label{exps}}
Phoretic and osmotic cross-interactions have been explored and quantified in various notable experiments -- so far mainly between active colloids and tracers (and hardly between different Janus colloids), which we will discuss in the following (see also the recent review \cite{wang2019interactions} on active particle - tracer interactions). 
\vsb \noindent
\underline{Phoretic response to imposed gradients:}
\\An early observation of the phoretic response of active colloids to an imposed phoretic field gradient 
has been reported in ref. \cite{hong2007chemotaxis} (2007). Here it has been observed that self-propelling 
$2\mu m$ long $AuPt$ rods perform an active random walk in an imposed $H_2 O_2$ concentration gradient and aggregate in regions where the concentration is high. 
This phenomenon has been termed 
''chemotaxis in a non-biological colloidal systems``.  
More recently it has also been observed that some Janus-colloids show ``artificial'' phototaxis and respond to external light gradients \cite{lozano2016phototaxis,moyses2016trochoidal,lozano2019propagating,jahanshahi2020realization,Yu_SM_Phototaxis_Active_2019} as well as thermotaxis \cite{Gittus_EPJE_Thermal_Orientation_2019,auschra2021thermotaxis,OlartePlata_PRE_Thermophoretic_Torque_2018}) and rheotaxis \cite{palacci2015artificial,Ren_AN_Rheotaxis_Bimetallic_2017,Baker_N_Fight_Flow_2019,Dwivedi_PF_Rheotaxis_Active_2021}. 
\vsb 
\underline{Cross-interactions:}
\\In ref. \cite{palacci2013living} the interactions between canted antiferromagnetic
hematite cubes, which have been immobilized by attaching them to a glass-substrate, %
and  
$1.5\mu m$-sized tracers made of polystyrene, silica or TPM have been measured under blue light illumination: here the tracer speed been measured as a function of the interparticle distance and has been fitted to an $1/r^2$-law as expected for particles acting as a point source of a concentration field. 
These experiments have ruled out hydrodynamic interactions to explain the tracer motion, because 
the tracers have been observed to move towards the Janus colloids from all directions, whereas the 
advective flow due to a Janus colloid must has zero divergence of course. In addition, 
heating studies have suggested that diffusiophoresis is more important in these experiments than thermo-osmosis.
\\Another setup where the $1/r^2$-law has been explicitly demonstrated in far-field is ref. \cite{singh2017non}. In these experiments UV-light driven active titania-silica Janus colloids attract passive colloids which has been quantified by measuring the interparticle 
speed between the passive and the active particles and fitting the data to the expected $1/r^2$-law for the phoretic far-field interactions. 
See further comparisons between the $1/r^2$-law and its screened variant see also refs. \cite{liebchen2019interactions,hauke2020clustering}. Here the results of the measurements of the distance between the Janus colloids and the tracers from \cite{palacci2013living} and \cite{singh2017non} has been compared with Eq. (\ref{yuk}) showing an even better agreement with the data than the unscreened $1/r^2$ law, suggesting that some effective screening is present in these experiments with an effective screening length of several $\mu m$. For self-electrophoretic Janus particles (bimetallic Au-Pt nanorods) phoretic interactions with passive tracers
have been experimentally quantified in \cite{wang2013catalytically} where it was shown that tracer particles aggregate at one end of the nanorod into seemingly close packed clusters. The tracer particles were seen to be almost at rest at distances more than a few micrometers away from the nanorod, and accelerated as they moved closer to the it. At a distance of about 0.5 $\mu$m away from the nanorod, the tracer speed reached a maximum and then decreased significantly as it moved even closer.
\\More recently, in ref. \cite{huang2020inverse} 
visible-light-driven $Ag/AgCl-$
based Janus particles
have been fixed to a glass substrate such that they do not move but create a phoretic field which acts on the particles in their vicinity. 
When embedding the Janus particles, here called ``active defects'', in a ``matrix'' of passive colloids at moderate density (where they form an amorphous state with a liquid-like structure), they repel the $SiO_2$ colloids and push them together which leads to crystallization. 
This repulsion has been attributed to electrophoretic repulsions due to the photochemical production of ions 
leading to a local chemical gradient around the immobilized Janus particles. 
The time-dependent relative distance between
$SiO_2$ beads and the immobilized $Ag/AgCl$-capped particles has been explored in more detail in ref. \cite{huang2020anisotropic} (and earlier also for the interactions of $SiO_2$ tracers (clusters of) non-immobilized $Ag/AgCl$-swimmers in ref. \cite{wang2018visible}).
This work has clearly shown that at moderate interparticle distances the passive particles are significantly stronger repelled from the catalytic $Ag/AgCl$ cap of the Janus particles than from 
the neutral side. Interestingly, when attaching the $Ag/AgCl$-capped particles to the substrate in a way preventing their translation but allowing them 
to rotate, the passive particles continuously redistribute in a way that they follows the orientational change of the Janus colloid. (This is an interesting consequence of the fact that phoretic (and osmotic) cross-interactions between Janus-colloids and tracers are non-reciprocal, leading to a strong response of the tracers to the Janus colloids, whereas the latter ones 
can freely turn (diffusively) without significantly experiencing a back-action due to the tracers). 
The strength of the repulsion in these experiments seems to be time-dependent leading to a non-monotonous 
size of the exclusion zone around the ``active defects'' 
making it difficult to compare the data for the tracer motion e.g. with the $1/r^2$ law (or an anisotropically generalized variant thereof).
\\Attractive phoretic interactions between photocatalytic Janus colloids ($Au-Ni-TiO_2$) and tracers ($PS$) have recently been applied to demonstrate 
``microplastic'' ($PS$) collection \cite{wang2019photocatalytic}.
\\Refs. \cite{aubret2018targeted,aubret2018diffusiophoretic}
have reported the targeted-formation of self-organized rotors made of Janus-dimers which synchronize with each other. 
Ref. \cite{aubret2018diffusiophoretic} in particular has shown that light patterns 
applied to photocatalytic particles (colloids containing haematite) allow designing (diffusio)phoretic interactions 
among the particles. The authors use 
``Highly Inclined Laminated Optical sheets microscopy'' as a tool to characterize diffusio-phoretic interactions among 
as well as among rotors made of several Janus-dimers. 
The authors have also developed a model leading to near quantitative agreement with 
their measurements; they stress that hydrodynamic 
flows do not seem to have a dominant effect in their experiment. 
\\Similarly to the above works, ref. \cite{katuri2021inferring}
has very recently experimentally studied the response of spherical $1-2\mu m$-sized silica tracers to 
active $5\mu m$ Pt/silica Janus particles in $H_2O_2$-solution near a substrate. The latter particles have been immobilized by attaching them in parallel or perpendicular orientation to the wall. In the latter case, the authors report attractions of the tracers from all directions (reaching tracer velocities $\sim 5\mu m/s$), which are strong enough to dominate thermal fluctuations, 
whereas in the former case, the tracers accumulate near the inert side and an exclusion zone around the $Pt$ side forms.
In these experiments tracers show a somewhat different response and tend to move towards the equator of the Janus colloid (rather than towards the center of the silica side) while still avoiding the $Pt$ side. 
An accompanying theoretic analysis qualitatively reproduces these results and indicates that for the present experiment the tracers would be mostly repelled from the Janus colloids in bulk (rather than being attracted by them). 
\\Other very recent experiments 
\cite{ketzetzi2021activity} have reported on experiments with small ensembles of 2$\mu m$-sized latex colloids half coated with $Pt$ 
in $H_2 O_2$ solution, moving in 1D along circular pillars. In these experiments it has been observed that the Janus colloids keep at a characteristic distance to each other which has been attributed to a competition between a dipolar fluid flow 
which is outwards directed along the swimmer axis and decays with $1/r^4$ and an inwards directed osmotic flow across the substrate which scales as $1/r^2$. 
\vsb \noindent
\underline{Other synthetic systems showing chemical/phoretic interactions:}
\\Besides Janus colloids and microorganisms, similar chemical interactions occur also in various other systems ranging from ion-exchange resins \cite{niu2017microfluidic,liebchen2018unraveling,moeller2021}, interfacial droplet swimmers \cite{maass2016swimming,jin2017chemotaxis,hokmabad2021emergence}, to camphor boats \cite{kohira2001synchronized,nakata2005characteristic,nakata2005synchronized,boniface2019self,gouiller2021mixing} and magnesium microparticles \cite{pavel2021cooperative}. 
\\To measure phoretic, or solute-mediated, (repulsive) cross interactions among synthetic droplets
ref. \cite{moerman2017solute} has identified a parameter regime where synthetic droplets release solutes (surfactant molecules) 
but do not self-propel, leading to a radially symmetric concentration profile. 
For such a system the interactions have been observed to be repulsive and have been measured in two different ways: 
(i) 
by using optical tweezers to bring two droplets close together before releasing them under the influence of the interaction 
and measuring their relative distance as a function of time and 
(ii) based on a balancing of the solute-mediated interactions and gravity. 
As a result, the authors find that the 
$1/r^2$ approximately holds true. However, they also point to effects due to the dissolution of the droplets and solute advection. 
\\Chemotaxis and chemical interactions have also been observed and analyzed in 
interfacial oil droplet swimmers which move by the Marangoni effect and respond to 
to micellar surfactant gradients and to empty micelles, which they leave in their wake \cite{maass2016swimming,jin2017chemotaxis,jin2018chemotactic}.
Very recently, for these droplets, the chemical and hydrodynamic fields have been explicitly visualized \cite{hokmabad2021emergence} and the droplets have been shown to exhibit a memory-driven irregular motion with an increase in the viscosity of the surrounding medium. 
More generally, recent theoretical results on the role of the Marangoni effect on phoretic particles trapped at fluid interfaces has been recently reviewed in \cite{Malgaretti_C_Phoretic_Colloids_2021}.
\\Another class of microswimmers for which 
osmotic cross-interactions have also been observed includes 
ion-exchange resins and tracers which form ion-exchange driven modular microswimmers and assemblies thereof \cite{niu2017microfluidic,niu2017controlled,liebchen2018unraveling,niu2017self,niu2018dynamics,niu2018modular,moeller2021} which we will discuss in more detail in section \ref{nonrec}.\\At larger scales also camphor boats  \cite{kohira2001synchronized,nakata2005characteristic,nakata2005synchronized,boniface2019self,gouiller2021mixing} show strong chemical self- and cross-interactions. These swimmers continuously dissolve and leave slowly diffusing 
chemical trails in their wake show chemical interactions and strong memory effects.
\\Another system in which chemical interactions have recently been explored are 
magnesium micro-particles \cite{pavel2021cooperative}. These particles create local pH gradients to which other particles in the system respond, leading to 
clusters which in turn show a chemotactic response to external pH gradients as induced by a 
corroding iron rod. That way, the magnesium particles can be used as a detector for corroding iron to which they autonomously move, resulting in a 
magnesium hydroxide layer which should slow down the corrosion. 
\\Let us finally mention that, 
to explain pattern formation in experiments with E.coli bacteria in which chemotaxis plays no role \cite{vsimkus2018phoretic}, the authors of ref. \cite{vsimkus2018phoretic} have suggested that 
phoretic interactions might also play a role for bacteria. Clearly, this far-reaching suggestion calls for further explorations. 
\vsb 
\underline{Phoretic osmotic or hydrodynamic interactions?\label{phos}}
\\In order to correctly model the interactions in a given experiment it is important to understand how one can distinguish between 
hydrodynamic, phoretic and osmotic interactions.
Clearly, hydrodynamic  interactions as induced e.g. by force free swimmers, such as force dipole swimmers, are divergence free and accordingly they are attractive and repulsive in different direction.
As opposed to this, both phoretic and osmotic interactions are expected to be either attractive or repulsive in all directions, at least in far-field, which has allowed it e.g. to rule out hydrodynamic interactions for the formation of living clusters in the experiments \cite{palacci2013living}.
Discriminating between phoretic and osmotic interactions is comparatively difficult: they both show the same symmetry, can both be attractive or repulsive, they are both expected to have a similar strength (as they both induce surface flows of a similar speed leading either to advection with the flow velocity or to a motion with a speed which is equal to the surface averaged flow) and they are both expected to have the same scaling with the interparticle distance. (For these reasons we 
discuss them together in the present section.)
\\What are the phenomenological differences between the two interactions which could be used to discriminate between these two types of interactions? 
First, osmotic interactions occur only near substrates and would be of little importance for swimmers far away from boundaries, whereas phoretic interactions would occur also in bulk (although their strength might change because walls displace or deform phoretic fields).
Second, the strength and direction of osmotic interactions is expected to strongly depend on the surface properties of the substrates. 
Electro-osmotic and diffusio-osmotic flows in charge neutral ionic solvents for example strongly depend the zeta potential of the substrate, such that their strength can be tuned via the choice of the substrate or by functionalizing the latter. 
Finally, since most relevant solvents
are essentially incompressible, an active colloid can only create an inward pointing far-field flow in all directions if the flow can stream away from the substrate (upwards) near the colloid; see e.g. \cite{niu2017microfluidic,liebchen2018unraveling} for the osmotic flows induced by an ion-exchange resin. 
Thus, for swimmers which are confined between a top and a bottom (glass) plate the strength of osmotic interactions is expected to strongly depend on the distance between the places so that systematically changing the latter can be used to evaluate their strength.

\subsection{Collective phenomena}
Let us now briefly discuss (collective) phenomena which have been attributed to phoretic (or osmotic) cross-interactions among different Janus colloids.
(Phenomena which hinge on the nonreciprocity of phoretic interactions such as active molecule formation will be discussed in section \ref{nonrec}.)
\vsb 
\begin{figure}
	\begin{center}
		\includegraphics[width=0.6\textwidth]{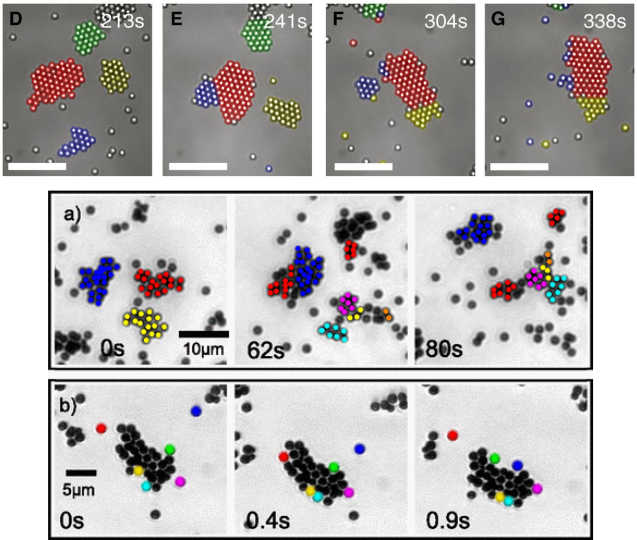}
		\caption{(top panel) Living crystals in light-activated colloidal surfers under blue-light illumination. The false colors show the time evolution of particles belonging to different clusters. The clusters are not static but rearrange, exchange particles, merge (D$\rightarrow$F), break apart (E$\rightarrow$F), or become unstable and explode (blue cluster, F$\rightarrow$G). (bottom panel) Dynamic clustering of self-propelled spherical gold colloids half-covered with platinum. Panel (a) illustrates the dynamics of the clusters with (false) colors indicating the dynamics of the clusters and (b) shows the dynamics inside a cluster. Figures reproduced 
		with permission 
		from \cite{palacci2013living,theurkauff2012dynamic}.}
		\label{palacci}
	\end{center}
\end{figure}
\noindent\underline{Experiments:}
\\One of the most celebrated phenomena which has been observe in several experiments with active colloids is dynamic (or living) clustering \cite{theurkauff2012dynamic, palacci2013living, buttinoni2013dynamical, ginot2018aggregation}. In these experiments the Janus colloids 
self-organize at remarkably low densities (typically less than 10\% area fraction) into clusters which spontaneously form at low density and dynamically grow, break-up and reform (see Fig.~\ref{palacci}). 
This aggregation has been attributed to chemical interactions of the Janus colloids \cite{theurkauff2012dynamic} and in 
ref. \cite{palacci2013living} the corresponding phoretic cross-interactions have even been measured explicitly as discussed above. 
(Notice that the dynamic clusters observed e.g. in \cite{theurkauff2012dynamic,ginot2018aggregation} show an approximately algebraic cluster size distribution even at area fractions of $<10\%$. They can therefore not be understood as a precursor of motility-induced phase separation within the active Brownian particle model, which would lead to an exponential cluster-size distribution at such low densities.)
\\Experiments exploring the collective behavior of self-thermophoretic microswimmers are still rather scarce in the literature. One exception where the collective behavior of thermophoretic Janus colloids has been explored is 
ref. \cite{io2017experimental} which has observed motile clusters of aligned particles which has been partly attributed to thermophoretic cross-interactions between the Janus colloids. 
\\Ref.~\cite{heckel2020active} has recently realized $BiVO_4$ swimmers with a spheroidal shape which self-propel without showing any obvious asymmetry. These swimmers significantly attract each other and form clusters of different conformation and speed. 
The clustering of these swimmers could be reproduced in simulations based on Langevin-equations with phoretic/osmotic cross-interactions essentially based on a generalization of 
Eqs. (\ref{langevin},\ref{chemdiff}) to spheroids with additional steric repulsions and with a distribution of ``reaction sides'' placed along one of the long sides of the spheroids. These simulations have allowed it to predict the speed of the clusters in close quantitative agreement with the 
experiments. 
\vsb 
\noindent
\underline{Theories and simulation studies}
\\A pioneering study of the collective behavior of phoretically interacting Janus colloids has been developed in ref. \cite{saha2014clusters} which has illustrated the complexity of phoretic interactions 
, in the case where two chemical fields, representing a fuel and a product species,
are relevant. 
This work starts with Langevin equations for the particle dynamics which are coupled to chemical diffusion equations representing the source and the sink species and derives a continuum theory based on which a catalogue of stationary and oscillatory linear instabilities are predicted, suggesting a rich panorama of patterns. 
\\Inspired by experiments observing dynamic clustering in Janus colloids \cite{theurkauff2012dynamic}, in parallel to \cite{saha2014clusters}, 
reference \cite{pohl2014dynamic,stark2018artificial} has developed a model 
of self-propelling Janus colloids with phoretic far-field interactions featuring a translational and a rotational component. By using particle based simulations and the far-field solution of the chemical field (in the stationary limit) (\ref{csol}) an interesting phase diagram has been predicted. In particular, 
if both the translational and the rotational contributions to the interactions are attractive, the expected collapse was observed (Keller-Segel instability). However, remarkably, when only the translational contribution is attractive and the rotational one is repulsive, dynamic clustering was found for a suitable strength of the interactions. Here, notably the predicted cluster size decays algebraically for small cluster sizes similar as in experiments \cite{ginot2018aggregation}. Oscillatory states have been observed in the same system for cases where the phoretic translation is repulsive and the 
rotations are attractive \cite{pohl2015self}.
\\Later, in \cite{liebchen2019interactions} it has been found that if the phoretic interactions are effectively screened (e.g. in the presence of bulk reactions), 
dynamic clustering can occur even without such a rotational contribution based on the competition of attractions and fluctuations only. In this work, the strength of the phoretic far-field interactions has been linked to the self-propulsion velocity showing that the phoretic interactions in (some) Janus colloids seem to just have the right strength to induce dynamic clustering. 
\\While the above works mostly focus on attractive phoretic interactions, refs. \cite{liebchen2015clustering,liebchen2017phoretic,liebchen2018synthetic}
have shown that also purely repulsive phoretic interactions can destabilize the uniform phase. The instability can occur either due to 
anisotropy in the chemical production of Janus colloids, leading to the so-called ``Janus-instability'', or due to memory (delay) effects 
which can create an oscillatory instability leading to traveling wave and traveling lattice patterns. 
Based on the model of \cite{liebchen2015clustering}, in \cite{mukherjee2018growth} it has been demonstrated that repulsive phoretic interactions 
and a logistic growth for particle (bacteria) concentration can lead to 
ring and spot patterns resembling experimental observations \cite{liu2011sequential}.
\\The collective behavior of self-thermophoretic active particles which interact via the temperature gradients induced by these particles has been theoretically studied in refs. \cite{golestanian2012collective,cohen2014emergent}. Here it was shown that thermorepulsive colloids, characterized by a positive Soret coefficient, show a depletion effect in a confined geometry and organize into hollow bands, tubes, or shells while thermoattractive (negative Soret coefficient) colloids feature a linear instability and collapse into a dense macrocluster \cite{golestanian2012collective}. 
If a light source heats up the particles from one side, such that the colloids create a shadow 
and prevent the light to reach other colloids, 
thermoattractive colloids can also spontaneously self-organize to form a long-lived swarm shaped like a comet \cite{cohen2014emergent}. 
Other 
clustering behaviors have been observed in simulations of thermophilic dimeric swimmers \cite{Wagner_EPJE_Collective_Behavior_2021} whereas thermophobic dimeric swimmers have been shown to display swarm like behaviors due to the interplay of attractive hydrodynamic and repulsive phoretic interactions \cite{Wagner_E_Hydrodynamic_Frontlike_2017}.

\subsection{Mixtures: Nonreciprocal interactions\label{nonrec}} 
\begin{figure}
\begin{center}
\includegraphics[width=0.4\textwidth]{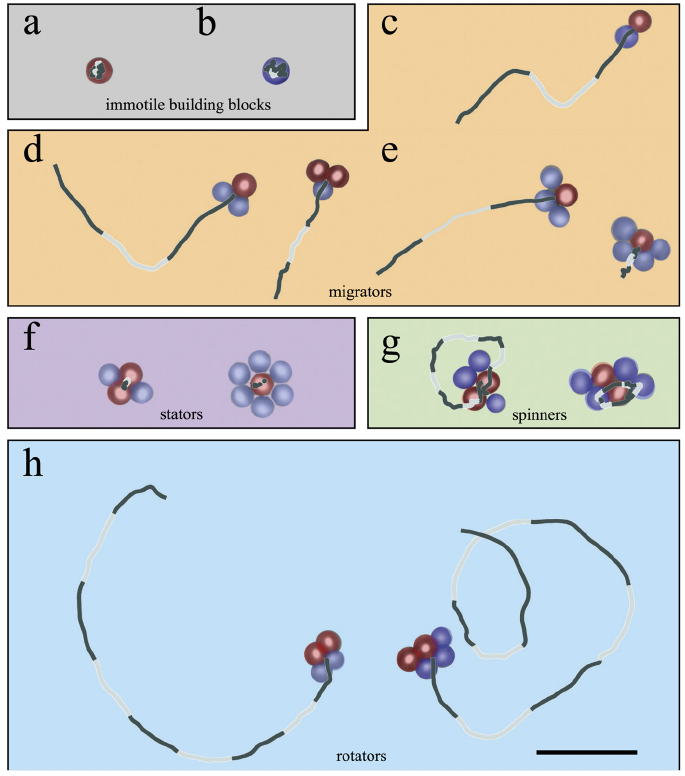}
\caption{Active colloidal molecules consisting of immobile building blocks (a,b) acquiring self-propulsion through 
nonreciprocal phoretic (and osmotic) interactions. The figure show experimental snapshots and trajectories of active molecules which occur in a variety of 
shapes, leading to characteristic motion patterns in the form of ballistically moving migrators (c,e), stators which move only diffusively for symmetry reasons, spinners, which rotate but hardly move ballistically (g) as well as rotators which self-propel and can be described as chiral active Brownian particles (h).
Reproduced 
with permission 
from \cite{schmidt2019light}.}
\label{expam}
\end{center}
\end{figure}

As one of their key properties both phoretic and osmotic cross-interactions can be non-reciprocal. 
The non-reciprocity occurs in particular in mixtures e.g. of active Janus particles and isotropic colloidal tracers. Here the former particle induces a phoretic field gradient attracting (or repelling) the latter particles, but not vice versa. 
At the level of the particles, the nonreciprocity spontaneously generates momentum, 
which can be observed in experiments showing active molecule formation made of non-motile components which spontaneously acquire motility through the nonreciprocal interaction of its components and show a persistent directed motion as we will discuss in detail in the next section. 
Since the fundamental equations (and the boundary equations) governing the system dynamics conserve momentum, the solvent has to compensate for the momentum-gain at the level of the particles. 
(This is similar as for most Brownian ratchets where the solvent has to compensate for the directed net motion of the particles.)
\vsb 
\underline{Active colloidal molecules:}
\\To exemplify how nonreciprocal phoretic (or osmotic) interactions come about, let us consider a specific experiment realized 
in ref. \cite{schmidt2019light}. In this experiment 
two species of isotropic and non-motile colloids have been placed in a near-critical water-lutidine solvent, only one of which contains iron-oxide and therefore efficiently absorbs light.  
Under light irradiation the solvent heats up in the vicinity of the light absorbing colloids beyond the critical point 
of the solvent such that water and lutidine unmix near these colloids. 
Thus each light-absorbing colloid evokes a gradient in the temperature field and the water-lutidine composition.
The other species moves 
by diffusio- and or thermophoresis ``up'' these gradients, towards the light-absorbing colloids to form a colloidal molecule, which persists over experimentally relevant timescales.  
Remarkably, even at close contact, within each of these molecules, the non-light absorbing colloids unidirectionally move towards the light-absorbing particles and push them forward, such that the colloidal molecule self-propels. 
The result is an ``active molecules'' (see Fig.~\ref{expam}) as predicted in a minimal model first in ref. \cite{soto2014self,soto2015self} and somewhat similarly for particles with nonreciprocal chemical interactions showing a predator-prey dynamics in ref. \cite{sengupta2011chemotactic}.
\\In the experiments \cite{schmidt2019light} the colloids self-organize into active molecules of different composition and shape, featuring a panorama of 
motion patterns such as ballistically moving migrators, circularly-swimming rotators as well as spinners, which rotate without moving ballistically (see Fig. \ref{expam}). Here, self-propulsion and self-rotations are emergent phenomena which are induced by the nonreciprocal cross-interactions of the particles. 
\vsb
A phenomenon which is closely related to active molecule formation is ``modular microswimming'' which also hinges on 
different non-motile components which interact nonreciprocally to create self-propulsion as an emergent phenomenon. 
The series of experiments \cite{niu2017controlled,niu2017microfluidic,niu2017assembly,niu2017self,niu2018modular,liebchen2018unraveling,moller2021shaping} uses ion-exchange resins which exchange certain ions (e.g. $K^+$) with a smaller one (e.g. $H^+$) which have a higher diffusion coefficient, resulting in a concentration gradient in both species, which has been explicitly measured very recently using three-channel micro-photometry \cite{moller2021shaping} and a spontaneous electric field. (This field occurs because the difference in the diffusivity of the exchanged ions evokes a charge imbalance leading to an internal electric field, balancing this imbalance by acting on the positive and negative ($Cl^-$) ions in the system.) 
The concentration gradient and the spontaneous electric field evoke neutral (chemiosmotic or diffusioosmotic) and electroosmotic flows along the substrate 
\cite{liebchen2018unraveling} which advect colloidal tracer particles towards the resin and forms a ``modular microswimmer''. 
However, unlike in the above example, the colloids do not get in close contact with the resin, but leave a characteristic exclusion zone in between, presumably because they phoretically move away from the resin, in competition to the effective osmotic attraction which weakens in the vicinity of the resin because the solvent has to escape into the third direction due to solvent incompressibility. The swimmer then moves presumably because of the osmotic flows induces along the surface of the stalled tracer colloids which effectively blow solvent towards the resin such that the entire ``module'' sees an asymmetric flow pattern and starts to move balistically. Here, self-propulsion is fueled by the ion-exchange process and emerges from the nonreciprocal osmotic interactions of the ion-exchange resin and the colloids. 
(See ref. \cite{liebchen2018unraveling} for more a more complete picture of the self-propulsion mechanism). 
\\When several modular swimmers involving cationic and anionic ion-exchange resins come together they show long-range interactions \cite{niu2017self} and can self-organize into active molecules similar to those described above and shown in ref.~\cite{niu2018dynamics}.
\\A further type of modular microswimmer which involves a similar self-propulsion mechanism is involved in the schooling of $AgCl$ particles \cite{ibele2009schooling,wang2015one}.
\\Other notable examples of active molecules, often comprising components which are active on their own have been realized e.g. in refs. 
\cite{zhang2016directed,ilday2017rich,singh2017non}.
For further discussions on active colloidal molecules and modular microswimming, see the recent reviews 
\cite{lowen2018active} and \cite{niu2018modular}, respectively. For discussions of active molecule formation in systems with combined hydrodynamic and phoretic interactions, see section \ref{chemhydropairs}.

\begin{figure}
	\begin{center}
		\includegraphics[width=1\textwidth]{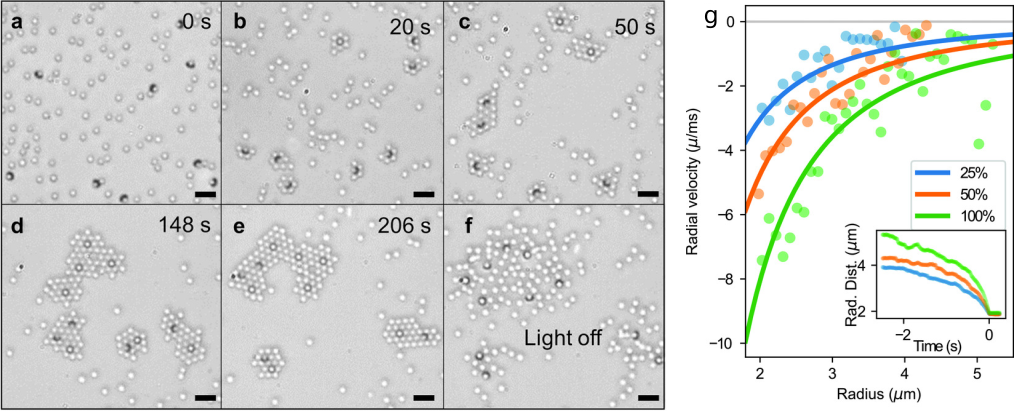}
		\caption{(a-e) Light-induced assembly of passive colloids around active colloidal seeds. (f) The colloidal crystallites dissociate once the light is turned off. (g) The radial velocities of passive tracer colloids as a function of the center-to-center distance from active seeds at different light intensities. The fits correspond to the $1/r^{2}$ law for the drift velocity discussed in the main text. The inset shows the mean radial position of the particles with time; see ref. \cite{singh2017non}. Reproduced 
		with permission 
		from \cite{singh2017non}.}
		\label{singh_fig}
	\end{center}
\end{figure}

\begin{figure}
	\begin{center}
		\includegraphics[width=0.5\textwidth]{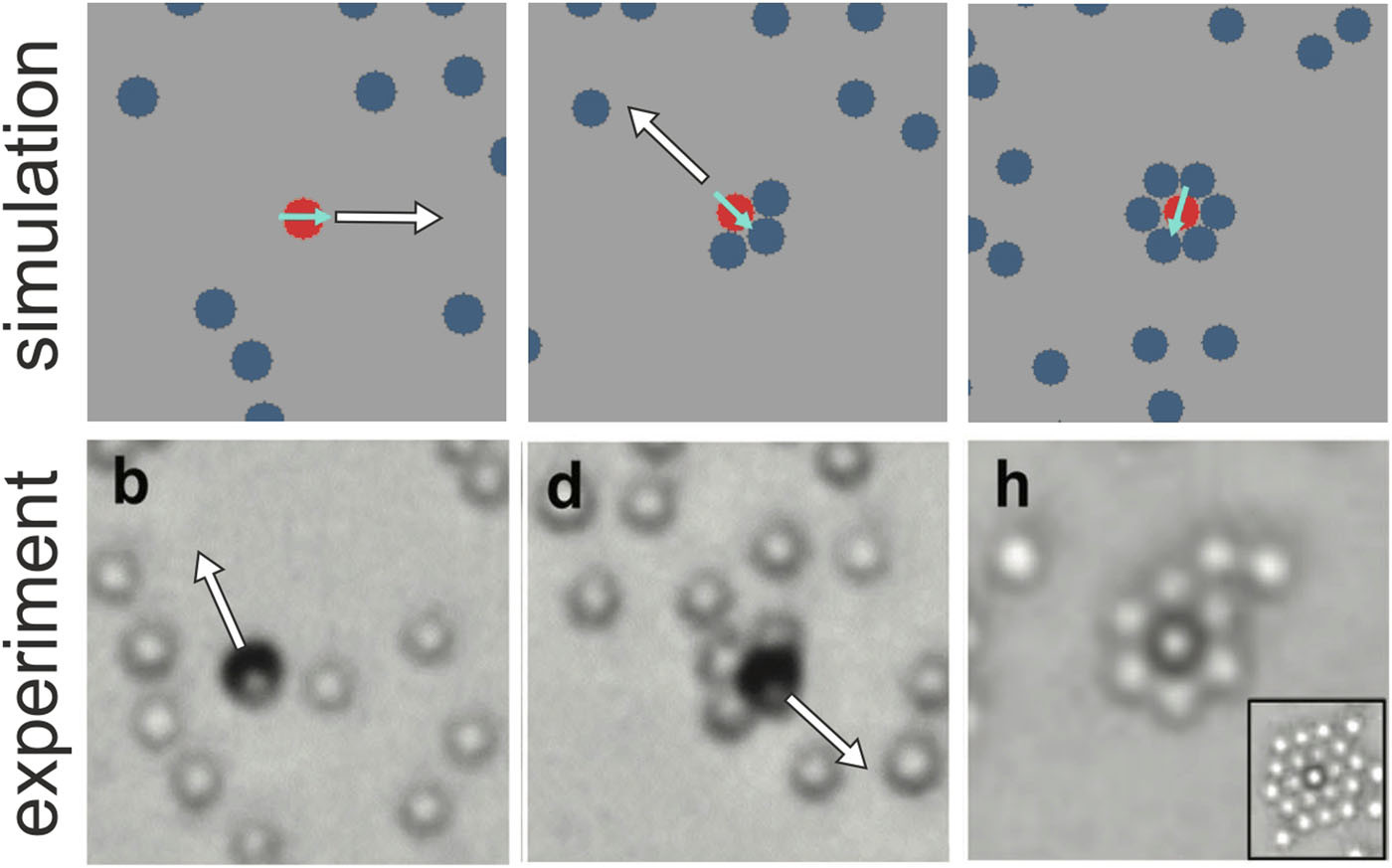}
		\caption{Velocity-reversals of active Janus particles (red/black disks) seen in experiments \cite{singh2017non} and simulations \cite{hauke2020clustering}. These reversals are 
		induced by nonreciprocal phoretic (and osmotic) interactions leading to the aggregation of passive colloids (blue/grey disks) at the cap of the Janus-colloids. 
		Even at close contact the passive colloids are nonreciprocally attracted by the active Janus colloids and push them against their self-propulsion direction. Reproduced 
		with permission 
		from \cite{hauke2020clustering}.}
		\label{hauke}
	\end{center}
\end{figure}

\begin{figure}
	\begin{center}
		\includegraphics[width=0.9\textwidth]{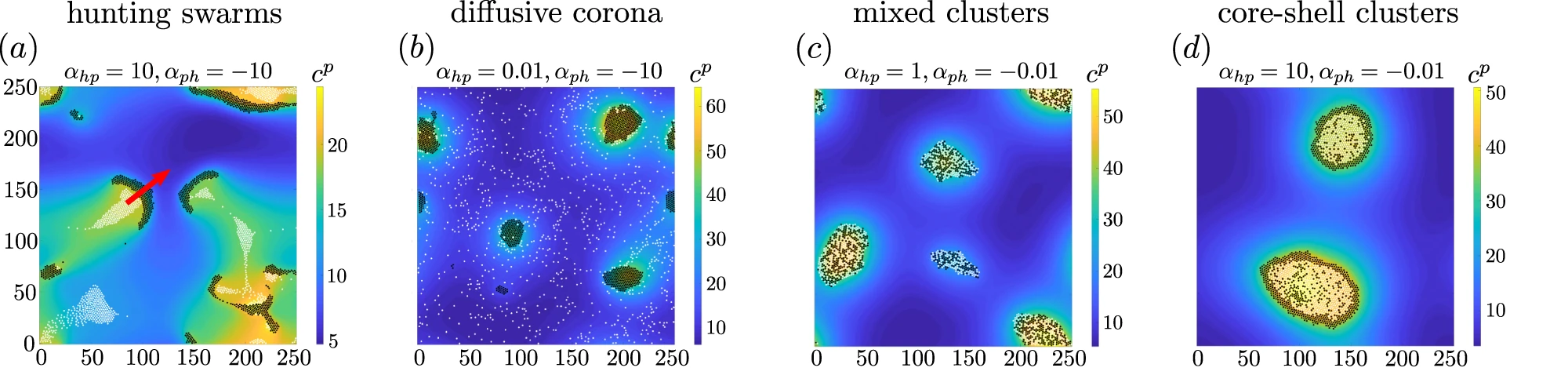}
		\caption{Snapshots from particle-based simulations of two chemically interacting particle species (represented by black and white dots respectively) which self-organize into a hunting swarm phase (a) as well as different types of clusters (b-d). The colormap shows the chemical field produced by the `black' particles. Reproduced 
		with permission 
		from \cite{grauer2020swarm}.}
		\label{grauer}
	\end{center}
\end{figure}

\vsb 
\underline{Clustering, swarming and velocity-reversals}
\\The large scale collective behavior of mixtures of many colloids with nonreciprocal phoretic (or osmotic) interactions has been experimentally explored e.g. in references \cite{ibele2009schooling,ibele2010emergent,duan2013transition,singh2017non} (see also the review \cite{wang2015one}) 
and theoretically/numerically in refs. \cite{ivlev2015statistical,agudo2019active,hauke2020clustering,grauer2020swarm,saha2020scalar}.
\\In ref. \cite{singh2017non} light driven mixtures of active Janus colloids which move with their catalytic cap ahead, 
and passive colloids have been explored. Here, the 
passive colloids are phoretically attracted by the active colloids which has been measured in ref. \cite{singh2017non} by quantifying the interparticle 
speed between the passive and the active particles and fitting the data to the (expected) $1/r^2$-law for the phoretic far-field interactions. As a consequence of these interactions, the passive particles have been observed to aggregate around the active seeds even at very low density (an active particle area fraction of $0.4\%$ is sufficient to initiate aggregation) (see Fig.~\ref{singh_fig}). 
\\This aggregation has been modeled in ref. \cite{sturmer2019chemotaxis} based on Langevin equations for the active and the passive particles with nonreciprocal phoretic far-field attractions of the passive colloids by the active ones. 
These simulations could successfully reproduce the aggregation dynamics with a near quantitative agreement of the time-dependent mean size of the largest cluster and broadly also the cluster speed as a function of cluster size. 
\\An interesting phenomenon in the experiments \cite{singh2017non} which has not been modeled in ref. \cite{sturmer2019chemotaxis}
is that the active passive clusters show intriguing velocity reversals as they grow (see Fig.~\ref{hauke}).
In the experiments small clusters self-propel in the same direction as the underlying Janus colloid (cap ahead), once a certain cluster size is reached the velocity of the aggregates changes direction and moves against the self-propulsion direction of the involved Janus colloid.
These velocity-reversals could be reproduced in a model for phoretically interacting active and passive particles which includes not only phoretic far-field interactions but also aspects of their near-field interactions \cite{hauke2020clustering}.
In this work, they have been attributed to the fact that the passive colloids predominantly aggregate at the phoretic cap of the Janus colloid, where the phoretic field is strongest. Thus, even at close contact, the passive colloids move up the chemical gradient, i.e. they are nonreciprocally attracted by the catalytic cap, and push the active colloid against its intrinsic self-propulsion direction \cite{hauke2020clustering}. (In addition, the passive colloids also deform, or displace, the phoretic field due to the Janus colloid, such that the latter one sees a modified concentration field which has an additional effect on the motion of the cluster.)
Besides providing a mechanism for the observed clustering-induced velocity-reversals in active-passive mixtures, ref. 
\cite{hauke2020clustering} also explores the collective behavior of active-passive mixtures with repulsive phoretic interactions. 
In such cases, in suitable parameter regime, the simulations predict traveling fronts of active particles pursued by passive ones coexisting with an active gas.
\\Other recent works studying the collective behavior of nonreciprocally interacting particles include ref.
\cite{agudo2019active} which considers two (and many species) and 
predicts different phase separation and collapse scenarios including states comprising self-propelling macroclusters where one species is preferentially in the front of the cluster and the other one is behind. 
Similarly as for the active molecules above, in this work the particles are originally not active and directed motion is caused by the nonreciprocal interactions. 
\\Cases where two particle species and two chemical fields are relevant have been considered in ref. 
 \cite{grauer2020swarm} which has used particle based simulations and a continuum theory to predict a phase diagram comprising a hunting swarm phase as well as different types of clustering and aggregation (see Fig.~\ref{grauer}). 
The hunting swarm phase occurs when the particles of (at least) one species effectively attract each other to form a ``pure'' cluster containing only one species, but interact nonreciprocally with particles of the other species in such a way that this second species gets attracted by the cluster but repel particles in the cluster. This 
leads to swarm like clusters which move ballistically and ``hunt'' each other. At late times these hunting swarms grow and feature a characteristic size which grows approximately as $L(t)\sim t^{0.55}$, which is significantly larger than diffusive growth due to Ostwald ripening. 
In a different parameter regime, 
ref. \cite{grauer2020swarm} predicts states where one species forms a cluster which is surrounded by a corona of the other species. Then, in the coarse of the dynamics, the species in the core of the cluster gets ejected from the cluster and the species which was originally at the rim of the cluster aggregates to form the cluster-core. 
Both the hunting swarms and the cluster ejections have been fully reproduced in  ref. \cite{grauer2020swarm} based on simulations of the continuum model developed in this work.
\\Another work focusing on nonreciprocal interactions (or of particles not obliging Newton's third law) is \cite{ivlev2015statistical}. This work has not specifically focused on phoretic or chemical interactions but 
has identified a class of nonreciprocal interactions (constant nonreciprocity) for which one can construct an effective Hamiltonian allowing to 
map the nonequilibrium dynamics of 
nonreciprocally interacting particles to an equilibrium problem. This leads to conservation laws for (pseudo)momentum and the (pseudo)energy of the resulting effective system. For other cases where the nonreciprocity depends on the interparticle distance such a mapping is not possible and the system generically heats up with the effective temperature growing asymptotically with time as $\propto t^{2/3}$. 
These theoretical predictions have been accompanied in experiments involving two quasi two-dimensional layers of complex plasmas leading to nonreciprocal interactions of one particle species with the wake of the other one. 
\\Very recently the Cahn-Hilliard model (model B) has been generalized to describe 
nonreciprocally interacting particles~\cite{saha2020scalar} and they reported an oscillating phase in which one component chases the other one. The strength of the asymmetry in the non-reciprocal interactions was shown to determine whether the steady state exhibits a macroscopic phase separation or a traveling density wave displaying global polar order.
\vsb 

\section{Hydrodynamic interactions}
Let us now briefly discuss hydrodynamic cross-interactions, which hinge on the action of the solvent flow as induced by a force-free microswimmer on other particles in the system and are fundamentally from hydrodynamic interactions in sedimenting or otherwise externally forced passive colloids. 
\\In the realm of low-Reynolds numbers where synthetic active colloids and biological microorganisms reside, the dynamics of the solvent is governed by the Stokes equation, which in contrast to the Navier-Stokes equation is instantaneous and linear. The latter means that flow field perturbations due to different microswimmers superimpose whereas the former one implies that inertia is absent, so that microswimmers only move forward at time-instances where they effectively transfer solvent into the opposite direction ensuring momentum conservation in the overall force-free system. 
The process by which microswimmers create the solvent flow which is required for self-propulsion determines the precise form of the hydrodynamic cross-interactions and depends on the type of microswimmer under consideration. 
In particular, while biological microswimmers often move through body-shape deformations (e.g. disturbing the solvent by rotating or beating certain flagella) which have to be nonreciprocal due to the absence of inertia to allow for self-propulsion (Scallop's theorem \cite{Purcell_AJP_Life_Low_1977}), in autophoretic Janus colloids the flow is caused by the action of some phoretic field(s) on the solvent within a (thin) interfacial layer, ideally leading to a continuous solvent slip over the rigid body of the colloids as described by (generalized) squirmer models as we will briefly discuss below. 
\\
Unlike in near-field where the precise form of the flow-field perturbations matter for the hydrodynamic-cross interactions among microswimmers, in far field, they show a certain degree of 
universality in that both the scaling of the interactions with interparticle distance and the orientational dependence of the interactions is 
identical for broad classes of microswimmers, whereas the details are relevant in far-field only for the coefficient of the interactions which determines the strength and the sign of the interactions. 
More specifically, 
hydrodynamic far-field interactions decay in microswimmers  
in leading order as $1/r^2$ (or as $1/r^3$) 
with the interparticle distance $r$ unless certain symmetries suppress corresponding distributions. 
This contrasts from hydrodynamic interactions 
in isotropic passive colloids moving under the influence of an external force  \cite{dhont1996introduction} or in an external phoretic field \cite{morrison1970electrophoresis} which lead to interactions which decay in far field as $1/r$ and $1/r^3$ respectively. While hydrodynamic interactions autophoretic Janus colloids and biological microswimmers show the same far-field scaling with the interparticle distance $r$\footnote{e.g. $1/r^2$ e.g. in deformable swimmers like E.coli bacteria or Chlamydomona algae and in Janus colloids with a heterogeneous surface mobility and
$1/r^3$ in nondeformable swimmers such as 
Janus colloids with a uniform surface mobility or in Paramecium \cite{lighthill1952squirming,blake1971spherical} which possess short flagella (``cilia'') which are almost uniformly distributed over their surface.}
Notably, however, 
the coefficients determining the strength and direction of these interactions can strongly differ from microswimmer to microswimmer. 
For example, while autophoretic colloids ideally induce a continuous slip velocity across their surface, 
Chlamydomona algaes \cite{guasto2010oscillatory} create a flow field which leads to $1/r^2$ interactions with an oscillatory sign 
\cite{guasto2010oscillatory}, which is caused by their swimming mechanism based on flagella arms which operate similarly to the arms of a human breast swimmer involving strokes (leading to forward motion) and recovery strokes (leading to backwards motion and a reverted flow field due to the instantaneous nature of the Stokes equation). 
\\In the following we first review some fundamental solutions of the Stokes equation for hydrodynamic point sources and point dipoles which are frequently used in the literature to represent far-field interactions among force-free microswimmers and then we specifically discuss hydrodynamic interactions in active colloids which are often modeled based on the squirmer model. We will keep the general considerations on hydrodynamic interactions comparatively short as the hydrodynamics of microswimmers and their interactions have been reviewed elsewhere; see e.g. \cite{lauga2009hydrodynamics,Koch_ARFM_Collective_Hydrodynamics_2011,elgeti2015physics,zottl2016emergent,bechinger2016active,yeomans2017hydrodynamics} and will mainly focus on the hydrodynamic interactions in squirmers and autophoretic colloids.
\subsection{Stokes equation}
\label{hydro_interactions}
Following their small sizes, microswimmers typically induce a fluid flow leading to a very small Reynolds number $Re=(\rho L u)/\eta \ll 1$, 
where $\rho,\eta$ are the density and the viscosity of the solvent and $L,u$ denote characteristic length and velocity scales. 
For E.coli bacteria in water for example, we have 
$L \sim 3 \mu 
m$, $v \sim 30 \mu m/s$, $\eta \approx  10^{-3} Pa s$ leading to $Re \sim 10^{-5}-10^{-4}$. 
Thus, 
viscous effects strongly dominate over inertial effects in the Navier-Stokes equation, which reduces, upon neglecting the latter contributions, to the purely algebraic Stokes equation which determines the velocity $u(\vec r,t)$ of the solvent
\begin{equation}\label{stokes}
\eta \nabla^{2} \vec{u}-\nabla p+\vec{f}=0,\quad \nabla \cdot \vec{u}=0.
\end{equation}
Here $\vec{f}(\vec r,t)$ is the force density acting on the solvent, $p(\vec r,t)$ is the pressure field and the source-free condition $\nabla \cdot \vec{u}$ ensures incompressibility of the solvent \cite{kim2013microhydrodynamics}. For autophoretic Janus colloids featuring a slip velocity, the force density is essentially zero outside their thin interfacial layer. Here, the effect of the interfacial forces driving the solvent flow is typically taken into account via the slip boundary condition at the particle surface.
\\More generally, one approach to fully describe hydrodynamic cross-interactions in deformable microswimmers is to 
explicitly solve the Stokes equation Eq.~(\ref{stokes}) together with appropriate boundary conditions on the surface of each microswimmer $\vec{u}(\vec{r}^s_i)=\vec{v}_i+\vec{\omega}_i \times \vec{r}^s_i + \vec{u}(\vec{r}^s_i), \quad \vec{r}^s_i \in S_i$ where $S_i$ is the (deforming) surface of the $i$-th microswimmer, $\vec r^s_i$ is a point on that surface, $\vec{u}(\vec{r}^s_i)$ denote the slip velocity and $i=1,2..N$ with $N$ denoting the number of microswimmers in the system. The velocity $\vec{v}_i$ and angular velocity $\vec{\omega}_i$ of each microswimmer can then be determined based on the stress tensor $\sigma=\eta\left(\nabla \otimes \vec{u}+(\nabla \otimes \vec{u})^{\mathrm{T}}\right)$, which in turn allows one to calculate the force $\vec{F}_i=\int_{S_i} \mathrm{~d} S_i \sigma(\vec{r}_i, t) \hat{n}_i$ and torque $\vec{T}_i=\int_{S_i} \mathrm{~d} S_i \vec{r}_i \times(\sigma(\vec{r}_i, t) \hat{n}_i)$ acting on each microswimmer. Here  $\mathrm{d}S_i$ denotes the differential element of the $i$-th microswimmer's surface and $\hat{n}_i$ is the surface normal on $S_i$. 
The force- and torque-free conditions ($\vec{F}_i=0, \vec{T}_i=0$) then determine the linear and angular velocity of each microswimmer $\vec{v}_i$ and $\vec{\omega}_i$.

\subsection{Far-field}
While the calculation of the full hydrodynamic cross-interactions in microswimmer ensembles requires an explicit 
solution of the Stokes equation with the detailed boundary conditions on the surface of each microswimmer (or of other equations like e.g. the Boltzmann equation carrying at least the same information), which is numerically demanding for very large systems, 
the interactions generically reduce to a much simpler form in far-field. In far-field the flow field due to each microswimmer is represented by the flow field which is created by 
a combination of point forces or point sources, three commonly used examples of which are the 
Stokeslet, the force dipole (Stresslet) and the source dipole solution which lead to a flow-field decaying as $r^{-1}$, $r^{-2}$  and $r^{-3}$ with distance $r$ from the singularity. 
\\The Stokeslet solution represents the flow field due to a point force 
of strength $f$ and direction $\hat e$ located at the origin and reads
	\begin{equation}
		\vec{u}_{\mathrm{PF}}(\vec{r})=\frac{f}{8 \pi \eta |\vec{r}|}[\hat{e}+(\hat{r} \cdot \hat{e}) \hat{r}]
	\end{equation}
	where $\hat{r}=\vec{r}/|\vec{r}|$. The Stokeslet is relevant to describe far-field interactions of particle which are subject to a net force such as passive colloids sedimenting under the influence of gravity. 
	Since microswimmers are force free, the simplest possible and most commonly used far-field flow is that of two 
 point forces of same strength $f$ (but oppositely directed) placed a small distance $l$ apart.
At distances $|\vec{r}|\gg l$ the flow field is given by 
	\begin{equation}
		\vec{u}_{\mathrm{FD}}(\vec{r})=\frac{f l}{8 \pi \eta |\vec{r}|^{2}}\left[3(\hat{e} \cdot \hat{r})^{2}-1\right] \hat{r}
	\end{equation} 
The sign of $f$ determines if the flow field corresponds to a pusher ($f>0$) or a puller ($f<0$). Analogous to the force dipole, the flow field at distances $|\vec{r}|\gg l$ due to two point sources (of solvent) with source density $Q$ but opposite signs, placed at a distance $l$ apart is given by the source dipole solution of the Stokes equation with a modified incompressibility condition $\nabla \cdot \vec u=s(\vec{r})$ with $s(\vec{r})$ denoting the source density. The source dipole solution has the form
	\begin{equation}
		\vec{u}_{\mathrm{SD}}(\vec{r})=\frac{Q l}{4 \pi |\vec{r}|^{3}}[3(\hat{e} \cdot \hat{r}) \hat{r}-\hat{e}]
	\end{equation}
These solutions have been shown to reproduce the far-field flow patterns of biological microswimmers; see e.g. ref. \cite{drescher2011fluid} for explicit measurements of the flow field of E.coli bacteria.
These solutions can also be readily amended to account for walls e.g. by using Blake tensors \cite{blake1972model,blake1974fundamental}. 
Specific expressions for the far-field flow in the presence of walls (or stress-free surfaces) have been provided in ref. \cite{spagnolie2012hydrodynamics} and shown to essentially agree with solutions of the Stokes equation down to particle-wall distances of 
only about one-tenth of a body length between the particle and the wall.
\vsb 
The discussed fundamental solutions have been used both as an ingredient for the development of continuum models for microswimmers with hydrodynamic far-field interactions  \cite{saintillan2007orientational,saintillan2008instabilities,Saintillan_JRSI_Emergence_Coherent_2012} and for particle based simulations \cite{Stenhammar_PRL_Role_Correlations_2017} (see also the review \cite{Koch_ARFM_Collective_Hydrodynamics_2011} and references therein). In particular, based on continuum theories, it has been shown that 
suspensions of pusher particles generically induce a correlated dynamics beyond a threshold pusher concentration and system size \cite{saintillan2007orientational,saintillan2008instabilities,Saintillan_JRSI_Emergence_Coherent_2012} somewhat similar to the large scale flows observed in bacterial suspensions \cite{Cisneros_EF_Fluid_Dynamics_2007,Sokolov_PRE_Enhanced_Mixing_2009}. 
 Particle based models to simulate the hydrodynamic interactions in such suspensions have also been developed e.g. in ref. \cite{Krishnamurthy_JFM_Collective_Motion_2015,Stenhammar_PRL_Role_Correlations_2017}. Even for moderate swimmer densities, collective motion has been predicted in such swimmer suspensions as a consequence of the long-range nature of the hydrodynamic interactions between the swimmers \cite{Stenhammar_PRL_Role_Correlations_2017}.

\subsection{Hydrodynamic interactions in autophoretic Janus colloids} 
\begin{figure}
	\begin{center}
		\includegraphics[width=0.9\textwidth]{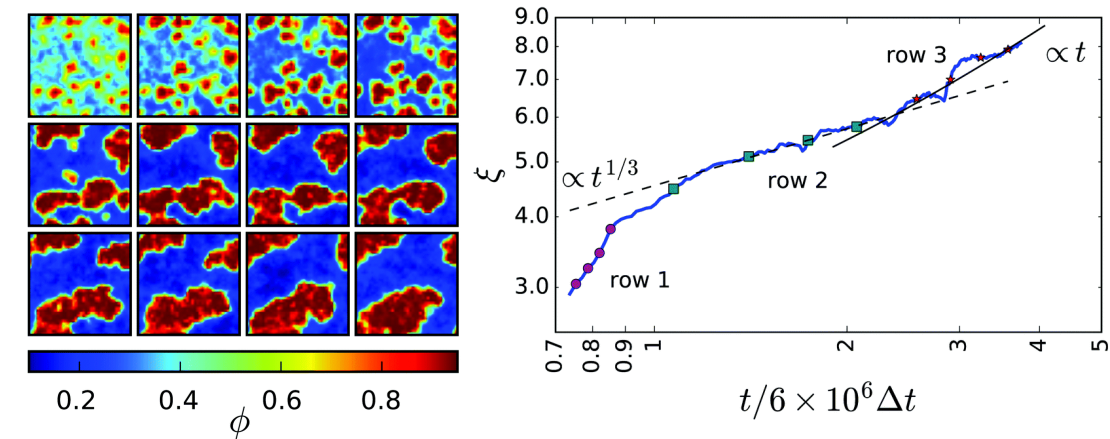}
		\caption{Left panel: Time-dependent snapshots of the spatially-resolved local squirmer density $\phi(x,y)$ showing the co-existence of a high-density (clusters) and a low-density phase. Right panel: Mean cluster size $\xi$ as a function of $t$. Figure reproduced 
		with permission 
		from \cite{blaschke2016phase}.}
		\label{blaschke}
	\end{center}
\end{figure}

Unlike most biological microswimmers, autophoretic Janus colloids have a rigid shape and move by inducing a gradient in a phoretic field which 
leads to directed motion in the same way as an 
externally imposed phoretic field gradient. In the simplest case this gradient creates a body force within a thin interfacial layer of the Janus colloids and drives a solvent dynamics across their own surface (slip velocity).
The fluid outside of the interfacial layer is then considered as force free and the particle motion is fully determined by the slip-velocity, 
leading to self-propulsion in the direction opposite to the surface averaged solvent flow \cite{golestanian2007designing,moran2017phoretic,anderson1989colloid}. 
For such swimmers, the hydrodynamic interactions 
either decay as $1/r^3$ or as $1/r^2$ in far-field 
and are often modeled based on the squirmer model as discussed below and are comparatively well explored. 
The $1/r^3$-behavior applies to the 
important special case of autophoretic 
Janus colloids with a uniform surface mobility and a radius of $R$ leading to the same flow field as an isotropic colloid in an external phoretic field. 
It reads at a point $\vec r$  
 well beyond the interfacial layer of the particle
$\vec{v}(\vec{r})=\frac{1}{2}\left(\frac{R}{r}\right)^{3}(3 \hat{r} \hat{r}-I) \cdot \vec{v}_{0}$ \cite{morrison1970electrophoresis}, 
where $r=|\vec r|$ and $\hat r = \vec r/r$. 
However, if the Janus colloid has a non-uniform surface mobility, the flow field features an additional $1/r^2$ far-field contribution, which scales for a half-coated Janus colloid with boundary conditions allowing to represent them as a squirmer as $|(\mu_1-\mu_2)/(\mu_1 + \mu_2)|$, where $\mu_1,\mu_2$ are the surface mobilities of the two hemispheres of the Janus-colloid under consideration \cite{popescu2018effective}. 
\vsb 
\underline{Simulations of interacting squirmers:}
\\The hydrodynamics of active Janus colloids which move by inducing a gradient in a phoretic field leading to a slip-velocity across their surface, outside of which the fluid can be considered as force-free (which does not apply to all active Janus colloids \cite{samin2015self}), 
can be captured by the squirmer model \cite{lighthill1952squirming,blake1971spherical,zottl2016emergent,yeomans2017hydrodynamics}. 
The squirmer model is based on a spherical solid particle, with a prescribed tangential flow velocity at each point $\vec{r}_s$ of its surface which reads in coordinate-free expression \cite{ishikawa2006hydrodynamic,lighthill1952squirming,blake1971spherical}
\begin{eqnarray}
	\vec{v}_s (\vec{r}_s)&=&\sum_{n=1} B_n \frac{2}{n(n+1)}\left[(\vec{e}\cdot \hat{r}_s)\hat{r}_s - \vec{e} \right] \frac{dP_n(\vec{e}\cdot \hat{r}_s)}{d(\vec{e}\cdot \hat{r}_s)}\\
	&=& B_1(1+\frac{B_2}{B_1} \vec{e}\cdot \hat{r}_s) \left[(\vec{e}\cdot \hat{r}_s)\hat{r}_s - \vec{e} \right] + ...
\end{eqnarray}
where $\vec{e}$ is the swimming direction of the squirmer, $\hat{r}_s=\vec{r}_s/|\vec{r}_s|$, $P_n(\vec{e}\cdot \hat{r}_s)$ is the $n$-th Legrendre polynomial and $B_n$ denotes the squirmer parameters characterizing the surface velocity modes. 
The parameters $B_1$ and $B_2$ have been shown to often capture the basic features of the microswimmers \cite{zottl2016emergent} and hence often the higher modes are ignored by setting $B_n=0$ for $n\geq 3$. Here $B_1$ is responsible for the source-dipole contribution to the flow-field due to the squirmer which decays as $r^{-3}$ \cite{lighthill1952squirming,blake1971spherical}, whereas the coefficient $B_2$ governs the strength of the force-dipole contribution decaying as $r^{-2}$. The swimming speed of the squirmer is given entirely in terms of the first mode $B_1$ as $v_0=2B_1/3$, whereas the second mode determines whether the swimmer is a pusher ($B_2<0$), a puller ($B_2>0$) or a neutral squirmer ($B_2=0$). The dimensionless squirmer parameter $\beta=B_2/B_1$ hence quantifies the ratio of the apolar active stresses generated by the squirmer to the squirmer polarity.
The squirmer model has been generalized in various ways e.g. to account for general-shapes \cite{shapere1989geometry}, non-symmetric slip-velocities \cite{thutupalli2018flow} (see above), deformable spheres \cite{felderhof1994small}, spheroidal shapes \cite{keller1977porous,Pohnl_JPCM_Axisymmetric_Spheroidal_2020}, squirmer dumbbells \cite{Clopes_SM_Hydrodynamic_Interactions_2020,Ishikawa_M_Stability_Dumbbell_2019} and squirmer rods \cite{Zantop_SM_Squirmer_Rods_2020}.
\vsb 
Detailed simulations of two squirmers using boundary element methods have been performed in \cite{ishikawa2006hydrodynamic} which have been complemented with analytical calculations confirming the expected form of the far-field interactions and in particular also the applicability of lubrication theory at interparticle distances which are short compared to the particle size \cite{ishikawa2006hydrodynamic}. The two-squirmer problem has recently been explored based on an exact analytical framework for certain cases \cite{Papavassiliou_JFM_Exact_Solutions_2017}
as well as based on 
lattice Boltzmann simulations \cite{Kuron_JCP_Lattice_Boltzmann_2019}.
Simulations of several squirmers based on Stokesian-dynamics \cite{ishikawa2007diffusion,ishikawa2007rheology,ishikawa2007diffusion}
have been used e.g. to explore the collective ensemble of monolayers which have been observed to form  coherent structures and other aggregates \cite{ishikawa2007diffusion} or to study the rheological properties of squirmer suspensions \cite{ishikawa2007rheology,Pagonabarraga_SM_Structure_Rheology_2013,Ishikawa_JFM_Rheology_Concentrated_2021}. Lattice-Boltzmann simulations of squirmers \cite{alarcon2013spontaneous} have shown intriguing polar ordering and flocking behavior in squirmer suspensions which can be be controlled solely via the squirmer parameter $\beta$. Multiparticle collision dynamics simulations in turn have been used to explore hydrodynamic interactions of squirmers confined between two plates so that they can move essentially only in 2D \cite{zottl2014hydrodynamics}. These simulations have shown that the collective behavior significantly 
depends on the type of swimmer which is considered (neutral squirmer, pusher or puller). Larger simulations of several thousand squirmers have been used later to explore the impact of hydrodynamic interactions on phase separation \cite{blaschke2016phase} (see Fig.~\ref{blaschke}). 
Other works studying squirmers in the presence of walls include refs. \cite{li2014hydrodynamic, Brumley_PRF_Stability_Arrays_2019,Kuhr_SM_Collective_Dynamics_2019,Kuhr_SM_Collective_Sedimentation_2017,Ruhle_EPJE_Emergent_Collective_2020,Ruhle_NJP_Gravityinduced_Dynamics_2018,Shen_PRF_Gravity_Induced_2019}.
An approximate method to characterize the near-field hydrodynamic interactions between squirmers via lubrication corrections has been developed in refs. \cite{Yoshinaga_EPJE_Hydrodynamic_Lubrication_2018,Yoshinaga_PRE_Hydrodynamic_Interactions_2017}, which has been further employed by the authors to uncover rich collective phases depending on the squirmer density and the squirmer parameter $\beta$. 
\vsb 
\noindent
\underline{Janus colloids not featuring a thin interfacial layer:}
\\Clearly, if the boundary layer is not thin compared to the size of the droplet, the concept of a slip velocity breaks down and the Janus colloids under consideration may create significant fluid flows also away from the surface. In this case, the self-generated fluid flows extend to fluid regions away from the surface \cite{samin2015self}, which has been argued to be relevant in particular for Janus-colloids in a near-critical water-lutidine mixtures\cite{samin2015self,gomez2017tuning}. 
For these swimmers, it has been observed in simulations that the flow-field far away from the surface is similar to 
a Stokes flow past a viscous droplet\cite{samin2015self}, suggesting a different form of the hydrodynamic interactions, which do not seem to be particularly strong in experiments \cite{buttinoni2013dynamical} whose collective behavior has been reasonably well described by the ABP model.

\vsb 
\subsection{Flow field measurements}
The flow field due to Janus colloids has been explicitly measured only for a few examples of active colloids, which we discuss in the present section. 
In ref. \cite{campbell2019experimental} the flow field has been explicitly measured for polystyrene colloids half coated with Pt in $H_2O_2$ solution based on tracers. While the surface reactions of these swimmers involve only neutral species, dissociation-association reaction in bulk create ions, 
allowing also electrophoretic effects to become relevant \cite{brown2014ionic,brown2017ionic} probably making the hydrodynamic flow fields more complicated. Surprisingly, in the experiments \cite{campbell2019experimental} no significant tracer motion was observed when using a uniformly coated particle rather than a Janus colloid, although one would expect tracer advection by isotropic osmotic flows induced by the chemically active colloid in its own interfacial layer as well as along the bottom substrate. \\Ref. \cite{bregulla2019flow} in turn has measured the 
flow fields caused by pinned $1\mu$-sized $Au-PS$ self-thermophoretic microswimmers in a water film between two parallel glass plates
(distance $1\mu m$ to $8\mu m$)
coated with triblock Pluronic F127. The flow fields were measured with 
thermophoretically inactive $250 nm$-sized 
gold tracers not directly responding to the temperature gradient induced by
the Janus particle. 
By performing numerical simulations of the flow field with different boundary conditions at the substrate (no-slip vs. thermo-osmotic slip), the authors have evaluated the 
relative contributions of the direct flow field caused by the action of the temperature gradient in the interfacial layer of the 
$Au-PS$ colloid (and the pinning force) and the osmotic contribution caused by the action of the temperature gradient on the substrate. 
This comparison unveils a significant 
thermo-osmotic flow along the substrate advecting the tracers. Notably, the variation of the wall spacing in these experiments 
clearly shows the decisive role played by the presence of a second wall on the osmotic flows which is generally expected due to the incompressibility of the flow, as mentioned above.
\\The role of boundaries has been explored rather systematically in a system of droplets in ref. \cite{thutupalli2018flow}
which we discuss here as the findings of these works may also be relevant to active colloids colloids. In this work experiments with $50\mu m$-sized active emulsions of monodisperse droplets of liquid crystal which create surface tension gradients at their interface driving active flows which are fueled by the free energy due to a slow dissolution of the droplets. 
These experiments 
have been combined with a model for a sphere with a prescribed surface slip velocity. The slip velocity has been constructed such that it
matches corresponding measurements of the flow outside of the droplets (obtained via particle image velocimetry and 200nm PS tracers) for various geometries (single bottom wall, Hele-shaw confinement with a gap size comparable to the particle size and much larger, air-water interface) leading to different characteristic flow fields and different aggregation phenomena, including the previously predicted 
flow-induced phase separation 
\cite{singh2016universal}. These experiments and simulations clearly show the decisive role played by the boundary conditions for the 
collective behavior of active particles.

\section{Combined phoretic and hydrodynamic interactions} 
We now discuss the interactions among autophoretic Janus colloids due to the combination of phoretic and hydrodynamic effects. 
In this section, we focus on autophoretic colloids in a single phoretic field (typically chemical concentration, temperature field or electric potential) which feature a thin interfacial layer and a slipping plane outside of which the solvent is force free. More general cases are briefly addressed in the section open challenges.

\subsection{General framework}
Unlike in the squirmer model which involves a prescribed slip velocity on the particle surface, in autophoretic colloidal microswimmers the slip velocity is determined by the gradient of a phoretic field $c(\vec{r},t)$ (or a combination of several fields) at the particle surface which is induced by the particle itself. 
The dynamics of the phoretic field is described by the (reaction)-advection-diffusion equation
\begin{equation}\label{diffusion}
	\frac{\partial c}{\partial t} + \vec{u}\cdot \nabla c = D_c\nabla^2 c
\end{equation}
where $D_c$ is the diffusion coefficient of the phoretic field, $\vec{u}(\vec r,t)$ is the solvent velocity and additional reaction terms can be incorporated via the boundary condition \cite{golestanian2007designing,Michelin_JFM_Phoretic_Selfpropulsion_2014}
\1 D_c\  \hat{n}_i\cdot \nabla c(\vec{r}^s_i,t)=-\mathcal{A}_i(\vec{r}^s_i,t) \label{cboundary} \2 
(or alternatively also directly in Eq. (\ref{diffusion}) via delta functions).  
Here $\vec{r}^s_i$ is a point on the surface of the $i$-th colloid ($i=1,2..N$) and $\hat n_i$ is the surface normal of that colloid and $\mathcal{A}_i(\vec r^s_i,t)$ is called the surface activity function which controls the production/consumption rate of the phoretic field at a point $(\vec r)$ and is nonzero only on the surface of a colloid. 
Gradients in the phoretic field drive a solvent flow and enter the Stokes equation (\ref{stokes}) as a boundary condition. In particular, the 
slip velocity on the surface of the $i$-th colloid reads \cite{Michelin_JFM_Phoretic_Selfpropulsion_2014,anderson1989colloid}:
\begin{equation}\label{slip_velocity}
	\vec{u}^s (\vec{r}^s_i)=\mu_i(\vec{r}^s_i)(I-\hat{n}_i\hat{n}_i)\nabla c(\vec{r}^s_i) \equiv \nabla_\parallel c(\vec{r}^s_i),
\end{equation}
where $\mu_i(\vec{r}_i^s)$ denotes the phoretic surface mobility of the $i$-th colloid, $I$ is the identity matrix and $\hat n_i \hat n_i$ is the tensor product of $\hat n_i$ with itself. 
The boundary condition for the Stokes equation (Eq.~\ref{stokes}) at the surface of the $i$-th particle then reads (see also  Sec.~\ref{hydro_interactions}) \cite{Michelin_JFM_Phoretic_Selfpropulsion_2014}
\1 
\vec{u}(\vec{r}^s_i)=\vec{v}_i+\vec{\omega}_i \times \vec{r}^s_i + \vec{u}^s (\vec{r}^s_i);\quad \vec u(|\vec r|\to \infty) =0 \label{stokesb}\2
where $\vec{v}_i$ and $\vec{\omega}_i$ are the translational and angular velocity of the $i$-th particle. 
In order to obtain the velocity $\vec{v}_i$ and the angular velocity $\vec{\omega}_i$ of the $N$ colloids in the system 
under the combined effect of phoretic and hydrodynamical interactions, one has to simultaneously solve for the phoretic field, Eq. (\ref{diffusion}), the solvent flow, Eq.~(\ref{stokes}) with $\vec f=0$, respectively complemented with the boundary conditions given by Eqs. (\ref{slip_velocity}) and (\ref{stokesb}), as well as for the dynamics of all colloids in the system, which are determined by the force- and torque free conditions$ \vec F_i=\vec T_i=0$
as detailed in Sec.~\ref{hydro_interactions}. 
These $6N+3+1$ equations serve as a complete system of equations to determine $\{\vec v_i,\vec \omega_i\},\vec u,c$.
\\In general we have a two-way coupling between the phoretic field and the solvent field, requiring a simultaneous solution of the Stokes equation and the advection-diffusion equation for the chemical field. 
A frequently used simplification is to focus on the  
regime of low P\'eclet number $Pe_c=R u/D_c$, which quantifies the relative importance of solute advection and diffusion for colloids of radius $R$
and a typical solvent velocity of $u$. Eqs.~(\ref{stokes},\ref{diffusion},\ref{slip_velocity},\ref{stokesb}) can be written in dimensionless form by choosing the units of length, mass and time as $R$, $\eta R^2 D_c/(AM)$ and $RD_c/(AM)$ with $\eta$ being the dynamic viscosity of the solvent and $A$ and $M$ denoting the maximum values of surface activity $\mathcal{A}_i(\vec{r}^s_i,t)$ and surface mobility $\mu_i(\vec{r}^s_i)$ of the colloids. Eq.~(\ref{diffusion}) can then be written as \cite{Michelin_PF_Spontaneous_Autophoretic_2013}
\begin{equation}\label{diffusion_nondim}
	Pe_c \left(\frac{\partial c^*}{\partial t^*} + \vec{u}^*\cdot \nabla c^* \right)= \nabla^2 c^*
\end{equation}
where $\vec{u}^*=\4{D_c}{AM}\vec{u}$, $t^*=\4{AM}{RD_c}t$ and $c^*=\4{D_c}{AR}c$ denote the dimensionless fluid velocity, time and phoretic field concentration respectively. This shows that both the advection term and the time-derivative of the phoretic field which is responsible for memory effects scale with the same parameter: $Pe_c$. For $Pe_c\ll 1$ the time derivative and the advection term in Eq. (\ref{diffusion}) can be neglected yielding the Laplace equation
\1 D_c\nabla^2 c=0.\label{dinst}\2
This equation is now independent of the solvent flow and can be solved separately. Together with boundary conditions (\ref{cboundary}) for axisymmetric colloids with a time-independent surface activity the solution reads in spherical polar coordinates \cite{golestanian2007designing,Kanso_JCP_Phoretic_Hydrodynamic_2019} 
\1
c(r, \theta)=\sum_{m=0}^{\infty} \frac{a A_{m}}{D(m+1)}\left(\frac{a}{r}\right)^{m+1} L_{m}(\cos \theta),
\2
where $A_{m}=-\frac{2 m+1}{2} \int_{-1}^{1} A L_{m} \sin \theta \mathrm{~d} \theta $ are the Legendre moments of the activity distribution $A(\cos \theta)$ and $L_m(\cos \theta)$ is the $m$-th Legendre polynomial.
\\For an isolated spherical half-coated Janus colloid one recovers the self-propulsion velocity given by Eq. (\ref{selfpr}) \cite{golestanian2007designing}. 
For two Janus colloids, exact analytical results in the regime $Pe_c\ll 1$ have been 
reported for certain special cases only fairly recently \cite{SharifiMood_JFM_Pair_Interaction_2016,Nasouri_JFM_Exact_Axisymmetric_2020,Nasouri_PRL_Exact_Phoretic_2020}.

\subsection{Hydro-phoretic interactions between a pair of colloids\label{chemhydropairs}}
\begin{figure}
	\begin{center}
		\includegraphics[width=1\textwidth]{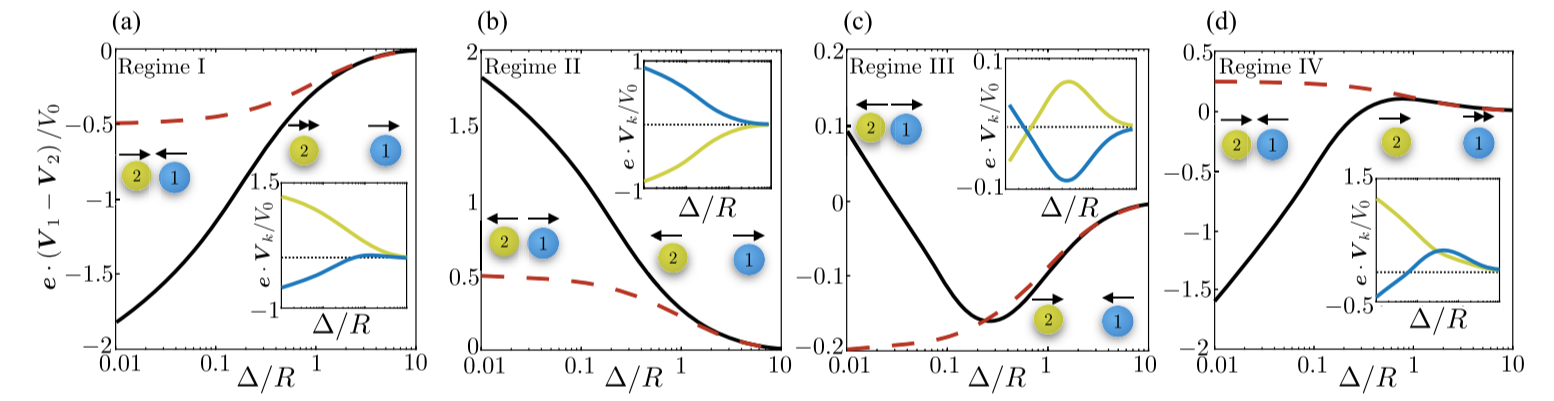}
		\caption{Relative velocity of two isotropic chemically active colloids as a function of their gap size in different parameter regimes. 
		The solid lines correspond to the exact relative motion caused by the exact phoretic interactions of the particles whereas the dashed lines represents the far-field approximation. The arrows indicate the velocity of each particle, the magnitude of which is quantified in the insets (dotted lines indicate zero speed). Figure reproduced 
		with permission 
		from \cite{nasouri2020exact}.}
		\label{nasouri}
	\end{center}
\end{figure}

\underline{Homogeneous pairs:}
\\A comparatively early exploration of the near-field hydro-phoretic interactions between  a pair of active Janus colloids is ref. \cite{SharifiMood_JFM_Pair_Interaction_2016} (2016). This work has shown that when the colloids are oriented such that their active sides are facing each other, the solute concentration increases between them leading to an increased self-propulsion. In contrast, when the passive sides face each other, the swimming velocities is reduced due to effective lubrication forces. The authors have also studied the effects of the relative area which is covered by the catalytic cap and the relative orientation of the colloids on their collective dynamics. Depending on these two parameters, they showed that the colloids can either swim towards each other and come into contact (assembly) or move away from each other (escape) (see Fig.~\ref{nasouri}). Refs. \cite{Nasouri_JFM_Exact_Axisymmetric_2020,nasouri2020exact}, have presented a generic solution for the relative velocity of two half-coated Janus colloids for certain symmetric orientations about their axes of approach. They showed that the relative velocity of a pair of neighboring Janus colloids can be decomposed into a sum of three geometric functions which represent self-propulsion, neighbor-induced motions and self-generated neighbor-reflected motion respectively. One key result of this study is that for small gap sizes between the colloids, the strength of the latter two near-field effects dominates, resulting in a non-monotonic pair interaction, which can change from attractive to repulsive as the interparticle distance changes. For smaller gap sizes in appropriate parameter regimes, the colloids were shown to either reach a stable equilibrium at a finite gap size or form a complex that can dissociate in the presence of sufficiently strong noise.
\\The individual contributions of the chemical and hydrodynamic fields in the total interaction between a pair of active colloids has also been studied in \cite{Popescu_EPJE_Effective_Squirmer_2018}. Here it was shown that, in particular, an effective squirmer model can correctly predict the velocity of active Janus colloids with radius $R$ approaching another one along the line of their symmetry axis for separation distances $H\gtrsim 4R$. 
Another analytical approach to study the hydro-phoretic pair interaction of diffusiophoretic colloids via the Method of Reflection (MoR) has been presented recently in ref. \cite{Varma_PRF_Modeling_Chemohydrodynamic_2019}. This method was shown to capture not only the instantaneous velocity of a pair of Janus colloids as a function of their gap size but also predicted the essential features of the long-term dynamics of the phoretic particles. This approach allows to explicitly distinguish between the direct chemical, direct hydrodynamic, and combined chemo-hydrodynamic interactions, thus allowing one to analyze the role and importance of each of these interaction components on the overall particle dynamics. 
\vsb 
\underline{Heterogeneous pairs:}
\\As discussed in section \ref{nonrec}, nonidentical phoretically active particles show nonreciprocal phoretic interactions leading to a center of mass motion often resulting in active molecule formation. Such situations have also been explored for particles with hydrodynamic and phoretic interactions. 
In particular, the hydrodynamic and phoretic interactions between a pair of uniform colloids which both create and respond to the phoretic field in the same way but which feature a different size has been explored in ref.  \cite{Michelin_EPJE_Autophoretic_Locomotion_2015}. 
Here, it has been found that at small interparticle distances, 
the two spheres moved together with the larger sphere ahead, whereas at large interparticle distances the center of mass velocity reversed. In the far field, the velocity decreased following an inverse square law with respect to the distance as expected based on the $1/r^2$ law (\ref{1r2law}). A similar setup of catalytic dimers has also been studied in ref. 
\cite{Reigh_SM_Catalytic_Dimer_2015} where two isotropic spheres of different radii have been linked by a rigid bond, only one of which 
catalyzes a chemical reaction on its surface, whereas the other one serves as a passive tracer. By using Molecular Dynamics (MD) and Multiparticle Collision Dynamics (MPCD), the authors have resolved the chemical concentration and hydrodynamic flow fields both in the near- and the far-field regime. In the far-field, the dimers were shown to produce hydrodynamic effects with a $r^{-2}$ decay law similar to those of biological swimmers and distinct from the $r^{-3}$ decay of spherical Janus motors with a uniform surface mobility. The bond lengths and sphere sizes were shown to be key controlling factors for the self-propulsion velocity.  The same setup but with non-bonded spheres has been studied in ref.  \cite{Reigh_SM_Diffusiophoretically_Induced_2018}. Here it was shown that the chemically inactive sphere moves by diffusiophoresis in the concentration gradient induced by the chemically active sphere. The phoretic motion causes a flow field dragging the chemically active sphere towards the passive one leading to a self-assembled self-propelling dimer motor.

\subsection{Collective effects due to hydro-phoretic interactions in bulk}
\begin{figure}
	\begin{center}
		\includegraphics[width=0.8\textwidth]{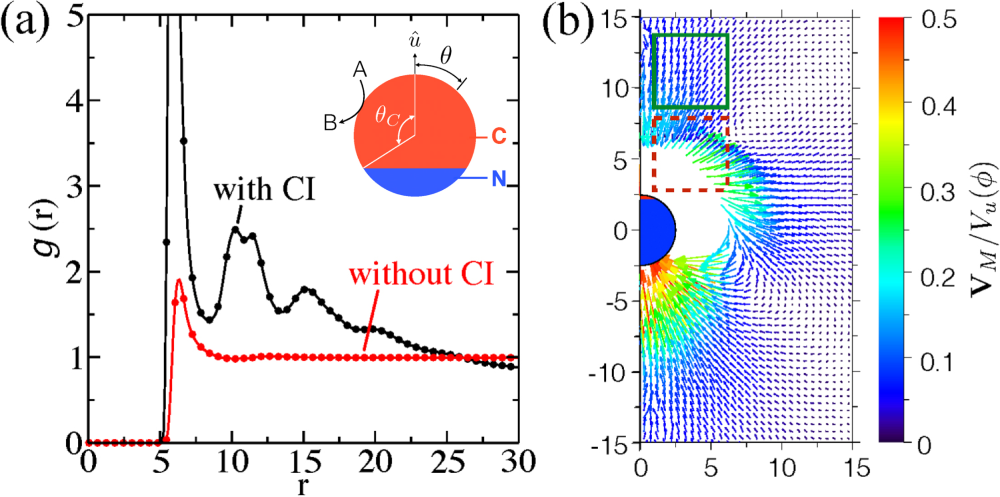}
		\caption{(a): Radial distribution function $g(r)$ of hydrodynamically interacting Janus motors with (black) and without the chemical interactions (red). The red and blue colors on the Janus motor (inset) denote catalytic and non-catalytic parts respectively. The near-absence of structures at large $r$ in the absence of chemical interactions 		
		shows that the chemical interactions are important for structure formation at the considered low densities. Panel (b) shows the motor velocity field when the chemical interactions are switched off. Figures adapted 
		with permission 
		from \cite{Huang_NJP_Chemotactic_Hydrodynamic_2017}.}
		\label{huang}
	\end{center}
\end{figure}

\begin{figure}
	\begin{center}
		\includegraphics[width=0.8\textwidth]{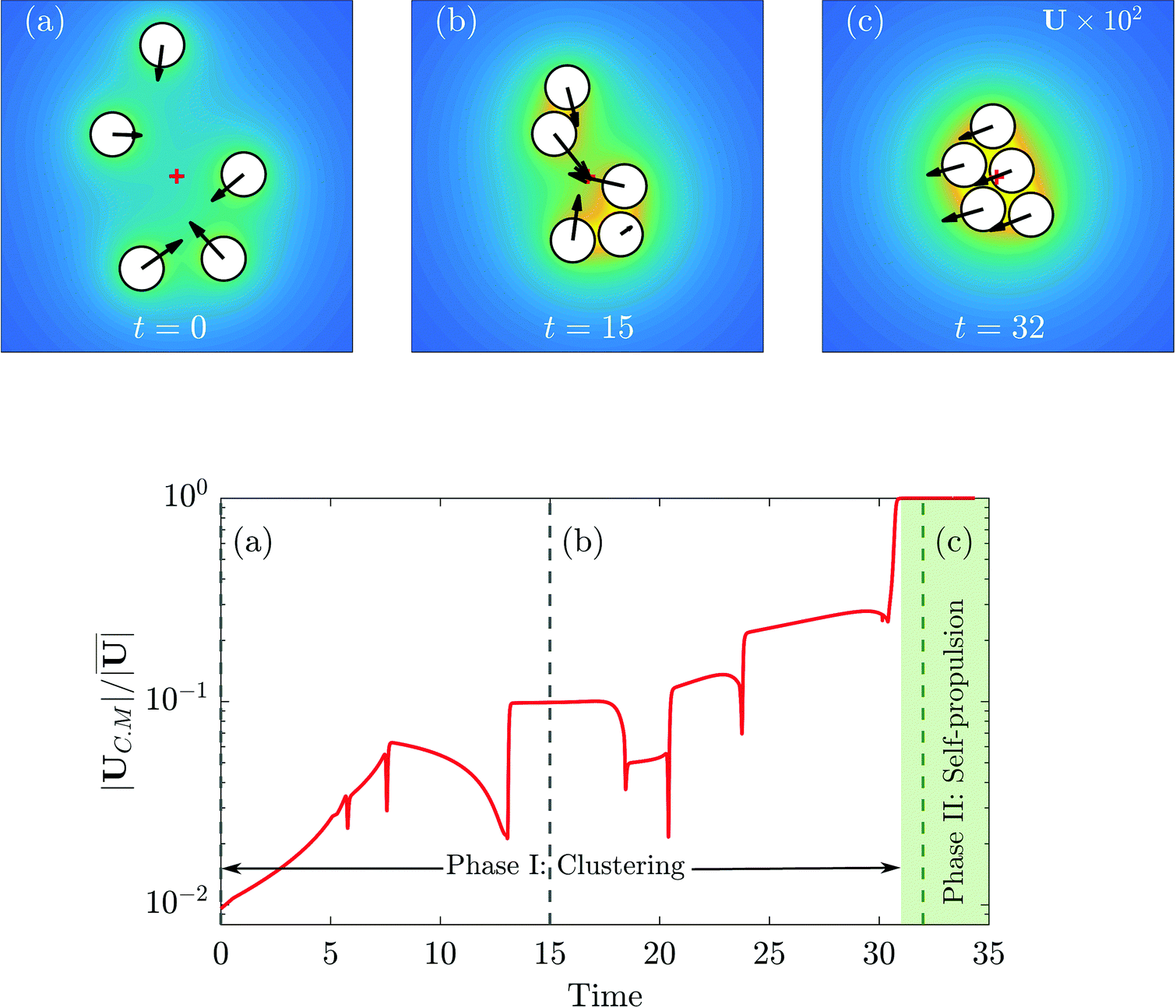}
		\caption{(a,b,c) Snapshots of the time-evolution 
of five chemically isotropic particles with phoretic and hydrodynamic interactions. The interactions induce geometrically anisotropic self-propelling clusters. The colormap denotes the strength (concentration) of the phoretic field and the black arrows on the particles denote the velocity of the particles. Bottom panel: Time-evolution of the ratio of the speed of the center of mass of the particles and the average individual speed. Figures reproduced 
		from \cite{Varma_SM_Clusteringinduced_Selfpropulsion_2018}.}
		\label{varma}
	\end{center}
\end{figure}

A comparatively early work exploring the combination of chemical interactions and hydrodynamics is provided by ref. \cite{lushi2012collective} which uses the framework of refs. \cite{saintillan2008instabilities,saintillan2013active} to explore how self-generated flows affect the collective dynamics of chemotactic run-and-tumble bacteria. This work has developed a continuum theory accounting for the impact of hydrodynamic far-field interactions on the collective particle dynamics. Numerical solutions of the field equations recover the Keller-Segel instability if the chemical interactions dominate but also lead to a new ``mixed phase'' if both chemical and hydrodynamic interactions are important. 
\\In ref. \cite{Huang_NJP_Chemotactic_Hydrodynamic_2017} multiparticle collision-dynamics simulations have been used to explore the collective behavior of diffusiophoretic Janus colloids. In particular, this work has calculated the radial distribution function $g(r)$ of 
the Janus colloids in the presence of their full chemical and hydrodynamic cross-interactions for different densities and coating geometries observing a large peak at close interparticle contact and several other peaks indicating structure formation at larger scales. 
Notably, when retaining only the hydrodynamic interactions and switching the chemical interactions off, the authors have found that $g(r)$ only features a small peak at close interparticle distance and almost no structure at larger interparticle distances (see Fig.~\ref{huang}), pointing to the key role of phoretic interactions at low particle density. 
\\A very recent study of the combined effect of phoretic and hydrodynamic interactions on the collective behavior of Janus colloids has been performed in ref. \cite{Scagliarini_SM_Unravelling_Role_2020} based on Lattice Boltzmann simulations coupled to a finite-difference solver for the phoretic field, it was shown that the active colloids exhibit a novel clustering phase when they are `chemoattractive': here, interestingly, hydrodynamic interactions have been found to suppress large clusters in parameter regimes where the same system without hydrodynamic interactions exhibited coarsening with the mean cluster size growing in time $t$ approximately as $t^{1/2}$.
\\In ref. 
\cite{Reigh_SM_Catalytic_Dimer_2015} the dynamics of dimers made of two differently sized colloids with hydrodynamic and phoretic interactions has been explored both analytically and based on multi-particle collision dynamics. The collective behavior of up to 
$6\times 10^3$ colloidal dimers (and $\sim 10^8$ fluid particles), here consisting of two rigidly linked spherical colloids, has been explored in ref. \cite{Colberg_JCP_Manybody_Dynamics_2017} based on multi-particle collision dynamics simulations. These simulations have 
explicitly modeled the chemical reaction kinetics on the dimer surface, dimer-dimer interactions, hydrodynamic interactions and thermal noise. The dimers were found to segregate into low-density gas-like and high-density disordered solid-like phases,
which were observed to coarsen in a nontrivial way, following a power law only at relatively early times and deviating from it significantly before the reached cluster sizes were comparable to the system size. 
\\In ref. \cite{Varma_SM_Clusteringinduced_Selfpropulsion_2018}, the authors showed that a collection of isolated chemically- and geometrically-isotropic particles can achieve self-propulsion by forming geometrically-anisotropic clusters via phoretic and hydrodynamic interactions (see Fig.~\ref{varma}). These particles have been observed to initially cluster which is followed by a directed motion (self-propulsion) of the cluster. It has been observed that larger particle numbers led to larger propulsion velocity.

\subsection{Hydro-phoretic interactions near walls or interfaces}
Walls and interfaces distort the shape of the phoretic field due to an autophoretic colloid as compared to the one produced by a colloid in bulk. In particular, while the field around an isolated symmetric Janus colloid in bulk is of course symmetric with respect to the symmetry axis of the latter, substrates and other confinements typically desymmetrize the flow field and the phoretic field. These fields act back onto the Janus colloid in that they advect the colloid and desymmetrize the slip-velocities along its surface, leading to an additional translation and rotation as compared to the motion of the colloid in bulk. In addition, the nonuniform phoretic field due a colloid in general also induces an osmotic flow along the walls, which additionally advects the colloids in the system. 
In the following we first briefly refer to works which have explored the impact of phoretic and hydrodynamic self-interactions (often not discussing osmotic self-interactions) 
and then turn our attention to the impact of confinement on the collective behavior of Janus-colloids with combined phoretic and hydrodynamic interactions (also to date often neglecting osmotic cross-interactions; see open challenges in section \ref{ochallenges}). 
\vsb
\underline{Self-interactions:}
\\Wall induced 
phoretic and hydrodynamic self-interactions can lead to a rich dynamics of individual Janus colloids including wall-attractions, wall-repulsions, interaction-induced particle reorientations as well as sliding states, where the particle
maintains a fixed height and orientation as it moves along the
wall, and hovering states, where the particle remains fixed in space but its orientation changes. In the following we specifically refer to several theoretical and numerical works which have exemplified these states in various situations.  
\\Ref. \cite{Popescu_JCP_Confinement_Effects_2009} has studied the effect of a spherical shell confinement on a diffusiophoretic Janus particle with a point-like catalytic site. In this work it has been found that the confinement enhances the steepness of the solute-gradients but also the viscous friction. The result of these counteracting effects is a possible increase of the particle velocity as compared to its in bulk value. 
\\In ref. \cite{Crowdy_JFM_Wall_Effects_2013}, the author has explored the self-propulsion of a diffusiophoretic two-dimensional Janus
in the presence of a straight no-slip wall. 
Also here, it has been found that the colloid's translation velocity can be enhanced by the confinement. 
It has also been observed in this work that the colloid is attracted by the wall when its active hemisphere faces the wall and that it is repelled by the wall for various other configurations. 
A similar setup involving an infinite no-slip wall but with a spherical diffusiophoretic Janus 3D colloid was studied in ref. \cite{Mozaffari_PoF_Selfdiffusiophoretic_Colloidal_2016}. In this work it has been shown that depending on initial orientation and the 
area of the catalytic coverage of the colloid different scenarios are possible. In some cases the Janus colloids are simply attracted to or repelled from the wall without any rotation, or even stop in front of it, whereas in other cases a comparatively rich variety of 
dynamical behaviors including a skimming motion along the surface has been observed. The skimming and stationary states were shown to persists even at higher P\'eclet number above $10^3$ in ref. \cite{Mozaffari_PRF_Selfpropelled_Colloidal_2018}.
The hydrodynamic and phoretic contributions to the colloid-wall interactions have also been studied separately in refs. \cite{Ibrahim_E_Dynamics_Selfphoretic_2015,Ibrahim_EPJST_How_Walls_2016} which showed that phoretic interactions led to significant 
wall attractions or repulsions whereas hydrodynamic interactions were found to modify the effective 
drag experienced by the colloid as well as a weak angular velocity leading to re-orientation of the colloid. 
In particular, colloids 
with low catalytic coverage and hence relatively weak phoretic wall-interactions were found to eventually move away from the wall, whereas for intermediate coverage they exhibited `bound states' sliding along the wall. For comparatively large areas of catalytic coverage the sliding motion ceased and led to a hovering motion near the wall. Similar dynamical behaviors have also been reported slightly earlier in ref. \cite{Uspal_SM_Selfpropulsion_Catalytically_2014} where the authors studied essentially the same problem based on detailed numerical calculations. 
\begin{figure}
	\begin{center}
		\includegraphics[width=0.80\textwidth]{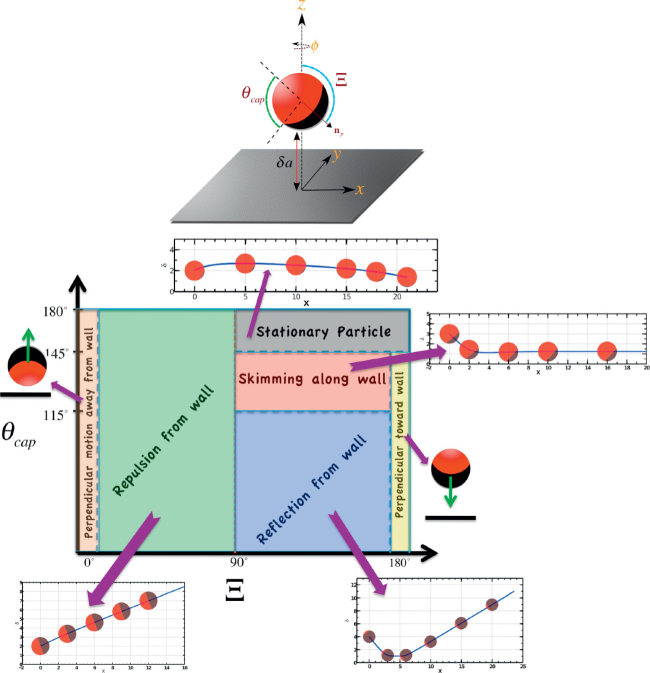}
		\caption{(Top panel) Schematic representation of a Janus colloid 
		whose self-propulsion direction features an angle  
		$\Xi$ to the wall normal. (Bottom panel) The different possible dynamics of Janus swimmers near a wall due to hydrodynamic and phoretic self-interactions depending on the size of the catalytic cap and the initial orientation. Figures reproduced 
		with permission 
		from \cite{Mozaffari_PoF_Selfdiffusiophoretic_Colloidal_2016}.}
		\label{mozafferi}
	\end{center}
\end{figure}
Very recently, also the behavior of an active colloid confined between two walls has also been studied in ref.  \cite{Choudhary__Selfpropulsion_2D_2021}. In addition to the sliding and hovering motion already noted in presence of a single wall, they reported two additional states characterized by damped and periodic oscillatory movement along the confined channel. In another study \cite{Yang_L_SelfDiffusiophoresis_Janus_2016}, the translocation of a single as well as a pair of active Janus colloids through a cylindrical pore has been shown to significantly depend on both phoretic and hydrodynamic interactions.
\vsb 
\underline{Cross-interactions:}
\\External walls and interfaces do not only lead to additional osmotic cross-interactions in active colloids, but they can also 
significantly modify their hydrodynamic and phoretic cross-interactions. We are just at the beginning of understanding the impact of wall-effects on the collective behavior of Janus colloids and refer in the following to some very recent works which have started to explore them. 
\\The collective effects due to competing chemical and hydrodynamic interactions among multiple active colloids in presence of a bottom substrate has been discussed in ref. \cite{Singh_JCP_Competing_Chemical_2019}. When the colloids are oriented perpendicular to the substrate, the chemical and hydrodynamic interactions can be expressed in terms of non-pairwise potentials. The potentials can be purely attractive or repulsive or can have a barrier, depending on the relative strengths of the monopolar and dipolar contributions to the surface activity of the colloids. Specifically, for
Janus colloids aligned normally to and stalled at the wall,
phoretic interactions have been found to scale as $1/r^2$ and hydrodynamic ones
as $1/r^4$. In such cases, 
many-body simulations predicted arrested phase separation, with a mean cluster size set by the competing chemical and hydrodynamic interactions.
The impact of a Hele-Shaw confinement on 
hydro-phoretic interactions has been studied in ref. \cite{Kanso_JCP_Phoretic_Hydrodynamic_2019}. 
In the particular limit which has been considered in this work it was found that the 
phoretic and hydrodynamic interactions feature the same form, and scale as $1/r^2$ with the interparticle distance in far-field, but act in opposite directions with the phoretic interactions being twice as strong as the hydrodynamic ones. 
This work has also studied the collective behavior of two and many particles based on linear stability analyses and numerical calculations, which have unveiled 
global swirling motion as well as stationary clusters. 
\\For active Janus colloids pinned on a substrate, ref. \cite{Robertson_C_Collective_Orientational_2018} has shown that the combined effect of 
hydrodynamic and phoretic interactions can lead to states with highly correlated particle orientations. Also in this work the 
phoretic interactions have been found to play the dominant role for the observed collective behavior.

\subsection{Simulation methods} \label{simulation}
Large ensembles of active particles with phoretic far-field interactions can be simulated in an efficient and relatively straightforward way e.g. by coupling Brownian dynamics simulations to finite difference solvers for the phoretic field \cite{liebchen2017phoretic} or by integrating out the 
phoretic field and performing 
particle-only simulations \cite{Liebchen_JCP_Which_Interactions_2019} both allowing to explore the long-time dynamics of large system sizes comprising $N \sim 10^5$ particles, which can be extended to even larger systems in the future.
However, to also account for hydrodynamic (far- and near-field) interactions as well as for phoretic (and osmotic) near-field interactions more sophisticated simulation schemes are required. 
In the following we briefly discuss some of the most remarkable and most commonly used schemes to simulate interacting active colloids (and other microswimmers) and 
briefly comment on their limitations with respect to the presently reachable system sizes in section \ref{ochallenges}.
\vsb 
\underline{Mesoscale simulation methods:}
\\Hydrodynamic interactions in active colloids and other microswimmers are often described based on mesoscale simulation methods involving a coarse-grained description of the solvent. These methods typically sacrifice 
some detail regarding the molecular interactions and focus on conservation laws (energy, mass, momentum).
Such methods comprise the Lattice-Boltzmann (LB) simulation method \cite{kruger2017lattice,chen1998lattice,desplat2001ludwig,Aidun_JSP_Lattice_Boltzmann_1995,Carenza_EPJE_Lattice_Boltzmann_2019} where a simplified and discretized version of the Boltzmann-equation is solved from which the Navier-Stokes equations can
be derived; dissipative particle dynamics (DPD) \cite{hoogerbrugge1992simulating,espanol1995statistical,pagonabarraga2001dissipative,espanol2017perspective} which uses `particles'
representing portions of fluids, e.g. based on a discretization of the Navier-Stokes equations using smoothed
particle hydrodynamics (SPH) \cite{monaghan2005smoothed} or by applying coarse-graining methods to link micro- and
mesoscales \cite{espanol2017perspective}, as well as direct simulation Monte Carlo \cite{alexander1997direct,oran1998direct}
and the related Multi-Particle Collision dynamics (MPCD) \cite{kapral2008multiparticle,malevanets1999mesoscopic,gompper2009multi} (also referred to as Stochastic Rotation Dynamics (SRD). The MPCD method constructs collisions
representing many microscopic collisions, based on the idea that individual collisions are unimportant, but
only their net effect.
\vsb
\underline{Hydrodynamically and phoretically interacting microswimmers:}
\\These methods have been widely applied to active matter. The 
Lattice Boltzmann (LB) method in particular has been used to study interactions in active particles such as the squirmer-squirmer interactions \cite{Kuron_JCP_Lattice_Boltzmann_2019,Kuron_SM_Hydrodynamic_Mobility_2019,Llopis_JoNFM_Hydrodynamic_Interactions_2010}, the rheological properties of squirmer suspensions \cite{Pagonabarraga_SM_Structure_Rheology_2013} and hydrodynamic interactions of squirmers near surfaces \cite{Li_PRE_Hydrodynamic_Interaction_2014}. A variant of the LB method 
allowing to describe very large ensembles of up to $\sim 10^6$ particles, at the expense of accounting for far-field interactions only has been developed in ref. \cite{nash2010run} and used in ref. 
\cite{stenhammar2017role,bardfalvy2020symmetric,vskultety2020swimming} to explore large ensembles of hydrodynamically interacting dilute particle suspensions. 
Very recently, the Lattice Boltzmann method has been generalized to resolve the near-field phoretic and hydrodynamic interactions of active colloids \cite{Scagliarini_SM_Unravelling_Role_2020}.
In this work the authors have 
observed that phoretic interactions are responsible for cluster formation (with the phoretic mobility being the key parameter), whereas hydrodynamic interactions inhibit cluster growth.
\\The MPCD method in turn has been used e.g. to explore hydrodynamic interactions in squirmers \cite{Gotze_PRE_Mesoscale_Simulations_2010,zottl2014hydrodynamics,blaschke2016phase,Zottl_EPJE_Simulating_Squirmers_2018} and has also been generalized to account also for the chemical reaction dynamics at the surface of active colloids and to study the combined effects of phoretic and hydrodynamic interactions in 
self-diffusiophoretic colloids \cite{Huang_NJP_Chemotactic_Hydrodynamic_2017,Buyl_N_Phoretic_Selfpropulsion_2013} as well as chemically powered nanomotors \cite{Colberg_JCP_Manybody_Dynamics_2017} and surface attached nanorotors \cite{Robertson_C_Collective_Orientational_2018}.
\\Another simulation method which is frequently used to study the interactions between (spherical) active particles is the 
Stokesian dynamics method \cite{brady1988stokesian}.
This method has been applied e.g. to 
study hydrodynamic interactions between large ensembles of (spherical) colloids \cite{Fiore_JFM_Fast_Stokesian_2019,Swan_PF_Modeling_Hydrodynamic_2011}. A variant of this method, the accelerated Stokesian dynamics approach \cite{Sierou_JFM_Accelerated_Stokesian_2001} has been recently been combined with existing methods to solve the many-body Laplace equation \cite{bonnecaze1990method}, to determine the  
distribution of the chemical concentration field in diffusiophoretic active colloids \cite{Yan_JCP_Behavior_Active_2016}. 
\\Other possible approaches to numerically study the combined hydrodynamic and phoretic interactions in ensembles of autophoretic active colloids have been formulated based on spectral expansions of the Laplace and Stokes equations \cite{Singh_JCP_Competing_Chemical_2019,Singh_JOSS_PyStokes_Phoresis_2020} 
as well as on the Force-Coupling Method (FCM) \cite{Delmotte_JoCP_Largescale_Simulation_2015}, which has been recently extended to the so-called diffusiophoretic Force-Coupling method \cite{RojasPerez_JFM_Hydrochemical_Interactions_2021}. 

\section{Open challenges\label{ochallenges}}
\subsection{Experimental characterization of cross-interactions}
Most measurements of phoretic and osmotic cross-interactions have so-far been performed based on tracking the 
motion of tracer particles under the influence of pinned or freely moving active colloids (see section \ref{exps}). 
However, detailed measurements quantifying the cross-interactions 
of two (and more) Janus colloids are hardly available. In addition, in most experiments measuring the interactions of active colloid and tracers, the relative importance of osmotic, phoretic and sometimes also hydrodynamic interactions remains largely unclear. 

This leads to the unfortunate situation that experiments using active colloids have observed a plethora of spectacular phenomena such as e.g. dynamic clustering which has also been predicted by various effective models, but where it remains largely unclear how these predictions compare with specific experiments. 
In fact, in view of the lack of sufficient experimental information on the strength of the various cross-interactions in active colloids, many modeling works choose to focus on predicting ``generic'' routes towards experimentally observed phenomena, often leaving it unclear how these routes actually compare to specific experiments. For example, a large variety of different mechanisms has been predicted to explain dynamic clustering 
but to date it is still under discussion which of them might explain the observations in the canonical experiments \cite{theurkauff2012dynamic,palacci2013living,buttinoni2013dynamical,ginot2018aggregation}.
This leaves a major gap in our understanding of the detailed mechanisms leading to some of the most spectacular collective phenomena in active colloids. 
\\Let us now discuss some open key points some of which future experiments might choose to address. 
\begin{itemize}
	\item Relative importance of phoretic, osmotic and hydrodynamic interactions: 
	\\Existing measurements have typically been performed near substrates and have measured the combined effect of osmotic, phoretic and hydrodynamic interactions. In some experiments \cite{palacci2013living,singh2017non} it has been observed that tracers are attracted from all directions to Janus particles, indicating that osmotic and phoretic interactions dominate over hydrodynamic ones in these experiments. 
	However, to date 
	it remains unclear in most experiments how strong the contributions of the individual interactions are, which makes it hard to formulate models reliably describing specific experiments (or a specific type of colloids).	\\(i) The strength of hydrodynamic interactions among tracers and Janus colloids (or among Janus-particles) in particular could in principle be separated from the other two interactions by quantifying the cross-interactions in different directions separately. 
	\\(ii) Since osmotic interactions hinge on the presence of external walls, an obvious way to separate them from phoretic interactions would be to compare measurements of the time evolution of the (ensemble averaged) relative distance between tracers and active colloids in 3D bulk and near substrates. 
	A perhaps more feasible approach could be to systematically vary the substrate properties and to measure how the overall interactions change.
	Finally experiments could in principle also exploit that the strength of osmotic interactions in colloids which are confined between parallel plates is expected to strongly depend on the distance between the plates
	due to solvent incompressibility (see also section \ref{phoros}). 
	\\(iii) First direct measurements of the flow field around Janus colloids \cite{bregulla2019flow,campbell2019experimental}
	and of the concentration gradient or the pH-field \cite{moeller2021} have 
	provided valuable information for future models of the cross-interactions among the colloids. While being experimentally demanding it would be interesting to understand how these fields change depending on the substrate properties and the type of colloid under consideration. 
	\item Near-field interactions: 
	\\Clearly, understanding the strength and functional of the near-field cross-interactions in active colloids would be highly important to test upcoming theories predicting their form and to understand collective behavior. This applies even to dilute colloids which often form clusters and other structures where the colloids closely approach each other. 
	\item Range of cross-interactions in active colloids: 
	\\The range of the pair-interactions among active colloids is still rather unclear. Phoretic interactions in particular are often called ``long-ranged''. 
	This is somewhat supported by fits of the 
	experimental measurements of the relative distance between Janus colloids and tracers to the $1/r^2$ law \cite{palacci2013living,singh2017non} as discussed in section \ref{phos}. 
	However, fits to a Yukawa/screened Coulomb interaction (\ref{yuk}) seem to lead to an even better agreement with the experimental data \cite{hauke2020clustering,liebchen2019interactions}, indicating that phoretic and osmotic cross-interactions in active colloids might actually have a finite range, indicating the presence of some mechanisms leading to effective screening e.g. due to bulk reactions. 
	These fits suggest that the range of phoretic (and osmotic) cross-interactions compares to a few times the size of the colloids. However, to reliably find out if phoretic cross-interactions are truly long-ranged and if not to reliably quantify their range, more precise measurements would be required. 
	\item Non-pairwise character: 
	\\Since phoretic interactions are expected to be non-pairwise, it would also be interesting to measure how the (phoretic) interactions between two particles change if a third particle is present. 
	Such measurements might be challenging but could provide important information to test detailed simulation schemes and the underlying models.
\end{itemize}

\subsection{Interactions for multiple phoretic fields and thick boundary layers}
Most theories describing phoretically interacting Janus colloids focus on relatively simple canonical models describing perfectly isotropic colloids featuring a thin interfacial layer (and a slip velocity) outside of which the fluid is essentially force free and on cases 
where only one (or two \cite{saha2014clusters,grauer2020swarm}) effective phoretic field(s) and one phoretic mechanism is (are) relevant. 
However, the various of types of Janus colloids used in different experiments show strong differences in their  
cross-interactions \cite{wang2020practical} which have not received much attention in existing modeling works and remain rather poorly understood. 
\\Specific open challenges include the development of a systematic understanding of the cross-interactions in active colloids in the following cases: 
\begin{itemize}
	\item Several phoretic fields: Although many works use only one effective phoretic field to describe the phoretic cross-interactions in Janus colloids it has not been generally shown when such a single effective phoretic field is representative for cases where several fields are present. 
	Clearly, using only one phoretic field is sensible for self-thermophoretic Janus colloids, but for diffusiophoretic ones at least one fuel and one product species are present and further species could play an important role, e.g. when bulk reactions take place. More extreme situations might occur for photocatalytic microswimmers where it is often unclear which chemical species are involved in their self-propulsion mechanism and hence in their interactions \cite{wang2020practical}. 
	\item Association-dissociation reactions: These reaction occur between neutral solutes and ions (or other bulk reactions) and lead to a characteristic chemical equilibrium. These reactions have been explored regarding their role for microswimmers such as Pt-coated polystyrene colloids in $H_2O_2$ solution \cite{brown2014ionic,brown2017ionic,ebbens2014electrokinetic} and are expected to cause 
	effective screening \cite{brown2017ionic} which should in turn strongly influence the far-field cross-interactions of the relevant colloids.  
	Similarly, for 
	microswimmers moving by diffusiophoresis in charge-neutral ionic solutes, both neutral diffusiophoresis (or chemiphoresis) and electrophoresis play a role \cite{prieve1984motion} and the precise form of the phoretic interactions among different swimmers is largely unknown.
	\item For swimmers not featuring a boundary layer which is thin compared to the size of the colloid, the concept of a slip velocity breaks down and the Janus colloids under consideration may induce significant flows also away from the surface \cite{samin2015self,gomez2017tuning}. Such scenarios would require a different theory for the phoretic interactions. 
	Likewise for swimmers moving by self-propulsion mechanisms such as the 
	``detonation'' of energetic molecules (like $H_2O_2$) near the surface of Janus colloids \cite{eloul2020reactive}, the cross-interactions between different swimmers have not yet been much discussed. 
\end{itemize}

\subsection{Memory and solute advection effects}
In general phoretic interactions are not instantaneous due to the time-derivative in 
Eq. (\ref{chemdiff}). While many works assume or 
argue that it is sufficient to solve the chemical diffusion equation in steady state,
due to the small size of the solutes (or the large diffusion coefficient in case of self-thermophoretic colloids), there is no general criterion explaining when memory effects due to chemical interactions can be safely neglected for the collective behavior of active colloids. 
In contrast, refs. \cite{liebchen2015clustering,liebchen2017phoretic} suggest that memory effects can in principle even induce pattern formation in parameter regimes where the disordered uniform phase would be stable in the absence of memory effects. 
Accordingly, it remains as an open challenge to generally understand when memory can be safely neglected and when they are important. 
In this connection, note that in contrast to the case of autophoretic colloids, for chemically interacting camphor boats and microorganisms memory-effects are commonly taken into account e.g. within the Keller-Segel model or more recently in  
refs.   
\cite{kranz2016effective, gelimson2016multicellular}. Ref. \cite{gelimson2016multicellular} in particular discusses memory effects in relation to experiments with \emph{Pseudomonas aeruginosa} cells which 
secrete exopolysaccharides (EPS) leading to slowly diffusing polymeric trails inducing strong memory effects. 
\\Like memory effects, most works have neglected also the advection of the phoretic  
field by the flow field induced by the colloids. It therefore remains as an open challenge to show how solute-advection change the cross-interactions in active colloids and to quantify e.g. how strongly the $1/r^2$ law would be violated by these commonly neglected effects.\\
Note also that most of the experimental and theoretical studies with active colloids focus on the behavior in purely viscous Newtonian fluids, 
whereas the natural environment for many active particles such as bacteria and cells are non-Newtonian viscoelastic fluids \cite{Houry_P_Bacterial_Swimmers_2012,Celli_P_Helicobacter_Pylori_2009}. 
Recent exceptions include the experimental works 
\cite{Narinder_EPJE_Active_Colloids_2021,Lozano_NM_Active_Particles_2019,GomezSolano_PRL_Dynamics_SelfPropelled_2016} which have provided important insights into the dynamics of active particles in viscoelastic media (so far mainly at the single particle level). 
In addition, the very recent theoretical work \cite{Choudhary_JFM_NonNewtonian_Effects_2020} has shown that 
viscoelastic effects can significantly modify the slip velocity on the active surface of the particles, which should have profound consequences not only for the single particle behavior but also for the (phoretic) cross-interactions of these particles.

\subsection{Large-scale simulations of active colloids including their full near-field interactions}
Following the enormous complexity of the (near-field) cross-interactions in active colloids, very large system sizes are still difficult to simulate. 
In particular, while powerful mesoscale simulation techniques 
such as multi-particle collision dynamics or the lattice Boltzmann method (see section \ref{simulation})
have recently been generalized to allow simulating $N\sim 10^3$ active colloids  \cite{Scagliarini_SM_Unravelling_Role_2020,Colberg_JCP_Manybody_Dynamics_2017}, 
systems of 
$N \gg 10^3$ active colloids with phoretic, osmotic and hydrodynamic near-field interactions 
are still hard to simulate with state-of-the art simulation techniques. 
However, simulations of such system sizes would be highly desirable e.g. to 
understand the 
large-scale collective behaviour of active colloids in three dimensions which has been hardly explored so far. 
It therefore remains as an open challenge to develop new approximate simulation methods, e.g. based on generalizations of particle based coarse graining techniques (such as e.g. force-matching 
\cite{ercolessi1994interatomic}) or, possibly, based on generalizations of very recent machine-learning based coarse-graining techniques \cite{wang2019machine,durumeric2019adversarial} to active colloids.

\subsection{When do minimal models apply?}
Given the overall complexity of the phoretic, osmotic and hydrodynamic
interactions, it would be desirable to
identify regimes or setups for which the overall complexity reduces and
minimal models involving only one or the other type of interaction
provide a representative description of the collective behavior of
active colloids.
A key challenge to achieve this is to understand the strength of the
individual interactions. Even in far-field it is in general not obvious
when phoretic, osmotic and hydrodynamic interactions dominate as the
strength of all these interactions can decay as $1/r^2$ with the
interparticle distance $r$. From a theoretical viewpoint estimating the
strength of the different interactions is hindered by the fact that key
parameters such as
surface mobility coefficients etc. are often unknown as described in
section \ref{sensitivity}, and also by the fact that simplified models
involving e.g. only one effective phoretic field or thin
boundary layers might not generally apply.
From an experimental viewpoint, explicitly measuring the strength of
the different interactions is clearly a major challenge and
explicit data quantifying and comparing the strength of the different
interactions are currently not available. (The results would likely lead
to results depending on the specific type of colloid under consideration
\cite{liebchen2019interactions,wang2020practical} and it would be very
interesting to identify
different types of Janus colloids allowing to neglect one or the other
type of interaction.)
Simulation studies and theoretical analyses in
turn suggest that
hydrodynamic cross-interactions tend to induce large structures in
initially uniform disordered suspensions of Janus colloids only at area
fractions beyond $\sim 30-50\%$ \cite{zottl2014hydrodynamics}, whereas
phoretic interactions tend to destabilize the uniform phase also at very
low density \cite{liebchen2017phoretic} in agreement with experiments
\cite{theurkauff2012dynamic,palacci2013living,ginot2018aggregation,singh2017non}.
Based on MPCD simulations \cite{Huang_NJP_Chemotactic_Hydrodynamic_2017}
it has been shown in addition that the radial distribution function
strongly changes when switching off phoretic interactions and keeping
only the hydrodynamic ones, leaving only a single small peak at
relatively short interparticle distances.
\\While the general situation is not yet clear,
there is one important special case for which phoretic
interactions dominate in far-field and a representative minimal model
can be formulated: for half-coated autophoretic Janus colloids in bulk,
with a nearly uniform surface mobility (and an arbitrarily heterogeneous
surface activity), the coefficient of the hydrodynamic $1/r^2$
interactions is significantly smaller than the one of the phoretic
interactions. (The ratio of the two interactions
scales with $\mu_r=(\mu_C-\mu_N)/(\mu_C + \mu_N)$ where $\mu_C,\mu_N$
are the surface mobilities of the two hemispheres.)
\cite{liebchen2019interactions}
For such setups (and parameter regimes) where the phoretic interactions
dominate in dilute suspensions, a minimal model, the ``active attractive
alignment'' (AAA) model can be derived, which can be viewed as a simple
extension of the active Brownian particle model.
In particular, for Janus colloids with a uniform surface mobility the
model simply reads (in 2D) \cite{liebchen2019interactions}:
\1 \dot {\vec r}_i = \textrm{Pe}{\vec p}_i - \4{4}{3}\textrm{Pe} \nu
\nabla u + \vec f_s(\vec r_i) ;\quad \dot \theta_i = \sqrt{2}\eta_i(t) \2
where $\nu=\pm 1$ accounts for attractive/repulsive phoretic
interactions and $\vec f_s$ for the (reduced) short-range forces such as
steric repulsions.
In these equations we have neglected translational diffusion for
simplicity and $u=\Sum{j=1,j\neq i}{N} \nabla_{{\vec r}_i} \4{{\rm
	e}^{-\alpha r_{ij}}}{r_{ij}}$ represents the effective
Coulomb interactions for $\alpha=0$ (which applies to cases where the
phoretic field does not ``decay'' or ``evaporate'') and effective Yukawa
pair interactions for $\alpha\neq 0$ (which applies if the phoretic
field decays e.g. due to bulk reactions).
Importantly, for such colloids, besides the screening number which is
expected to be zero in some cases such as self-thermophoretic active
colloids, the collective behavior of the AAA model is determined by the
same two parameters as the collective behavior of the minimalistic ABP
model - which are the P\'eclet number and the density. In particular,
the strength of the phoretic interactions is not a free parameter in
this model, but is directly linked to the P\'eclet number. Hence, as opposed to phenomenological models of active particles with attractions \cite{Rein_EPJE_Applicability_Effective_2016,Mognetti_PRL_Living_Clusters_2013,Alarcon_SM_Morphology_Clusters_2017,Redner_PRE_Reentrant_Phase_2013,Prymidis_SM_Selfassembly_Active_2015,Bauerle_NC_Selforganization_Active_2018}, the AAA model explicitly relates the interaction strength to the P\'eclet number.
\\For colloids with a non-ideally uniform
surface mobility, in leading order we have $\dot \theta_i =
\4{3}{2}\textrm{Pe} \mu_r {\vec p}_i \times \nabla u + \sqrt{2}\eta_i(t)$where $\times$ is the 2D cross product, yielding for two vectors $\vec a,\vec b$, $\vec a \times \vec b =
a_1 b_2 - a_2 b_1$. Note that for a small but finite
$\mu_r$ there are also hydrodynamic $1/r^2$ interactions affecting the
translational motion of the colloids.
\\Interestingly, unlike the popular active Brownian particle (ABP) model
which neglects phoretic (and hydrodynamic) interactions altogether, the
AAA model naturally leads to dynamic clustering at low density with a
near-algebraic cluster size distribution
\cite{liebchen2019interactions}, similarly as in experiments
\cite{theurkauff2012dynamic,palacci2013living,ginot2018aggregation}.
This suggests that the interplay of effectively screened phoretic
attractions, self-propulsion and (rotational) diffusion is enough to
create dynamic clusters at low density. That is, the phoretic
interactions in the AAA model seem to have just the right strength to
lead to
dynamic clustering at densities similar to those where dynamic
clustering is seen in experiments
\cite{theurkauff2012dynamic,palacci2013living,ginot2018aggregation}.
However, it should be stressed that while the AAA model might be
well-suited to understand the onset of structure formation in certain
dilute suspensions of active colloids, it only accounts for far-field
interactions and can of course not be expected to accurately describe
the interactions at the close interparticle distances which occur in
dynamic clustering experiments. Regarding the applicability regime of the AAA model, it should also be
stressed that it is expected to apply specifically to dilute suspensions of
active colloids with a near-uniform surface mobility. For colloids with
a significantly heterogeneous surface mobility (see e.g.
\cite{campbell2019experimental}), it might still provide useful
information, e.g. if hydrodynamic interactions are ``screened'' due to
the presence of a substrate, but should be expected to be
rather inaccurate.

\section{Acknowledgements}
We acknowledge financial support by the Deutsche Forschungsgemeinschaft (DFG, German Research Foundation) through project number 233630050 (TRR-146).

\bibliographystyle{iopart-num}
\bibliography{ref}

\providecommand{\newblock}{}
\begin{thebibliography}{100}
\expandafter\ifx\csname url\endcsname\relax
  \def\url#1{{\tt #1}}\fi
\expandafter\ifx\csname urlprefix\endcsname\relax\def\urlprefix{URL }\fi
\providecommand{\eprint}[2][]{\url{#2}}

\bibitem{verwey1948theory}
Verwey E~J~W, Overbeek J~T~G and Van~Nes K 1948 {\em Theory of the stability of
  lyophobic colloids: the interaction of sol particles having an electric
  double layer\/} (Elsevier Publishing Company)

\bibitem{derjaguin1987surface}
Derjaguin B~V, Churaev N~V and Muller V~M 1987 {\em Surface forces\/}
  (Springer)

\bibitem{langbein1974theory}
Langbein D 1974 {\em Theory of van der Waals Attraction\/} (Springer (Tracts in
  modern physics), Berlin)

\bibitem{mahanty1976dispersion}
Mahanty J and Ninham B~W 1976 {\em Dispersion forces\/} vol~1 (Academic Press)

\bibitem{napper1983polymeric}
Napper D~H 1983 {\em Polymeric stabilization of colloidal dispersions\/} vol~3
  (Academic Press)

\bibitem{farinato1999colloid}
Farinato R~S and Dubin P 1999 {\em Colloid-polymer interactions\/} (Wiley)

\bibitem{parsegian2005van}
Parsegian V~A 2005 {\em Van der Waals forces: a handbook for biologists,
  chemists, engineers, and physicists\/} (Cambridge University Press)

\bibitem{butt2010surface}
Butt H~J, Kappl M {\em et~al.\/} 2010 {\em Surface and interfacial forces\/}
  (Wiley Online Library)

\bibitem{lekkerkerker2011depletion}
Lekkerkerker H~N and Tuinier R 2011 {\em Colloids and the depletion interaction
  (Lecture Notes in Physics)\/} (Springer)

\bibitem{israelachvili2015intermolecular}
Israelachvili J~N 2015 {\em Intermolecular and surface forces\/} (Academic
  press)

\bibitem{russel1991colloidal}
Russel W~B, Russel W, Saville D~A and Schowalter W~R 1991 {\em Colloidal
  dispersions\/} (Cambridge university press)

\bibitem{evans1999colloidal}
Evans D~F and Wennerstr{\"o}m H 1999 {\em The colloidal domain: where physics,
  chemistry, biology, and technology meet\/} (Wiley-Vch, New York)

\bibitem{hunter2001foundations}
Hunter R~J 2001 {\em Foundations of colloid science\/} (Oxford university
  press)

\bibitem{morrison2002colloidal}
Morrison I~D and Ross S 2002 {\em Colloidal dispersions: suspensions,
  emulsions, and foams\/} (Wiley-Interscience, New York)

\bibitem{lyklema2005fundamentals}
Lyklema J 2005 {\em Fundamentals of interface and colloid science: soft
  colloids\/} vol~5 (Elsevier)

\bibitem{de2005food}
De~Kruif C and Singh H 2005 {\em Food Colloids: Interactions, microstructure
  and processing\/} vol 298 (Royal Society of Chemistry)

\bibitem{everett2007basic}
Everett D~H 2007 {\em Basic principles of colloid science\/} (Royal society of
  chemistry)

\bibitem{berg2010introduction}
Berg J~C 2010 {\em An introduction to interfaces \& colloids: the bridge to
  nanoscience\/} (World Scientific)

\bibitem{paxton2004catalytic}
Paxton W~F, Kistler K~C, Olmeda C~C, Sen A, St~Angelo S~K, Cao Y, Mallouk T~E,
  Lammert P~E and Crespi V~H 2004 {\em Journal of the American Chemical
  Society\/} {\bf 126} 13424--13431

\bibitem{marchetti2013hydrodynamics}
Marchetti M~C, Joanny J~F, Ramaswamy S, Liverpool T~B, Prost J, Rao M and Simha
  R~A 2013 {\em Reviews of Modern Physics\/} {\bf 85} 1143

\bibitem{bechinger2016active}
Bechinger C, Di~Leonardo R, L{\"o}wen H, Reichhardt C, Volpe G and Volpe G 2016
  {\em Reviews of Modern Physics\/} {\bf 88} 045006

\bibitem{eisenbach2007chemotaxis}
Eisenbach M 2007 {\em Wiley Encyclopedia of Chemical Biology\/}  1--8

\bibitem{gerisch1982chemotaxis}
Gerisch G 1982 {\em Annual Review of Physiology\/} {\bf 44} 535--552

\bibitem{van1982signal}
van HAASTERT P~J and Konijn T~M 1982 {\em Molecular and Cellular
  Endocrinology\/} {\bf 26} 1--17

\bibitem{devreotes1988chemotaxis}
Devreotes P~N and Zigmond S~H 1988 {\em Annual Review of Cell Biology\/} {\bf
  4} 649--686

\bibitem{nichols2015chemotaxis}
Nichols J~M, Veltman D and Kay R~R 2015 {\em Current opinion in cell biology\/}
  {\bf 36} 7--12

\bibitem{king2009chemotaxis}
King J~S and Insall R~H 2009 {\em Trends in cell biology\/} {\bf 19} 523--530

\bibitem{mesibov1972chemotaxis}
Mesibov R and Adler J 1972 {\em Journal of Bacteriology\/} {\bf 112} 315--326

\bibitem{tso1974negative}
Tso W~W and Adler J 1974 {\em Journal of Bacteriology\/} {\bf 118} 560--576

\bibitem{berg2008coli}
Berg H~C 2008 {\em E. coli in Motion\/} (Springer Science \& Business Media)

\bibitem{budrene1991complex}
Budrene E~O and Berg H~C 1991 {\em Nature\/} {\bf 349} 630--633

\bibitem{budrene1995dynamics}
Budrene E~O and Berg H~C 1995 {\em Nature\/} {\bf 376} 49--53

\bibitem{laganenka2016chemotaxis}
Laganenka L, Colin R and Sourjik V 2016 {\em Nature Communications\/} {\bf 7}
  1--11

\bibitem{laganenka2018autoinducer}
Laganenka L and Sourjik V 2018 {\em Applied and Environmental Microbiology\/}
  {\bf 84}

\bibitem{liebchen2018synthetic}
Liebchen B and Löwen H 2018 {\em Acc. Chem. Res.\/} {\bf 51} 2982

\bibitem{wang2020practical}
Wang W, Lv X, Moran J~L, Duan S and Zhou C 2020 {\em Soft matter\/} {\bf 16}
  3846--3868

\bibitem{poon2013clarkia}
Poon W 2013 {\em Proc. Int. Sch. Phys. Enrico Fermi\/} {\bf 184} 317--386

\bibitem{tindall2008overview}
Tindall M~J, Maini P~K, Porter S~L and Armitage J~P 2008 {\em Bulletin of
  Mathematical Biology\/} {\bf 70} 1570

\bibitem{liebchen2020modeling}
Liebchen B and Löwen H 2020 {\em Modeling Chemotaxis of Microswimmers: From
  Individual to Collective Behavior\/} (World Scientific) chap Chapter 20, pp
  493--516

\bibitem{anderson1989colloid}
Anderson J~L 1989 {\em Annual review of fluid mechanics\/} {\bf 21} 61--99

\bibitem{schweitzer1994clustering}
Schweitzer F and Schimansky-Geier L 1994 {\em Physica A: Statistical Mechanics
  and its Applications\/} {\bf 206} 359--379

\bibitem{ben1997snowflake}
Ben-Jacob E 1997 {\em Contemporary Physics\/} {\bf 38} 205--241

\bibitem{romanczuk2008beyond}
Romanczuk P, Erdmann U, Engel H and Schimansky-Geier L 2008 {\em The European
  Physical Journal Special Topics\/} {\bf 157} 61--77

\bibitem{romanczuk2012active}
Romanczuk P, B{\"a}r M, Ebeling W, Lindner B and Schimansky-Geier L 2012 {\em
  The European Physical Journal Special Topics\/} {\bf 202} 1--162

\bibitem{meyer2014active}
Meyer M, Schimansky-Geier L and Romanczuk P 2014 {\em Phys. Rev. E\/} {\bf 89}
  022711

\bibitem{patlak1953random}
Patlak C~S 1953 {\em The Bulletin of Mathematical Biophysics\/} {\bf 15}
  311--338

\bibitem{alt1980biased}
Alt W 1980 {\em Journal of mathematical biology\/} {\bf 9} 147--177

\bibitem{stevens1997aggregation}
Stevens A and Othmer H~G 1997 {\em SIAM Journal on Applied Mathematics\/} {\bf
  57} 1044--1081

\bibitem{othmer2000diffusion}
Othmer H~G and Hillen T 2000 {\em SIAM Journal on Applied Mathematics\/} {\bf
  61} 751--775

\bibitem{othmer2002diffusion}
Othmer H~G and Hillen T 2002 {\em SIAM Journal on Applied Mathematics\/} {\bf
  62} 1222--1250

\bibitem{horstmann20031970}
Horstmann D 2003 {\em Jahresbericht der Deutschen Mathematiker-Vereinigung\/}
  {\bf 105} 103--165 ISSN 0012-0456

\bibitem{Sengupta_PRE_Chemotactic_Predatorprey_2011}
Sengupta A, Kruppa T and L{\"o}wen H 2011 {\em Physical Review E\/} {\bf 83}
  031914

\bibitem{tsori2004self}
Tsori Y and De~Gennes P~G 2004 {\em EPL (Europhysics Letters)\/} {\bf 66} 599

\bibitem{grima2005strong}
Grima R 2005 {\em Physical review letters\/} {\bf 95} 128103

\bibitem{sengupta2009dynamics}
Sengupta A, van Teeffelen S and L{\"o}wen H 2009 {\em Physical Review E\/} {\bf
  80} 031122

\bibitem{taktikos2011modeling}
Taktikos J, Zaburdaev V and Stark H 2011 {\em Physical Review E\/} {\bf 84}
  041924

\bibitem{murray2001mathematical}
Murray J 2001 {\em Mathematical biology II: spatial models and biomedical
  applications\/} vol~3 (Springer-Verlag)

\bibitem{keller1971traveling}
Keller E~F and Segel L~A 1971 {\em Journal of theoretical biology\/} {\bf 30}
  235--248

\bibitem{kalinin2009logarithmic}
Kalinin Y~V, Jiang L, Tu Y and Wu M 2009 {\em Biophysical journal\/} {\bf 96}
  2439--2448

\bibitem{lapidus1976model}
Lapidus I~R and Schiller R 1976 {\em Biophysical journal\/} {\bf 16} 779--789

\bibitem{dahlquist1972quantitative}
Dahlquist F, Lovely P and Koshland D 1972 {\em Nature New Biology\/} {\bf 236}
  120--123

\bibitem{segev2000generic}
Segev R and Ben-Jacob E 2000 {\em Neural Networks\/} {\bf 13} 185--199

\bibitem{ford1991analysis}
Ford R~M and Lauffenburger D~A 1991 {\em Bulletin of mathematical biology\/}
  {\bf 53} 721--749

\bibitem{ford1991measurement1}
Ford R~M, Phillips B~R, Quinn J~A and Lauffenburger D~A 1991 {\em Biotechnology
  and bioengineering\/} {\bf 37} 647--660

\bibitem{ford1991measurement2}
Ford R~M and Lauffenburger D~A 1991 {\em Biotechnology and Bioengineering\/}
  {\bf 37} 661--672

\bibitem{liebchen2019interactions}
Liebchen B and L{\"o}wen H 2019 {\em J. Chem. Phys.\/} {\bf 150} 061102

\bibitem{hauke2020clustering}
Hauke F, L{\"o}wen H and Liebchen B 2020 {\em J. Chem. Phys.\/} {\bf 152}
  014903

\bibitem{brown2014ionic}
Brown A and Poon W 2014 {\em Soft matter\/} {\bf 10} 4016--4027

\bibitem{ebbens2014electrokinetic}
Ebbens S, Gregory D, Dunderdale G, Howse J, Ibrahim Y, Liverpool T and
  Golestanian R 2014 {\em EPL (Europhysics Letters)\/} {\bf 106} 58003

\bibitem{brown2017ionic}
Brown A~T, Poon W~C, Holm C and de~Graaf J 2017 {\em Soft Matter\/} {\bf 13}
  1200--1222

\bibitem{banigan2016self}
Banigan E~J and Marko J~F 2016 {\em Physical Review E\/} {\bf 93} 012611

\bibitem{keller1970initiation}
Keller E~F and Segel L~A 1970 {\em Journal of Theoretical Biology\/} {\bf 26}
  399--415

\bibitem{keller1971model}
Keller E~F and Segel L~A 1971 {\em Journal of Theoretical Biology\/} {\bf 30}
  225--234

\bibitem{adler1966chemotaxis}
Adler J 1966 {\em Science\/} {\bf 153} 708--716

\bibitem{adler1966effect}
Adler J 1966 {\em Journal of Bacteriology\/} {\bf 92} 121--129

\bibitem{rein2016collective}
Rein M, Hein{\ss} N, Schmid F and Speck T 2016 {\em Physical review letters\/}
  {\bf 116} 058102

\bibitem{dean1996langevin}
Dean D~S 1996 {\em Journal of Physics A: Mathematical and General\/} {\bf 29}
  L613

\bibitem{chavanis2010stochastic}
Chavanis P~H 2010 {\em Communications in Nonlinear Science and Numerical
  Simulation\/} {\bf 15} 60--70

\bibitem{newman2004many}
Newman T and Grima R 2004 {\em Physical Review E\/} {\bf 70} 051916

\bibitem{chavanis2007kinetic}
Chavanis P~H and Sire C 2007 {\em Physica A: Statistical Mechanics and its
  Applications\/} {\bf 384} 199--222

\bibitem{lushnikov2010critical}
Lushnikov P~M 2010 {\em Physics Letters A\/} {\bf 374} 1678--1685

\bibitem{liebchen2016pattern}
Liebchen B, Cates M~E and Marenduzzo D 2016 {\em Soft Matter\/} {\bf 12}
  7259--7264

\bibitem{herrero1996chemotactic}
Herrero M~A and Vel{\'a}zquez J~J 1996 {\em Journal of Mathematical Biology\/}
  {\bf 35} 177--194

\bibitem{levine1998dynamics}
Levine H 1998 {\em Physica A: Statistical Mechanics and its Applications\/}
  {\bf 249} 53--63

\bibitem{lauffenburger1979effects}
Lauffenburger D and Keller K~H 1979 {\em Journal of Theoretical Biology\/} {\bf
  81} 475--503

\bibitem{nossal1973analysis}
Nossal R and Weiss G~H 1973 {\em Journal of Theoretical Biology\/} {\bf 41}
  143--147

\bibitem{theurkauff2012dynamic}
Theurkauff I, Cottin-Bizonne C, Palacci J, Ybert C and Bocquet L 2012 {\em
  Physical review letters\/} {\bf 108} 268303

\bibitem{palacci2013living}
Palacci J, Sacanna S, Steinberg A~P, Pine D~J and Chaikin P~M 2013 {\em
  Science\/} {\bf 339} 936--940

\bibitem{hillen2009user}
Hillen T and Painter K~J 2009 {\em Journal of Mathematical Biology\/} {\bf 58}
  183--217

\bibitem{painter2019mathematical}
Painter K~J 2019 {\em Journal of theoretical biology\/} {\bf 481} 162--182

\bibitem{arumugam2021keller}
Arumugam G and Tyagi J 2021 {\em Acta Applicandae Mathematicae\/} {\bf 171}
  1--82

\bibitem{liebchen2015clustering}
Liebchen B, Marenduzzo D, Pagonabarraga I and Cates M~E 2015 {\em Physical
  Review Letters\/} {\bf 115} 258301

\bibitem{mahdisoltani2021nonequilibrium}
Mahdisoltani S, Zinati R~B~A, Duclut C, Gambassi A and Golestanian R 2021 {\em
  Physical Review Research\/} {\bf 3} 013100

\bibitem{gelimson2015collective}
Gelimson A and Golestanian R 2015 {\em Physical review letters\/} {\bf 114}
  028101

\bibitem{moerman2017solute}
Moerman P~G, Moyses H~W, Van Der~Wee E~B, Grier D~G, Van~Blaaderen A, Kegel
  W~K, Groenewold J and Brujic J 2017 {\em Physical Review E\/} {\bf 96} 032607

\bibitem{jiang2010active}
Jiang H~R, Yoshinaga N and Sano M 2010 {\em Physical review letters\/} {\bf
  105} 268302

\bibitem{kroy2016hot}
Kroy K, Chakraborty D and Cichos F 2016 {\em The European Physical Journal
  Special Topics\/} {\bf 225} 2207--2225

\bibitem{saha2014clusters}
Saha S, Golestanian R and Ramaswamy S 2014 {\em Phys. Rev. E\/} {\bf 89} 062316

\bibitem{liebchen2017phoretic}
Liebchen B, Marenduzzo D and Cates M~E 2017 {\em Phys. Rev. Lett.\/} {\bf 118}
  268001

\bibitem{uspal2015self}
Uspal W, Popescu M~N, Dietrich S and Tasinkevych M 2015 {\em Soft Matter\/}
  {\bf 11} 434--438

\bibitem{liebchen2018unraveling}
Liebchen B, Niu R, Palberg T and L{\"o}wen H 2018 {\em Phys. Rev. E\/} {\bf 98}
  052610

\bibitem{thutupalli2018flow}
Thutupalli S, Geyer D, Singh R, Adhikari R and Stone H~A 2018 {\em Proceedings
  of the National Academy of Sciences\/} {\bf 115} 5403--5408

\bibitem{bickel2014polarization}
Bickel T, Zecua G and W{\"u}rger A 2014 {\em Physical Review E\/} {\bf 89}
  050303

\bibitem{bickel2013flow}
Bickel T, Majee A and W{\"u}rger A 2013 {\em Physical Review E\/} {\bf 88}
  012301

\bibitem{golestanian2007designing}
Golestanian R, Liverpool T and Ajdari A 2007 {\em New Journal of Physics\/}
  {\bf 9} 126

\bibitem{samin2015self}
Samin S and van Roij R 2015 {\em Physical review letters\/} {\bf 115} 188305

\bibitem{gomez2017tuning}
Gomez-Solano J~R, Samin S, Lozano C, Ruedas-Batuecas P, van Roij R and
  Bechinger C 2017 {\em Scientific reports\/} {\bf 7} 1--12

\bibitem{eloul2020reactive}
Eloul S, Poon W~C, Farago O and Frenkel D 2020 {\em Physical Review Letters\/}
  {\bf 124} 188001

\bibitem{burelbach2018unified}
Burelbach J, Frenkel D, Pagonabarraga I and Eiser E 2018 {\em The European
  Physical Journal E\/} {\bf 41} 1--12

\bibitem{burelbach2019determining}
Burelbach J and Stark H 2019 {\em The European Physical Journal E\/} {\bf 42}
  1--9

\bibitem{piazza2004thermal}
Piazza R 2004 {\em Journal of Physics: Condensed Matter\/} {\bf 16} S4195

\bibitem{dhont2008single}
Dhont J~K and Briels W~J 2008 {\em The European Physical Journal E\/} {\bf 25}
  61--76

\bibitem{taktikos2012collective}
Taktikos J, Zaburdaev V and Stark H 2012 {\em Physical Review E\/} {\bf 85}
  051901

\bibitem{pohl2015self}
Pohl O and Stark H 2015 {\em The European Physical Journal E\/} {\bf 38} 1--11

\bibitem{saha2019pairing}
Saha S, Ramaswamy S and Golestanian R 2019 {\em New Journal of Physics\/} {\bf
  21} 063006

\bibitem{Varma_PRF_Modeling_Chemohydrodynamic_2019}
Varma A and Michelin S 2019 {\em Phys. Rev. Fluids\/} {\bf 4} 124204

\bibitem{nasouri2020exact}
Nasouri B and Golestanian R 2020 {\em Physical review letters\/} {\bf 124}
  168003

\bibitem{uspal2016guiding}
Uspal W, Popescu M~N, Dietrich S and Tasinkevych M 2016 {\em Physical review
  letters\/} {\bf 117} 048002

\bibitem{popescu2017chemically}
Popescu M, Uspal W and Dietrich S 2017 {\em Journal of Physics: Condensed
  Matter\/} {\bf 29} 134001

\bibitem{heidari2020self}
Heidari M, Bregulla A, Landin S~M, Cichos F and von Klitzing R 2020 {\em
  Langmuir\/} {\bf 36} 7775--7780

\bibitem{heidari2021}
Heidari M, Jakob F, Liebchen B and von Klitzing R 2021 {\em (under review)\/}

\bibitem{wang2019interactions}
Wang L and Simmchen J 2019 {\em Condensed Matter\/} {\bf 4} 78

\bibitem{hong2007chemotaxis}
Hong Y, Blackman N~M, Kopp N~D, Sen A and Velegol D 2007 {\em Physical Review
  Letters\/} {\bf 99} 178103

\bibitem{lozano2016phototaxis}
Lozano C, Ten~Hagen B, L{\"o}wen H and Bechinger C 2016 {\em Nature
  Communications\/} {\bf 7} 1--10

\bibitem{moyses2016trochoidal}
Moyses H, Palacci J, Sacanna S and Grier D~G 2016 {\em Soft Matter\/} {\bf 12}
  6357--6364

\bibitem{lozano2019propagating}
Lozano C, Liebchen B, Ten~Hagen B, Bechinger C and L{\"o}wen H 2019 {\em Soft
  matter\/} {\bf 15} 5185

\bibitem{jahanshahi2020realization}
Jahanshahi S, Lozano C, Liebchen B, L{\"o}wen H and Bechinger C 2020 {\em Comm.
  Phys.\/} {\bf 3} 1

\bibitem{Yu_SM_Phototaxis_Active_2019}
Yu N, Lou X, Chen K and Yang M 2019 {\em Soft Matter\/} {\bf 15} 408--414

\bibitem{Gittus_EPJE_Thermal_Orientation_2019}
Gittus O~R, {Olarte-Plata} J~D and Bresme F 2019 {\em The European Physical
  Journal E\/} {\bf 42} 90

\bibitem{auschra2021thermotaxis}
Auschra S, Bregulla A, Kroy K and Cichos F 2021 {\em The European Physical
  Journal E\/} {\bf 44} 90

\bibitem{OlartePlata_PRE_Thermophoretic_Torque_2018}
{Olarte-Plata} J, Rubi J~M and Bresme F 2018 {\em Physical Review E\/} {\bf 97}
  052607

\bibitem{palacci2015artificial}
Palacci J, Sacanna S, Abramian A, Barral J, Hanson K, Grosberg A~Y, Pine D~J
  and Chaikin P~M 2015 {\em Science advances\/} {\bf 1} e1400214

\bibitem{Ren_AN_Rheotaxis_Bimetallic_2017}
Ren L, Zhou D, Mao Z, Xu P, Huang T~J and Mallouk T~E 2017 {\em ACS Nano\/}
  {\bf 11} 10591--10598

\bibitem{Baker_N_Fight_Flow_2019}
Baker R, Kauffman J~E, Laskar A, Shklyaev O~E, Potomkin M, {Dominguez-Rubio} L,
  Shum H, {Cruz-Rivera} Y, Aranson I~S, Balazs A~C and Sen A 2019 {\em
  Nanoscale\/} {\bf 11} 10944--10951

\bibitem{Dwivedi_PF_Rheotaxis_Active_2021}
Dwivedi P, Shrivastava A, Pillai D and Mangal R 2021 {\em Physics of Fluids\/}
  {\bf 33} 082108

\bibitem{singh2017non}
Singh D~P, Choudhury U, Fischer P and Mark A~G 2017 {\em Advanced Materials\/}
  {\bf 29} 1701328

\bibitem{wang2013catalytically}
Wang W, Duan W, Sen A and Mallouk T~E 2013 {\em Proceedings of the National
  Academy of Sciences\/} {\bf 110} 17744--17749

\bibitem{huang2020inverse}
Huang T, Misko V~R, Gobeil S, Wang X, Nori F, Sch{\"u}tt J, Fassbender J,
  Cuniberti G, Makarov D and Baraban L 2020 {\em Advanced Functional
  Materials\/} {\bf 30} 2003851

\bibitem{huang2020anisotropic}
Huang T, Gobeil S, Wang X, Misko V, Nori F, De~Malsche W, Fassbender J, Makarov
  D, Cuniberti G and Baraban L 2020 {\em Langmuir\/} {\bf 36} 7091--7099

\bibitem{wang2018visible}
Wang X, Baraban L, Misko V~R, Nori F, Huang T, Cuniberti G, Fassbender J and
  Makarov D 2018 {\em Small\/} {\bf 14} 1802537

\bibitem{wang2019photocatalytic}
Wang L, Kaeppler A, Fischer D and Simmchen J 2019 {\em ACS applied materials \&
  interfaces\/} {\bf 11} 32937--32944

\bibitem{aubret2018targeted}
Aubret A, Youssef M, Sacanna S and Palacci J 2018 {\em Nature Physics\/} {\bf
  14} 1114--1118

\bibitem{aubret2018diffusiophoretic}
Aubret A and Palacci J 2018 {\em Soft Matter\/} {\bf 14} 9577--9588

\bibitem{katuri2021inferring}
Katuri J, Uspal W~E, Popescu M~N and S{\'a}nchez S 2021 {\em Science
  Advances\/} {\bf 7} eabd0719

\bibitem{ketzetzi2021activity}
Ketzetzi S, Rinaldin M, Dr{\"o}ge P, de~Graaf J and Kraft D~J 2021 {\em arXiv
  preprint arXiv:2103.07335v1\/}

\bibitem{niu2017microfluidic}
Niu R, Kreissl P, Brown A~T, Rempfer G, Botin D, Holm C, Palberg T and De~Graaf
  J 2017 {\em Soft Matter\/} {\bf 13} 1505--1518

\bibitem{moeller2021}
Möller N, Liebchen B and Palberg T 2021 {\em Eur. Phys. J. E\/} {\bf 44} 41

\bibitem{maass2016swimming}
Maass C~C, Kr{\"u}ger C, Herminghaus S and Bahr C 2016 {\em Annual Review of
  Condensed Matter Physics\/} {\bf 7} 171--193

\bibitem{jin2017chemotaxis}
Jin C, Kr{\"u}ger C and Maass C~C 2017 {\em Proceedings of the National Academy
  of Sciences\/} {\bf 114} 5089--5094

\bibitem{hokmabad2021emergence}
Hokmabad B~V, Dey R, Jalaal M, Mohanty D, Almukambetova M, Baldwin K~A, Lohse D
  and Maass C~C 2021 {\em Physical Review X\/} {\bf 11} 011043

\bibitem{kohira2001synchronized}
Kohira M~I, Hayashima Y, Nagayama M and Nakata S 2001 {\em Langmuir\/} {\bf 17}
  7124--7129

\bibitem{nakata2005characteristic}
Nakata S and Matsuo K 2005 {\em Langmuir\/} {\bf 21} 982--984

\bibitem{nakata2005synchronized}
Nakata S, Doi Y and Kitahata H 2005 {\em The Journal of Physical Chemistry B\/}
  {\bf 109} 1798--1802

\bibitem{boniface2019self}
Boniface D, Cottin-Bizonne C, Kervil R, Ybert C and Detcheverry F 2019 {\em
  Physical Review E\/} {\bf 99} 062605

\bibitem{gouiller2021mixing}
Gouiller C, Raynal F, Maquet L, Bourgoin M, {Cottin-Bizonne} C, Volk R and
  Ybert C 2021 {\em Physical Review Fluids\/} {\bf 6} 014501

\bibitem{pavel2021cooperative}
Pavel I, Salinas G, Mierzwa M, Arnaboldi S, Garrigue P and Kuhn A 2021 {\em
  ChemPhysChem\/}

\bibitem{jin2018chemotactic}
Jin C, Hokmabad B~V, Baldwin K~A and Maass C~C 2018 {\em Journal of Physics:
  Condensed Matter\/} {\bf 30} 054003

\bibitem{Malgaretti_C_Phoretic_Colloids_2021}
Malgaretti P and Harting J 2021 {\em ChemNanoMat\/} {\bf 7} 1073--1081

\bibitem{niu2017controlled}
Niu R, O{\u{g}}uz E~C, M{\"u}ller H, Reinm{\"u}ller A, Botin D, L{\"o}wen H and
  Palberg T 2017 {\em Physical Chemistry Chemical Physics\/} {\bf 19}
  3104--3114

\bibitem{niu2017self}
Niu R, Palberg T, Speck T {\em et~al.\/} 2017 {\em Physical Review Letters\/}
  {\bf 119} 028001

\bibitem{niu2018dynamics}
Niu R, Fischer A, Palberg T and Speck T 2018 {\em ACS Nano\/} {\bf 12}
  10932--10938

\bibitem{niu2018modular}
Niu R and Palberg T 2018 {\em Soft Matter\/} {\bf 14} 7554--7568

\bibitem{vsimkus2018phoretic}
{\v{S}}imkus R, Me{\v{s}}kien{\.e} R, Au{\v{c}}ynait{\.e} A, Ledas {\v{Z}},
  Baronas R and Me{\v{s}}kys R 2018 {\em Royal Society open science\/} {\bf 5}
  180008

\bibitem{buttinoni2013dynamical}
Buttinoni I, Bialk{\'e} J, K{\"u}mmel F, L{\"o}wen H, Bechinger C and Speck T
  2013 {\em Physical review letters\/} {\bf 110} 238301

\bibitem{ginot2018aggregation}
Ginot F, Theurkauff I, Detcheverry F, Ybert C and Cottin-Bizonne C 2018 {\em
  Nature communications\/} {\bf 9} 1--9

\bibitem{io2017experimental}
Io C~W, Chen T~Y, Yeh J~W and Cai S~C 2017 {\em Physical Review E\/} {\bf 96}
  062601

\bibitem{heckel2020active}
Heckel S, Grauer J, Semmler M, Gemming T, Löwen H, Liebchen B and Simmchen J
  2020 {\em Langmuir\/} {\bf 36} 12473

\bibitem{pohl2014dynamic}
Pohl O and Stark H 2014 {\em Physical Review Letters\/} {\bf 112} 238303

\bibitem{stark2018artificial}
Stark H 2018 {\em Accounts of Chemical Research\/} {\bf 51} 2681--2688

\bibitem{mukherjee2018growth}
Mukherjee M and Ghosh P 2018 {\em Physical Review E\/} {\bf 97} 012413

\bibitem{liu2011sequential}
Liu C, Fu X, Liu L, Ren X, Chau C~K, Li S, Xiang L, Zeng H, Chen G, Tang L~H
  {\em et~al.\/} 2011 {\em Science\/} {\bf 334} 238--241

\bibitem{golestanian2012collective}
Golestanian R 2012 {\em Physical Review Letters\/} {\bf 108} 038303

\bibitem{cohen2014emergent}
Cohen J~A and Golestanian R 2014 {\em Physical Review Letters\/} {\bf 112}
  068302

\bibitem{Wagner_EPJE_Collective_Behavior_2021}
Wagner M, {Roca-Bonet} S and Ripoll M 2021 {\em The European Physical Journal
  E\/} {\bf 44} 43

\bibitem{Wagner_E_Hydrodynamic_Frontlike_2017}
Wagner M and Ripoll M 2017 {\em EPL (Europhysics Letters)\/} {\bf 119} 66007

\bibitem{schmidt2019light}
Schmidt F, Liebchen B, L{\"o}wen H and Volpe G 2019 {\em J. Chem. Phys.\/} {\bf
  150} 094905

\bibitem{soto2014self}
Soto R and Golestanian R 2014 {\em Physical Review Letters\/} {\bf 112} 068301

\bibitem{soto2015self}
Soto R and Golestanian R 2015 {\em Physical Review E\/} {\bf 91} 052304

\bibitem{sengupta2011chemotactic}
Sengupta A, Kruppa T and L{\"o}wen H 2011 {\em Physical Review E\/} {\bf 83}
  031914

\bibitem{niu2017assembly}
Niu R, Botin D, Weber J, Reinmüller A and Palberg T 2017 {\em Langmuir\/} {\bf
  33} 3450--3457

\bibitem{moller2021shaping}
M{\"o}ller N, Liebchen B and Palberg T 2021 {\em The European Physical Journal
  E\/} {\bf 44} 1--17

\bibitem{ibele2009schooling}
Ibele M, Mallouk T~E and Sen A 2009 {\em Angewandte Chemie\/} {\bf 121}
  3358--3362

\bibitem{wang2015one}
Wang W, Duan W, Ahmed S, Sen A and Mallouk T~E 2015 {\em Accounts of Chemical
  Research\/} {\bf 48} 1938--1946

\bibitem{zhang2016directed}
Zhang J, Yan J and Granick S 2016 {\em Angewandte Chemie\/} {\bf 128}
  5252--5255

\bibitem{ilday2017rich}
Ilday S, Makey G, Akguc G~B, Yavuz {\"O}, Tokel O, Pavlov I, G{\"u}lseren O and
  Ilday F~{\"O} 2017 {\em Nature communications\/} {\bf 8} 1--10

\bibitem{lowen2018active}
L{\"o}wen H 2018 {\em EPL (Europhysics Letters)\/} {\bf 121} 58001

\bibitem{ibele2010emergent}
Ibele M~E, Lammert P~E, Crespi V~H and Sen A 2010 {\em ACS Nano\/} {\bf 4}
  4845--4851

\bibitem{duan2013transition}
Duan W, Liu R and Sen A 2013 {\em Journal of the American Chemical Society\/}
  {\bf 135} 1280--1283

\bibitem{ivlev2015statistical}
Ivlev A~V, Bartnick J, Heinen M, Du C~R, Nosenko V and L{\"o}wen H 2015 {\em
  Physical Review X\/} {\bf 5} 011035

\bibitem{agudo2019active}
Agudo-Canalejo J and Golestanian R 2019 {\em Physical Review Letters\/} {\bf
  123} 018101

\bibitem{grauer2020swarm}
Grauer J, L{\"o}wen H, Be’er A and Liebchen B 2020 {\em Sci. Rep.\/} {\bf 10}
  1

\bibitem{saha2020scalar}
Saha S, Agudo-Canalejo J and Golestanian R 2020 {\em Physical Review X\/} {\bf
  10} 041009

\bibitem{sturmer2019chemotaxis}
St{\"u}rmer J, Seyrich M and Stark H 2019 {\em The Journal of chemical
  physics\/} {\bf 150} 214901

\bibitem{Purcell_AJP_Life_Low_1977}
Purcell E~M 1977 {\em Am. J. Phys.\/} {\bf 45} 3--11

\bibitem{dhont1996introduction}
Dhont J~K 1996 {\em An introduction to dynamics of colloids\/} (Elsevier)

\bibitem{morrison1970electrophoresis}
Morrison~Jr F~A 1970 {\em Journal of Colloid and Interface Science\/} {\bf 34}
  210--214

\bibitem{lighthill1952squirming}
Lighthill M 1952 {\em Communications on pure and applied mathematics\/} {\bf 5}
  109--118

\bibitem{blake1971spherical}
Blake J~R 1971 {\em Journal of Fluid Mechanics\/} {\bf 46} 199--208

\bibitem{guasto2010oscillatory}
Guasto J~S, Johnson K~A and Gollub J~P 2010 {\em Physical review letters\/}
  {\bf 105} 168102

\bibitem{lauga2009hydrodynamics}
Lauga E and Powers T~R 2009 {\em Rep. Prog. Phys.\/} {\bf 72} 096601

\bibitem{Koch_ARFM_Collective_Hydrodynamics_2011}
Koch D~L and Subramanian G 2011 {\em Annu. Rev. Fluid Mech.\/} {\bf 43}
  637--659

\bibitem{elgeti2015physics}
Elgeti J, Winkler R~G and Gompper G 2015 {\em Rep. Prog. Phys.\/} {\bf 78}
  056601

\bibitem{zottl2016emergent}
Z{\"o}ttl A and Stark H 2016 {\em J. Phys. Cond. Matter\/} {\bf 28} 253001

\bibitem{yeomans2017hydrodynamics}
Yeomans J~M 2017 {\em La Rivista del Nuovo Cimento\/} {\bf 40} 1--31

\bibitem{kim2013microhydrodynamics}
Kim S and Karrila S~J 2013 {\em Microhydrodynamics: principles and selected
  applications\/} (Courier Corporation)

\bibitem{drescher2011fluid}
Drescher K, Dunkel J, Cisneros L~H, Ganguly S and Goldstein R~E 2011 {\em
  Proceedings of the National Academy of Sciences\/} {\bf 108} 10940--10945

\bibitem{blake1972model}
Blake J 1972 {\em Journal of Fluid Mechanics\/} {\bf 55} 1--23

\bibitem{blake1974fundamental}
Blake J and Chwang A 1974 {\em Journal of Engineering Mathematics\/} {\bf 8}
  23--29

\bibitem{spagnolie2012hydrodynamics}
Spagnolie S~E and Lauga E 2012 {\em Journal of Fluid Mechanics\/} {\bf 700}
  105--147

\bibitem{saintillan2007orientational}
Saintillan D and Shelley M~J 2007 {\em Physical review letters\/} {\bf 99}
  058102

\bibitem{saintillan2008instabilities}
Saintillan D and Shelley M~J 2008 {\em Physical Review Letters\/} {\bf 100}
  178103

\bibitem{Saintillan_JRSI_Emergence_Coherent_2012}
Saintillan D and Shelley M~J 2012 {\em J. R. Soc. Interface\/} {\bf 9} 571--585

\bibitem{Stenhammar_PRL_Role_Correlations_2017}
Stenhammar J, Nardini C, Nash R~W, Marenduzzo D and Morozov A 2017 {\em Phys.
  Rev. Lett.\/} {\bf 119} 028005

\bibitem{Cisneros_EF_Fluid_Dynamics_2007}
Cisneros L~H, Cortez R, Dombrowski C, Goldstein R~E and Kessler J~O 2007 {\em
  Exp Fluids\/} {\bf 43} 737--753

\bibitem{Sokolov_PRE_Enhanced_Mixing_2009}
Sokolov A, Goldstein R~E, Feldchtein F~I and Aranson I~S 2009 {\em Phys. Rev.
  E\/} {\bf 80} 031903

\bibitem{Krishnamurthy_JFM_Collective_Motion_2015}
Krishnamurthy D and Subramanian G 2015 {\em J. Fluid Mech.\/} {\bf 781}
  422--466

\bibitem{moran2017phoretic}
Moran J~L and Posner J~D 2017 {\em Annual Review of Fluid Mechanics\/} {\bf 49}
  511--540

\bibitem{popescu2018effective}
Popescu M, Uspal W, Eskandari Z, Tasinkevych M and Dietrich S 2018 {\em The
  European Physical Journal E\/} {\bf 41} 1--24

\bibitem{ishikawa2006hydrodynamic}
Ishikawa T, Simmonds M and Pedley T~J 2006 {\em Journal of Fluid Mechanics\/}
  {\bf 568} 119

\bibitem{shapere1989geometry}
Shapere A and Wilczek F 1989 {\em Journal of Fluid Mechanics\/} {\bf 198}
  557--585

\bibitem{felderhof1994small}
Felderhof B and Jones R 1994 {\em Physica A: Statistical Mechanics and its
  Applications\/} {\bf 202} 119--144

\bibitem{keller1977porous}
Keller S~R and Wu T~Y 1977 {\em Journal of Fluid Mechanics\/} {\bf 80} 259--278

\bibitem{Pohnl_JPCM_Axisymmetric_Spheroidal_2020}
P{\"o}hnl R, Popescu M~N and Uspal W~E 2020 {\em J. Phys.: Condens. Matter\/}
  {\bf 32} 164001

\bibitem{Clopes_SM_Hydrodynamic_Interactions_2020}
Clop{\'e}s J, Gompper G and Winkler R~G 2020 {\em Soft Matter\/} {\bf 16}
  10676--10687

\bibitem{Ishikawa_M_Stability_Dumbbell_2019}
Ishikawa T 2019 {\em Micromachines\/} {\bf 10} 33

\bibitem{Zantop_SM_Squirmer_Rods_2020}
Zantop A~W and Stark H 2020 {\em Soft Matter\/} {\bf 16} 6400--6412

\bibitem{Papavassiliou_JFM_Exact_Solutions_2017}
Papavassiliou D and Alexander G~P 2017 {\em J. Fluid Mech.\/} {\bf 813}
  618--646

\bibitem{Kuron_JCP_Lattice_Boltzmann_2019}
Kuron M, St{\"a}rk P, Burkard C, {de Graaf} J and Holm C 2019 {\em J. Chem.
  Phys.\/} {\bf 150} 144110

\bibitem{ishikawa2007diffusion}
Ishikawa T and Pedley T 2007 {\em Journal of Fluid Mechanics\/} {\bf 588} 437

\bibitem{ishikawa2007rheology}
Ishikawa T and Pedley T 2007 {\em Journal of Fluid Mechanics\/} {\bf 588} 399

\bibitem{Pagonabarraga_SM_Structure_Rheology_2013}
Pagonabarraga I and Llopis I 2013 {\em Soft Matter\/} {\bf 9} 7174--7184

\bibitem{Ishikawa_JFM_Rheology_Concentrated_2021}
Ishikawa T, Brumley D~R and Pedley T~J 2021 {\em J. Fluid Mech.\/} {\bf 914}

\bibitem{alarcon2013spontaneous}
Alarc{\'o}n F and Pagonabarraga I 2013 {\em Journal of Molecular Liquids\/}
  {\bf 185} 56--61

\bibitem{zottl2014hydrodynamics}
Z{\"o}ttl A and Stark H 2014 {\em Physical Review Letters\/} {\bf 112} 118101

\bibitem{blaschke2016phase}
Blaschke J, Maurer M, Menon K, Z{\"o}ttl A and Stark H 2016 {\em Soft matter\/}
  {\bf 12} 9821--9831

\bibitem{li2014hydrodynamic}
Li G~J and Ardekani A~M 2014 {\em Physical Review E\/} {\bf 90} 013010

\bibitem{Brumley_PRF_Stability_Arrays_2019}
Brumley D~R and Pedley T~J 2019 {\em Phys. Rev. Fluids\/} {\bf 4} 053102

\bibitem{Kuhr_SM_Collective_Dynamics_2019}
Kuhr J~T, R{\"u}hle F and Stark H 2019 {\em Soft Matter\/} {\bf 15} 5685--5694

\bibitem{Kuhr_SM_Collective_Sedimentation_2017}
Kuhr J~T, Blaschke J, R{\"u}hle F and Stark H 2017 {\em Soft Matter\/} {\bf 13}
  7548--7555

\bibitem{Ruhle_EPJE_Emergent_Collective_2020}
R{\"u}hle F and Stark H 2020 {\em Eur. Phys. J. E\/} {\bf 43} 26

\bibitem{Ruhle_NJP_Gravityinduced_Dynamics_2018}
R{\"u}hle F, Blaschke J, Kuhr J~T and Stark H 2018 {\em New J. Phys.\/} {\bf
  20} 025003

\bibitem{Shen_PRF_Gravity_Induced_2019}
Shen Z and Lintuvuori J~S 2019 {\em Phys. Rev. Fluids\/} {\bf 4} 123101

\bibitem{Yoshinaga_EPJE_Hydrodynamic_Lubrication_2018}
Yoshinaga N and Liverpool T~B 2018 {\em Eur. Phys. J. E\/} {\bf 41} 76

\bibitem{Yoshinaga_PRE_Hydrodynamic_Interactions_2017}
Yoshinaga N and Liverpool T~B 2017 {\em Phys. Rev. E\/} {\bf 96} 020603

\bibitem{campbell2019experimental}
Campbell A~I, Ebbens S~J, Illien P and Golestanian R 2019 {\em Nature Comm.\/}
  {\bf 10} 1--8

\bibitem{bregulla2019flow}
Bregulla A~P and Cichos F 2019 {\em The Journal of chemical physics\/} {\bf
  151} 044706

\bibitem{singh2016universal}
Singh R and Adhikari R 2016 {\em Physical review letters\/} {\bf 117} 228002

\bibitem{Michelin_JFM_Phoretic_Selfpropulsion_2014}
Michelin S and Lauga E 2014 {\em J. Fluid Mech.\/} {\bf 747} 572--604

\bibitem{Michelin_PF_Spontaneous_Autophoretic_2013}
Michelin S, Lauga E and Bartolo D 2013 {\em Phys. Fluids\/} {\bf 25} 061701

\bibitem{Kanso_JCP_Phoretic_Hydrodynamic_2019}
Kanso E and Michelin S 2019 {\em J. Chem. Phys.\/} {\bf 150} 044902

\bibitem{SharifiMood_JFM_Pair_Interaction_2016}
{Sharifi-Mood} N, Mozaffari A and {C{\'o}rdova-Figueroa} U~M 2016 {\em J. Fluid
  Mech.\/} {\bf 798} 910--954

\bibitem{Nasouri_JFM_Exact_Axisymmetric_2020}
Nasouri B and Golestanian R 2020 {\em J. Fluid Mech.\/} {\bf 905} A13

\bibitem{Nasouri_PRL_Exact_Phoretic_2020}
Nasouri B and Golestanian R 2020 {\em Phys. Rev. Lett.\/} {\bf 124} 168003

\bibitem{Popescu_EPJE_Effective_Squirmer_2018}
Popescu M~N, Uspal W~E, Eskandari Z, Tasinkevych M and Dietrich S 2018 {\em
  Eur. Phys. J. E\/} {\bf 41} 145

\bibitem{Michelin_EPJE_Autophoretic_Locomotion_2015}
Michelin S and Lauga E 2015 {\em Eur. Phys. J. E\/} {\bf 38} 7

\bibitem{Reigh_SM_Catalytic_Dimer_2015}
Reigh S~Y and Kapral R 2015 {\em Soft Matter\/} {\bf 11} 3149--3158

\bibitem{Reigh_SM_Diffusiophoretically_Induced_2018}
Reigh S~Y, Chuphal P, Thakur S and Kapral R 2018 {\em Soft Matter\/} {\bf 14}
  6043--6057

\bibitem{lushi2012collective}
Lushi E, Goldstein R~E and Shelley M~J 2012 {\em Physical Review E\/} {\bf 86}
  040902

\bibitem{saintillan2013active}
Saintillan D and Shelley M~J 2013 {\em Comptes Rendus Physique\/} {\bf 14}
  497--517

\bibitem{Huang_NJP_Chemotactic_Hydrodynamic_2017}
Huang M~J, Schofield J and Kapral R 2017 {\em New J. Phys.\/} {\bf 19} 125003

\bibitem{Scagliarini_SM_Unravelling_Role_2020}
Scagliarini A and Pagonabarraga I 2020 {\em Soft Matter\/} {\bf 16} 8893--8903

\bibitem{Colberg_JCP_Manybody_Dynamics_2017}
Colberg P~H and Kapral R 2017 {\em J. Chem. Phys.\/} {\bf 147} 064910

\bibitem{Varma_SM_Clusteringinduced_Selfpropulsion_2018}
Varma A, {Montenegro-Johnson} T~D and Michelin S 2018 {\em Soft Matter\/} {\bf
  14} 7155--7173

\bibitem{Popescu_JCP_Confinement_Effects_2009}
Popescu M~N, Dietrich S and Oshanin G 2009 {\em J. Chem. Phys.\/} {\bf 130}
  194702

\bibitem{Crowdy_JFM_Wall_Effects_2013}
Crowdy D~G 2013 {\em J. Fluid Mech.\/} {\bf 735} 473--498

\bibitem{Mozaffari_PoF_Selfdiffusiophoretic_Colloidal_2016}
Mozaffari A, {Sharifi-Mood} N, Koplik J and Maldarelli C 2016 {\em Physics of
  Fluids\/} {\bf 28} 053107

\bibitem{Mozaffari_PRF_Selfpropelled_Colloidal_2018}
Mozaffari A, {Sharifi-Mood} N, Koplik J and Maldarelli C 2018 {\em Phys. Rev.
  Fluids\/} {\bf 3} 014104

\bibitem{Ibrahim_E_Dynamics_Selfphoretic_2015}
Ibrahim Y and Liverpool T~B 2015 {\em EPL\/} {\bf 111} 48008

\bibitem{Ibrahim_EPJST_How_Walls_2016}
Ibrahim Y and Liverpool T 2016 {\em Eur. Phys. J. Spec. Top.\/} {\bf 225}
  1843--1874

\bibitem{Uspal_SM_Selfpropulsion_Catalytically_2014}
Uspal W~E, Popescu M~N, Dietrich S and Tasinkevych M 2014 {\em Soft Matter\/}
  {\bf 11} 434--438

\bibitem{Choudhary__Selfpropulsion_2D_2021}
Choudhary A, Chaithanya K~V~S, Michelin S and Pushpavanam S 2021 {\em The
  European Physical Journal E\/} {\bf 44} 97

\bibitem{Yang_L_SelfDiffusiophoresis_Janus_2016}
Yang F, Qian S, Zhao Y and Qiao R 2016 {\em Langmuir\/} {\bf 32} 5580--5592

\bibitem{Singh_JCP_Competing_Chemical_2019}
Singh R, Adhikari R and Cates M~E 2019 {\em J. Chem. Phys.\/} {\bf 151} 044901

\bibitem{Robertson_C_Collective_Orientational_2018}
Robertson B, Stark H and Kapral R 2018 {\em Chaos\/} {\bf 28} 045109

\bibitem{Liebchen_JCP_Which_Interactions_2019}
Liebchen B and L{\"o}wen H 2019 {\em J. Chem. Phys.\/} {\bf 150} 061102

\bibitem{kruger2017lattice}
Kr{\"u}ger T, Kusumaatmaja H, Kuzmin A, Shardt O, Silva G and Viggen E~M 2017
  {\em Springer International Publishing\/} {\bf 10} 4--15

\bibitem{chen1998lattice}
Chen S and Doolen G~D 1998 {\em Annual review of fluid mechanics\/} {\bf 30}
  329--364

\bibitem{desplat2001ludwig}
Desplat J~C, Pagonabarraga I and Bladon P 2001 {\em Computer Physics
  Communications\/} {\bf 134} 273--290

\bibitem{Aidun_JSP_Lattice_Boltzmann_1995}
Aidun C~K and Lu Y 1995 {\em J Stat Phys\/} {\bf 81} 49--61

\bibitem{Carenza_EPJE_Lattice_Boltzmann_2019}
Carenza L~N, Gonnella G, Lamura A, Negro G and Tiribocchi A 2019 {\em Eur.
  Phys. J. E\/} {\bf 42} 81

\bibitem{hoogerbrugge1992simulating}
Hoogerbrugge P and Koelman J 1992 {\em EPL (Europhysics Letters)\/} {\bf 19}
  155

\bibitem{espanol1995statistical}
Espanol P and Warren P 1995 {\em EPL (Europhysics Letters)\/} {\bf 30} 191

\bibitem{pagonabarraga2001dissipative}
Pagonabarraga I and Frenkel D 2001 {\em The Journal of Chemical Physics\/} {\bf
  115} 5015--5026

\bibitem{espanol2017perspective}
Espanol P and Warren P~B 2017 {\em The Journal of chemical physics\/} {\bf 146}
  150901

\bibitem{monaghan2005smoothed}
Monaghan J~J 2005 {\em Reports on progress in physics\/} {\bf 68} 1703

\bibitem{alexander1997direct}
Alexander F~J and Garcia A~L 1997 {\em Computers in Physics\/} {\bf 11}
  588--593

\bibitem{oran1998direct}
Oran E, Oh C and Cybyk B 1998 {\em Annual Review of Fluid Mechanics\/} {\bf 30}
  403--441

\bibitem{kapral2008multiparticle}
Kapral R 2008 {\em Advances in Chemical Physics\/} {\bf 140} 89

\bibitem{malevanets1999mesoscopic}
Malevanets A and Kapral R 1999 {\em The Journal of chemical physics\/} {\bf
  110} 8605--8613

\bibitem{gompper2009multi}
Gompper G, Ihle T, Kroll D and Winkler R 2009 {\em Advanced computer simulation
  approaches for soft matter sciences III\/}  1--87

\bibitem{Kuron_SM_Hydrodynamic_Mobility_2019}
Kuron M, St{\"a}rk P, Holm C and de~Graaf J 2019 {\em Soft Matter\/} {\bf 15}
  5908--5920

\bibitem{Llopis_JoNFM_Hydrodynamic_Interactions_2010}
Llopis I and Pagonabarraga I 2010 {\em Journal of Non-Newtonian Fluid
  Mechanics\/} {\bf 165} 946--952

\bibitem{Li_PRE_Hydrodynamic_Interaction_2014}
Li G~J and Ardekani A~M 2014 {\em Phys. Rev. E\/} {\bf 90} 013010

\bibitem{nash2010run}
Nash R, Adhikari R, Tailleur J and Cates M 2010 {\em Physical Review Letters\/}
  {\bf 104} 258101

\bibitem{stenhammar2017role}
Stenhammar J, Nardini C, Nash R~W, Marenduzzo D and Morozov A 2017 {\em
  Physical Review Letters\/} {\bf 119} 028005

\bibitem{bardfalvy2020symmetric}
B{\'a}rdfalvy D, Anjum S, Nardini C, Morozov A and Stenhammar J 2020 {\em
  Physical Review Letters\/} {\bf 125} 018003

\bibitem{vskultety2020swimming}
{\v{S}}kult{\'e}ty V, Nardini C, Stenhammar J, Marenduzzo D and Morozov A 2020
  {\em Physical Review X\/} {\bf 10} 031059

\bibitem{Gotze_PRE_Mesoscale_Simulations_2010}
G{\"o}tze I~O and Gompper G 2010 {\em Phys. Rev. E\/} {\bf 82} 041921

\bibitem{Zottl_EPJE_Simulating_Squirmers_2018}
Z{\"o}ttl A and Stark H 2018 {\em Eur. Phys. J. E\/} {\bf 41} 61

\bibitem{Buyl_N_Phoretic_Selfpropulsion_2013}
de~Buyl P and Kapral R 2013 {\em Nanoscale\/} {\bf 5} 1337--1344

\bibitem{brady1988stokesian}
Brady J~F and Bossis G 1988 {\em Annual review of fluid mechanics\/} {\bf 20}
  111--157

\bibitem{Fiore_JFM_Fast_Stokesian_2019}
Fiore A~M and Swan J~W 2019 {\em J. Fluid Mech.\/} {\bf 878} 544--597

\bibitem{Swan_PF_Modeling_Hydrodynamic_2011}
Swan J~W, Brady J~F and Moore R~S 2011 {\em Phys. Fluids\/} {\bf 23} 071901

\bibitem{Sierou_JFM_Accelerated_Stokesian_2001}
Sierou A and Brady J~F 2001 {\em J. Fluid Mech.\/} {\bf 448} 115--146

\bibitem{bonnecaze1990method}
Bonnecaze R and Brady J 1990 {\em Proceedings of the Royal Society of London.
  Series A: Mathematical and Physical Sciences\/} {\bf 430} 285--313

\bibitem{Yan_JCP_Behavior_Active_2016}
Yan W and Brady J~F 2016 {\em J. Chem. Phys.\/} {\bf 145} 134902

\bibitem{Singh_JOSS_PyStokes_Phoresis_2020}
Singh R and Adhikari R 2020 {\em J. Open Source Softw.\/} {\bf 5} 2318

\bibitem{Delmotte_JoCP_Largescale_Simulation_2015}
Delmotte B, Keaveny E~E, Plourabou{\'e} F and Climent E 2015 {\em Journal of
  Computational Physics\/} {\bf 302} 524--547

\bibitem{RojasPerez_JFM_Hydrochemical_Interactions_2021}
{Rojas-P{\'e}rez} F, Delmotte B and Michelin S 2021 {\em J. Fluid Mech.\/} {\bf
  919}

\bibitem{prieve1984motion}
Prieve D, Anderson J, Ebel J and Lowell M 1984 {\em J. Fluid Mech\/} {\bf 148}
  247--269

\bibitem{kranz2016effective}
Kranz W~T, Gelimson A, Zhao K, Wong G~C and Golestanian R 2016 {\em Physical
  review letters\/} {\bf 117} 038101

\bibitem{gelimson2016multicellular}
Gelimson A, Zhao K, Lee C~K, Kranz W~T, Wong G~C and Golestanian R 2016 {\em
  Physical review letters\/} {\bf 117} 178102

\bibitem{Houry_P_Bacterial_Swimmers_2012}
Houry A, Gohar M, Deschamps J, Tischenko E, Aymerich S, Gruss A and Briandet R
  2012 {\em Proceedings of the National Academy of Sciences\/} {\bf 109}
  13088--13093

\bibitem{Celli_P_Helicobacter_Pylori_2009}
Celli J~P, Turner B~S, Afdhal N~H, Keates S, Ghiran I, Kelly C~P, Ewoldt R~H,
  McKinley G~H, So P, Erramilli S and Bansil R 2009 {\em Proceedings of the
  National Academy of Sciences\/} {\bf 106} 14321--14326

\bibitem{Narinder_EPJE_Active_Colloids_2021}
Narinder N, Zhu W~j and Bechinger C 2021 {\em The European Physical Journal
  E\/} {\bf 44} 28

\bibitem{Lozano_NM_Active_Particles_2019}
Lozano C, {Gomez-Solano} J~R and Bechinger C 2019 {\em Nature Materials\/} {\bf
  18} 1118--1123

\bibitem{GomezSolano_PRL_Dynamics_SelfPropelled_2016}
{Gomez-Solano} J~R, Blokhuis A and Bechinger C 2016 {\em Physical Review
  Letters\/} {\bf 116} 138301

\bibitem{Choudhary_JFM_NonNewtonian_Effects_2020}
Choudhary A, Renganathan T and Pushpavanam S 2020 {\em Journal of Fluid
  Mechanics\/} {\bf 899}

\bibitem{ercolessi1994interatomic}
Ercolessi F and Adams J~B 1994 {\em EPL\/} {\bf 26} 583--588

\bibitem{wang2019machine}
Wang J, Olsson S, Wehmeyer C, P{\'e}rez A, Charron N~E, De~Fabritiis G, No{\'e}
  F and Clementi C 2019 {\em ACS Cent. Sci.\/} {\bf 5} 755--767

\bibitem{durumeric2019adversarial}
Durumeric A~E and Voth G~A 2019 {\em J. Chem. Phys.\/} {\bf 151} 124110

\bibitem{Rein_EPJE_Applicability_Effective_2016}
Rein M and Speck T 2016 {\em Eur. Phys. J. E\/} {\bf 39} 84

\bibitem{Mognetti_PRL_Living_Clusters_2013}
Mognetti B~M, {\v S}ari{\'c} A, {Angioletti-Uberti} S, Cacciuto A, Valeriani C
  and Frenkel D 2013 {\em Phys. Rev. Lett.\/} {\bf 111} 245702

\bibitem{Alarcon_SM_Morphology_Clusters_2017}
Alarc{\'o}n F, Valeriani C and Pagonabarraga I 2017 {\em Soft Matter\/} {\bf
  13} 814--826

\bibitem{Redner_PRE_Reentrant_Phase_2013}
Redner G~S, Baskaran A and Hagan M~F 2013 {\em Phys. Rev. E\/} {\bf 88} 012305

\bibitem{Prymidis_SM_Selfassembly_Active_2015}
Prymidis V, Sielcken H and Filion L 2015 {\em Soft Matter\/} {\bf 11}
  4158--4166

\bibitem{Bauerle_NC_Selforganization_Active_2018}
B{\"a}uerle T, Fischer A, Speck T and Bechinger C 2018 {\em Nat Commun\/} {\bf
  9} 3232

\end{thebibliography}

\end{document}